\documentclass[11pt,a4paper]{article}
\pdfoutput=1

\usepackage{graphicx}
\usepackage{float}
\usepackage{afterpage}
\usepackage{epsfig,cite}
\usepackage{amssymb}
\usepackage{amsmath}
\usepackage{dsfont}
\usepackage{multirow}
\usepackage{url}
\usepackage{hyperref}
\usepackage{booktabs}
\usepackage{rotating}
\usepackage{morefloats}
\usepackage{xcolor}
\usepackage{graphicx}
\usepackage{lipsum}

\newcommand{\be}{\begin{equation}}
\newcommand{\ee}{\end{equation}}
\newcommand{\lp}{\left(}
\newcommand{\rp}{\right)}
\def\gsim{\mathrel{\rlap{\lower4pt\hbox{\hskip1pt$\sim$}}
    \raise1pt\hbox{$>$}}}         
\def\lsim{\mathrel{\rlap{\lower4pt\hbox{\hskip1pt$\sim$}}
    \raise1pt\hbox{$<$}}}         

\begin{document}

\begin{flushright}
  OUTP-16-15P\\
  Nikhef/2016-046\\
  Cavendish-HEP-16/17\\
  Aachen-TTK-16-49\\
\end{flushright}
\vspace{.3cm}

\begin{center}
{\LARGE\bf Pinning down the large-$x$ gluon with\\ 
NNLO top-quark pair differential distributions\\}
\vspace{0.5cm}
Micha{\l}~Czakon$^{1}$, Nathan~P.~Hartland$^{3,4}$, Alexander~Mitov$^{5}$, Emanuele~R.~Nocera$^{2}$\\ and Juan~Rojo$^{3,4}$\\
\vspace{0.2cm}
       {\it
         ~$^{1}$Institut f\"ur Theoretische Teilchenphysik und Kosmologie,\\
         RWTH Aachen University, D-52056 Aachen, Germany \\[3pt]
         ~$^{2}$Rudolf Peierls Centre for Theoretical Physics, 1 Keble Road\\
         University of Oxford, OX1 3NP, Oxford, United Kingdom\\[3pt]
         ~$^{3}$ Department of Physics and Astronomy, VU University Amsterdam,\\
         De Boelelaan 1081, NL-1081, HV Amsterdam, The Netherlands\\[3pt]
         ~$^{4}$Nikhef, Science Park 105, NL-1098 XG Amsterdam, The Netherlands\\[3pt]
         ~$^{5}$Cavendish Laboratory, University of Cambridge, Cambridge CB3 0HE, United Kingdom
       }\\
       \vspace{1.5cm}

{\bf \large Abstract}
\end{center}

Top-quark pair production at the LHC is directly 
sensitive to the gluon PDF at large $x$.
While total cross-section data is already included in several PDF 
determinations, differential distributions are not, because the
corresponding NNLO calculations
have become available only recently.
In this work we study the impact on the large-$x$ gluon of top-quark pair differential 
distributions measured by ATLAS and CMS 
at $\sqrt{s}=8$ TeV.
Our analysis, performed in the NNPDF3.0 
framework at NNLO accuracy, allows us to identify the optimal combination of LHC top-quark pair 
measurements that maximize the constraints on the gluon, as well as
to assess the compatibility between ATLAS and  CMS data.
We find that differential distributions from top-quark pair production provide significant
constraints on the large-$x$ gluon, comparable to those
obtained from inclusive jet production data, and thus should become
an important ingredient for the next generation of global PDF fits.

\clearpage

\tableofcontents

\section{Introduction}
\label{sec:intro}

The accurate determination of the parton distribution functions (PDFs) of the 
proton~\cite{Forte:2010dt,Ball:2012wy,Forte:2013wc,Butterworth:2015oua}
is an essential requirement for the precision phenomenology 
program at the Large Hadron Collider (LHC).
Traditionally, the bulk of the available experimental information on PDFs
came from deep-inelastic scattering (DIS) and fixed-target Drell-Yan (DY) data.
In recent years, however, the data from the LHC has
provided a wealth of new information on the structure of the 
proton, see {\it e.g.}~\cite{Rojo:2015acz} and references therein.
LHC measurements on inclusive electroweak vector boson  
and jet production are routinely included in most of the modern PDF
determinations~\cite{Ball:2014uwa,Alekhin:2013nda,Dulat:2015mca,
  Harland-Lang:2014zoa}.
Furthermore, several dedicated analyses have demonstrated
the constraining power of many other LHC processes, including 
 $W$+charm production~\cite{Stirling:2012vh,Chatrchyan:2013uja},
the transverse momentum distribution of $W$ and $Z$
bosons~\cite{Malik:2013kba,Chatterjee:2016sxt},
prompt photons~\cite{d'Enterria:2012yj} and charm
production in the forward region~\cite{Gauld:2015yia,Zenaiev:2015rfa,Cacciari:2015fta,Gauld:2016kpd}.

In the case of top-quark pair production, the next-to-next-to leading order 
(NNLO) QCD corrections to the total cross-section were computed in 
2013~\cite{Czakon:2013goa,Czakon:2012pz,Baernreuther:2012ws}.
This development allowed for the consistent inclusion
of the Tevatron and LHC inclusive top-quark pair measurements
into a NNLO global PDF fit~\cite{Czakon:2013tha}, which demonstrated
how this data could help reducing the rather sizable uncertainty of the 
gluon PDF for $x\gsim 0.1$.
With this motivation, top-quark total cross-sections were included in the 
latest updates of some PDF fits, specifically 
NNPDF3.0 and MMHT14
(see also the ABM12 fit for related studies).

Last year, the calculation of NNLO corrections to inclusive 
top-quark pair production was extended to fully differential
distributions for stable
tops~\cite{Czakon:2015owf,Czakon:2016dgf,Czakon:2014xsa,Czakon:2016ckf}.
It is therefore natural to investigate how the constraints
upon the large-$x$ gluon PDF obtained from inclusive measurements
are improved once the additional information contained in the differential
distributions is accounted for in a global NNLO analysis
(see~\cite{Guzzi:2014wia}
for a first study based on approximate NNLO).
Such a program is enabled by the availability of
precision measurements 
of top-quark pair differential cross-sections at $\sqrt{s}=8$ TeV  
from ATLAS~\cite{Aad:2015mbv} and CMS~\cite{Khachatryan:2015oqa},
provided with the full information on the breakdown
of experimental statistical and systematic uncertainties.

Given that DIS structure functions and DY production 
provide only a rather loose constraint upon the gluon PDF, particularly at
large $x$, inclusive jet production data has been traditionally used to
obtain additional information~\cite{Rojo:2014kta}.
While the NLO QCD corrections to jet
production at hadron colliders
have been available for more than two
decades~\cite{Ellis:1990ek,Ellis:1992en,Nagy:2001fj},
the corresponding NNLO corrections (in the leading
color approximation) have been computed
only very recently~\cite{Currie:2016bfm},
building on the partial results of Refs.~\cite{Ridder:2013mf,Currie:2013dwa}.
Since the results of~\cite{Currie:2016bfm} are not yet publicly available,
in this work we will not include collider jet data,
so that we can use exact NNLO theory for all the processes included in the global PDF fit.

The PDF constraints provided by
the ATLAS and CMS top-quark differential distributions at $\sqrt{s}=8$ TeV
will be investigated
by means of the NNPDF global analysis framework~\cite{Ball:2014uwa,Ball:2016neh}.
For the baseline PDF fit, the input dataset will be
largely the same as in NNPDF3.0, with two
main differences: the HERA-II structure function
data from H1 and ZEUS has been replaced by the final
HERA combination~\cite{Abramowicz:2015mha,Rojo:2015nxa}
and inclusive jet production measurements
from CDF, ATLAS and CMS have been excluded.
In order to achieve the computational speed required for the PDF fit,
we generate theoretical calculations of NLO top-quark pair
production with {\tt Sherpa}~\cite{Gleisberg:2008ta}
interfaced to {\tt MCgrid}~\cite{DelDebbio:2013kxa} and dynamical scales as in Ref.~\cite{Czakon:2016dgf}.
These NLO calculations
are then supplemented with NNLO/NLO bin-by-bin $K$-factors
consistently derived using the theory settings of~\cite{Czakon:2016dgf}.

Including the LHC differential distributions from top-quark pair production
into the NNPDF global analysis allows us to quantitatively tackle
a number of important issues.
In particular, we investigate the compatibility between the ATLAS and CMS measurements;
how the constraints provided by the differential measurements compare to
those obtained from inclusive cross-sections;
whether it is advantageous to use normalized or absolute
distributions; and which is the optimal combination of LHC top-quark
measurements to be included in the global PDF fit.
We then demonstrate how differential
distributions from top-quark pair production lead to a significant reduction
of the gluon PDF uncertainty at large $x$,
and that their impact is comparable to that obtained
from inclusive jet measurements.
The resulting improved gluon will have direct beneficial
implications for searches of new physics beyond the Standard Model (BSM) in final states
involving top quarks and in general for gluon-initiated processes.

This paper is organized as follows. In Sect.~\ref{sec:data} we describe
the available LHC top-quark
pair production data and the treatment
of their experimental uncertainties.
In Sect.~\ref{sec:comparison} we discuss the
calculation of the NNLO theoretical predictions for
top-quark pair differential cross-sections
and provide a systematic 
comparison between them and the LHC data.
In Sect.~\ref{sec:results} we present NNLO fits including top-quark
differential distributions, assess the agreement between data and theory, and
discuss their impact in the determination of the large-$x$ gluon.
In Sect.~\ref{sec:summary} we summarize and comment on possible
future developments.
Further investigations on the compatibility between the ATLAS and CMS
data are presented in appendix~\ref{sec:compatibility}.

\section{Experimental data}
\label{sec:data}

In this section we describe the top-quark pair production
data from ATLAS and CMS that  will be used as input in the PDF fit.
First, we describe the features of the various differential distributions 
available, including a comparison between absolute and normalized measurements,
and then  we review the total inclusive cross-sections that will be
included alongside the normalized differential distributions.

\subsection{Top-quark pair differential distributions from the LHC}

In this work we consider the most recent differential cross-section measurements on
top-quark pair production at $\sqrt{s}=8$  TeV from ATLAS~\cite{Aad:2015mbv}
and CMS~\cite{Khachatryan:2015oqa} in the lepton+jets final state.
These datasets correspond to an integrated luminosity 
of $20.3$ fb$^{-1}$ and $19.7$ fb$^{-1}$, respectively.
In this channel, the $t\bar{t}$ pair is reconstructed from its decays 
into $W^+bW^-\bar{b}$, with one $W$ boson decaying hadronically and the other into
an electron or muon and the associated neutrino.
We do not consider earlier measurements at
$\sqrt{s}=7$ TeV~\cite{Aad:2012hg,Aad:2014zka,Aad:2015eia,Chatrchyan:2012saa},
which are affected by larger uncertainties and are not provided
with the full breakdown of systematic error sources.

The ATLAS and CMS top-quark production measurements of Refs.~\cite{Khachatryan:2015oqa,Aad:2015mbv}
are
provided in both the fiducial phase space,
with observables reconstructed in terms of final-state leptonic
and jet variables,
and in the full phase space, in terms of the top or 
top-pair kinematic variables.
In our analysis, we are restricted to using the latter as the NNLO calculations 
are available only for stable top quarks.
Ongoing work into extending 
these calculations
to include top-quark decays will eventually overcome this restriction.
Among the available distributions,
we will focus on the 
transverse momentum $p_T^t$ and the rapidity $y_t$ of the top quark or 
antiquark, and on the rapidity $y_{t\bar{t}}$ and the invariant mass $m_{t\bar{t}}$ 
of the top-quark pair system.
We will not consider
the transverse momentum
of the top-quark pair $p_T^{t\bar{t}}$, for which a complete NNLO theoretical 
description is not available.
The binning and kinematical cuts for each distribution
are the same in the ATLAS and CMS measurements,
a feature which simplifies the
benchmarking of results between the two experiments and their
comparison with the theoretical predictions.

In Table~\ref{tab:unc} we summarize the features of each kinematical
distribution, indicating whether it is an absolute or
a normalized distribution; which of the correlated systematic
errors are common; the number of
data points $N_{\rm dat}$; and their kinematic coverage.
All systematic uncertainties are treated as multiplicative, and
absolute distributions share 
the luminosity uncertainty across each experiment.
Moreover, the absolute distributions from CMS also
share the same systematic uncertainties of their corresponding normalized 
distributions (see below).
Wherever asymmetric uncertainties are provided, they are 
symmetrized according to~\cite{DelDebbio:2004xtd}.
To avoid double counting, for each experiment
only one of the distributions listed in Table~\ref{tab:unc} 
can be included in a PDF fit, due to
the unavailability of the statistical correlations between 
different distributions within the same experiment.
 One of the goals of this study is therefore to identify the
 combination of ATLAS and CMS top-quark pair 
 measurements that maximizes the constraints on the gluon.

\begin{table}[t]
  \footnotesize
\centering
\begin{tabular}{ll|c|c|c}
\toprule
Exp. & 
Dataset & 
Sys. Unc. & 
$N_{\rm dat}$ & Kinematics \\[0.05cm]
\midrule
\multirow{8}{*}{ATLAS} & ATLAS $d\sigma/dp_T^t$
      & a
      & 8
      & $0<p_T^t<500$ GeV \\[0.05cm]
      & ATLAS $d\sigma/d|y_t|$
           & a
      & 5
      & $0<|y_t|<2.5$  \\[0.05cm]
      & ATLAS $d\sigma/d|y_{t\bar{t}}|$
           & a
      & 5
      & $0<|y_{t\bar{t}}|<2.5$  \\[0.05cm]
      & ATLAS $d\sigma/dm_{t\bar{t}}$
           & a
      & 7
      & $345<m_{t\bar{t}}<1600$ GeV  \\[0.05cm]
      & ATLAS $(1/\sigma)d\sigma/dp_T^t$
           & 
      & 8
      & $0<p_T^t<500$ GeV \\[0.05cm]
      & ATLAS $(1/\sigma)d\sigma/d|y_t|$
           &  
      & 5
      & $0<|y_t|<2.5$  \\[0.05cm]
      & ATLAS $(1/\sigma)d\sigma/d|y_{t\bar{t}}|$
           & 
      & 5
      & $0<|y_{t\bar{t}}|<2.5$  \\[0.05cm]
      & ATLAS $(1/\sigma)d\sigma/dm_{t\bar{t}}$
           & 
      & 7
      & $345<m_{t\bar{t}}<1600$ GeV \\[0.05cm]
\midrule
\multirow{8}{*}{CMS} & CMS $d\sigma/dp_T^t$
           & b, f
      & 8
      & $0<p_T^t<500$ GeV \\[0.05cm]
      & CMS $d\sigma/dy_t$
           & c, f
      & 10
      & $-2.5<y_t<2.5$  \\[0.05cm]
      & CMS $d\sigma/dy_{t\bar{t}}$
           & d, f
      & 10
      & $-2.5<y_{t\bar{t}}<2.5$  \\[0.05cm]
      & CMS $d\sigma/dm_{t\bar{t}}$
           & e, f
      & 7
      & $345<m_{t\bar{t}}<1600$ GeV \\[0.05cm]
      & CMS $(1/\sigma)d\sigma/dp_T^t$
           & b
      & 8
      & $0<p_T^t<500$ GeV \\[0.05cm]
      & CMS $(1/\sigma)d\sigma/dy_t$
           & c 
      & 10
      & $-2.5<y_t<2.5$  \\[0.05cm]
      & CMS $(1/\sigma)d\sigma/dy_{t\bar{t}}$
           & d
      & 10
      & $-2.5<y_{t\bar{t}}<2.5$  \\[0.05cm]
      & CMS $(1/\sigma)d\sigma/dm_{t\bar{t}}$
        & e
      & 7
      & $345<m_{t\bar{t}}<1600$ GeV \\[0.05cm]
\bottomrule
\end{tabular}
\caption{\small The ATLAS and
  CMS 
  top-quark pair distributions at $\sqrt{s}=8$ TeV
  used in this work.
  For each distribution we indicate the number of data points 
  and their kinematic  coverage.
  In the second column, distributions that are labeled with the same
  letter have common experimental systematic uncertainties.
}
\label{tab:unc}
\end{table}

In addition to these 8 TeV lepton+jets kinematical
distributions,
ATLAS and CMS have presented other differential measurements
of top-quark pair production.
To begin with, differential distributions at $\sqrt{s}=8$
TeV for the dilepton final state are
available from both ATLAS and
CMS~\cite{Khachatryan:2015oqa,Aaboud:2016iot},
which in the latter case are also presented in the form of
 double-differential
normalized cross-sections~\cite{CMS-PAS-TOP-16-007}.
In addition,
measurements of differential distributions of high-$p_T$ boosted top quarks
from ATLAS~\cite{Aad:2015hna} and CMS~\cite{Khachatryan:2016gxp} at
$\sqrt{s}=8$ TeV  have also been published,
although their interpretation requires an
assessment of electroweak corrections~\cite{Pagani:2016caq}.
Finally, results on differential distributions
at $\sqrt{s}=13$ TeV in the
lepton+jets channel  by ATLAS~\cite{ATLAS-CONF-2016-040}
and in the dilepton~\cite{CMS-PAS-TOP-16-007}
and lepton+jets channels~\cite{Khachatryan:2016mnb} from CMS
are now also available.
In this first exploratory work, we concentrate on the most precise data available,
the lepton+jets distributions at 8 TeV, but future studies should include
also these other available top-quark differential measurements.

\subsection{Absolute versus normalized distributions}

As indicated in Table~\ref{tab:unc},
 ATLAS and CMS have presented their measurements
of top-quark pair differential distributions in two different ways.
  In the first case, each distribution is
  normalized to the sum over the cross-sections in each bin, in a way that
  it then integrates to one by construction.
  This procedure is motivated by the
  partial cancellation of uncertainties, such as the luminosity, 
  that takes place in the ratio.
  However, some PDF-sensitive information describing the overall normalization 
  of the gluon PDF might be
  lost in this procedure.
  In order to compensate for this,
  the PDF fit should include both total
  inclusive cross-sections and normalized differential distributions.
  Typically, the mutual correlation between
  the two is small and can be neglected.
  
  On the other hand, top-quark pair
  differential measurements are also
  provided as absolute distributions.
  In this case, experimental uncertainties are larger than
  in the normalized case, though this way one also maintains
  a handle on the overall magnitude of the gluon.
  Note that for absolute distributions, the
  simultaneous inclusion of total and differential
  measurements would result in a double counting.
  While constraints arising from the use
  of either normalized distributions supplemented with total
  cross-sections or absolute distributions
  should be equivalent, it turns out
  that the former are somewhat
  more stringent than the latter
  (see Sect.~\ref{sec:results}).
  
ATLAS  has released measurements for both
normalized and absolute distributions, and provided the corresponding 
full breakdown of systematic uncertainties separately.
The former are affected by an additional $2.8\%$ fully correlated uncertainty 
from the integrated luminosity at 8 TeV.
The CMS measurements are available only for the normalized
distributions, from which the absolute differential
distributions can be reconstructed  by means of the
corresponding total cross-section measurement~\cite{CMS:2015uoa}.
  In this procedure,
statistical uncertainties from the normalized 
distribution and the total inclusive cross-section are added in quadrature.
Two additional sources of systematics are retained on the absolute
differential distribution, which originate respectively from the total 
systematic and the luminosity 
uncertainties of the inclusive cross-section.\footnote{
  We thank the conveners of the CMS Top Quark Physics group
  for providing us with this recommendation.}

\subsection{Total inclusive cross-section measurements}

The LHC measurements of normalized top-quark pair differential
distributions benefit from reduced experimental uncertainties
as compared to their absolute counterparts, but
consequently they might also lose some sensitivity on the overall magnitude of the gluon.
It is therefore important to 
supplement the normalized distributions included
in the PDF fits with the corresponding measurements
of the inclusive cross-section in order to obtain a complete picture. 

\begin{table}[t]
\footnotesize
\centering
\begin{tabular}{lccccc}
\toprule
Exp. &  $\sqrt{s}$ [TeV] & 
Fin. st. & $\mathcal{L}$ [fb$^{-1}$]& 
$\sigma^{\rm tot}(t\bar{t})$ [pb] & Ref. \\[0.1cm]
\midrule
\multirow{3}{*}{ATLAS} & 7  & $l$+jets & 4.6  &
$182.9 \pm 3.1~{\rm (stat)} \pm 4.2~{\rm (sys)}
\pm 3.6~{\rm (lumi)} \pm 3.3~{\rm (bm)}$
&  \cite{Aad:2014kva} \\[0.1cm]
& {\bf 8} & {\bf $\bf\it l$+jets}    & {\bf 20.3} &
$\bf 242.4 \pm 1.7~{\rm \bf (stat)} \pm 5.5~{\rm \bf (sys)}
\pm 7.5~{\rm \bf (lumi)} \pm 4.2~{\rm \bf (bm)}$
&  \cite{Aad:2014kva} \\[0.1cm]
& 13 &  $l$+jets & 3.2 &  
$818 \pm 8~{\rm (stat)} \pm 27~{\rm (sys)} \pm 19~{\rm (lumi)} \pm 12~{\rm (bm)}$
& \cite{Aaboud:2016pbd}\\[0.1cm]
\midrule
\multirow{3}{*}{CMS} & 7  &   $l$+jets & 5.0 &
$173.6 \pm 2.1~{\rm (stat)}^{+4.5}_{-4.0}~{\rm (sys)} \pm 3.8~{\rm (lumi)}$
&\cite{Khachatryan:2016mqs}  \\[0.1cm]
& {\bf 8} &
{\bf $\bf \it l$+jets} & {\bf 19.7}   &
{\bf $\bf 244.9 \pm 1.4~{\rm \bf (stat)} ^{+6.3}_{-5.5}~{\rm
    \bf (sys)} \pm 6.4~{\rm  \bf (lumi)}$}
&\cite{Khachatryan:2016mqs} \\[0.1cm]
& 13 &  $l$+jets & 2.2 &  
$792 \pm 8~{\rm (stat)} \pm 37~{\rm (sys)} \pm 21~{\rm (lumi)}$
& \cite{Khachatryan:2016kzg}\\[0.1cm]
\bottomrule
\end{tabular}
\caption{\small Summary of the
  most precise ATLAS and CMS measurements of the total inclusive
  $t\bar{t}$ cross-sections at 7, 8 and 13 TeV.
  We indicate the final state,
  the integrated luminosity, the breakdown of statistical
  and systematic uncertainties (where  ``lumi'' stands for the
  luminosity  and ``bm'' stands for
  the beam energy).
  The measurements in boldface are those
  used in the fits of this work.
}
\label{tab:uncTotalXsec}
\end{table}

In Table~\ref{tab:uncTotalXsec} we collect the results for the
most precise ATLAS and CMS measurements of the total inclusive
top-quark pair cross-section at various center-of-mass energies.
In each case, we indicate the final state,
the integrated luminosity, the value of the total
cross-section with the breakdown of statistical
and systematic uncertainties (where ``lumi'' stands for the 
luminosity and ``bm'' stands for
the beam energy), and the corresponding publication reference.
These measurements (with the exception of the 13 TeV measurement) have a total 
experimental uncertainty of only a few percent.
The 8 TeV cross-sections are notably limited by the luminosity uncertainty,
which amounts to 2.8\% and 2.6\% for ATLAS and CMS respectively.

As a general rule,
in a global fit it is advantageous to include as many PDF-sensitive
observables as possible.
In the particular case of fits including top-quark production data,
one should then add all the 
total cross-sections listed in 
Table~\ref{tab:uncTotalXsec} as well as available measurements in
 other final states.
However, one of the aims of this work is to compare the impact 
on the large-$x$ gluon of top-quark pair production at 8 TeV,
arising from either absolute distributions or from the normalized
ones supplemented with the corresponding total cross-sections.
To perform such a comparison consistently, we include here only
the total cross-sections at 8 TeV from the lepton+jets final state, 
highlighted in boldface in Table~\ref{tab:uncTotalXsec}.
Therefore, in the following,  whenever one of the ATLAS or CMS normalized
differential listed in Table~\ref{tab:unc} is included in the PDF
fit, it will always be supplemented by the corresponding 8 TeV total cross-section 
from Table~\ref{tab:uncTotalXsec}.

\section{Comparison between NNLO theory and LHC data}
\label{sec:comparison}

In this section, first we provide details on the theory settings
used for the  calculation of NNLO differential
distributions in top-quark pair production.
Then we perform a qualitative comparison between the 
predictions obtained from various NNLO PDF sets and the 8 TeV ATLAS and CMS
data, for both absolute and normalized distributions.
Finally, we quantify these comparisons by means of
a $\chi^2$ estimator.

\subsection{Differential top-quark pair production at NNLO}
\label{sec:theory}

The calculation of the NNLO QCD corrections to differential distributions 
in top-quark pair production has been recently
completed~\cite{Czakon:2015owf,Czakon:2016dgf}.
This
calculation is however not yet 
available in a format suitable for its direct inclusion
during a global fit, which requires the evaluation
of hadronic cross-sections for different input PDFs a
large number of times during the minimization procedure.
Therefore, in order to include this data into the global NNLO PDF fit,
we begin by using fast calculations of NLO matrix elements with
NNLO DGLAP evolution and $\alpha_s(Q)$ running.
These fast NLO calculations are
based upon precomputing the partonic matrix elements
in such a way that the 
standard numerical convolution with generic input PDFs can be reliably 
approximated by means of interpolation techniques.

There exist two main frameworks for the implementation of this fast convolution procedure,
{\tt APPLgrid}~\cite{Carli:2010rw} and {\tt FastNLO}~\cite{Wobisch:2011ij}.
In this work we will utilize the  former, which has been interfaced to various
codes of common use for calculations in PDF fits, such
as {\tt NLOjet++}~\cite{Nagy:2003tz},
{\tt MCFM}~\cite{Campbell:2002tg},
{\tt MadGraph5\_aMC@NLO}/{\tt aMCfast}~\cite{Alwall:2014hca,amcfast}
and {\tt SHERPA}~\cite{Gleisberg:2008ta}.
In particular, here we will use {\tt SHERPA} 
interfaced to {\tt APPLgrid} using
the {\tt MCgrid} code~\cite{DelDebbio:2013kxa} and
the {\tt Rivet}~\cite{Buckley:2010ar} analysis package,
with {\tt OpenLoops}~\cite{Cascioli:2011va} for the NLO matrix elements.
The calculations have been performed with Monte Carlo
integration statistics sufficiently large in order to ensure that the residual
fluctuations are at the few permille level at most.
The NLO {\tt SHERPA}/{\tt MCgrid} results have been benchmarked with the
corresponding calculation using the code of~\cite{Czakon:2016dgf},
finding excellent agreement for all kinematic distributions.

An important aspect of the NNLO calculation
is the choice of central renormalization
and factorization scales, $\mu_R$ and $\mu_F$.
Following Ref.~\cite{Czakon:2016dgf},
the following optimized settings are adopted
in this work.
For the differential distributions in the top (or anti-top) quark rapidity 
$y_t$ and in the top-quark pair rapidity $y_{t\bar{t}}$ and invariant mass 
$m_{t\bar{t}}$ we use
\be
\label{eq:scale1}
\mu_R=\mu_F=\mu=H_T/4 \, , \qquad H_T \equiv \sqrt{m_t^2+\lp {p_T^t}\rp^2}
+ \sqrt{m_t^2+\lp {p_T^{\bar{t}}}\rp^2} \, ,
\ee
where $m_t=173.3$ GeV is the PDG world average for the
top-quark pole mass~\cite{Agashe:2014kda},
and $p_T^t$ ($p_T^{\bar{t}}$) is the top (anti-top) transverse momentum.
For the top-quark transverse momentum distribution,
constructed
from the average of the
distributions for the top and the anti-top quarks,
it has been
found that the optimal choice of dynamical scales for the former
case is
\be
\label{eq:scale12}
\mu_R^\prime=\mu_F^\prime=\mu^\prime=\sqrt{m_t^2+\lp {p_T^t}\rp^2}\Big/2 \, ,
\ee
with an analogous expression for anti-top quarks (replacing
$p_T^t$ by $p_T^{\bar{t}}$).
This scale choice
leads to an improvement in the convergence of the perturbative series.

The resulting NLO calculations are then
supplemented by bin-by-bin
$\mathcal{C}$-factors~\cite{Ball:2011uy}, defined as
the ratio of the NNLO to NLO calculations, 
\begin{equation}
\mathcal{C}
=
\frac{\widetilde{\sigma}^{\rm nnlo}\otimes \mathcal{L}^{\rm nnlo}}
{\widetilde{\sigma}^{\rm nlo}\otimes \mathcal{L}^{\rm nnlo}}
\,\mbox{,}
\label{eq:cfacts}
\end{equation} 
where $\widetilde\sigma^{\rm nnlo}$ ($\widetilde\sigma^{\rm nlo}$) is the partonic 
cross-section computed with NNLO (NLO) matrix elements and $\mathcal{L}^{\rm nnlo}$ is the
corresponding parton luminosity evaluated with a reference set of NNLO PDFs.
The numerator and the denominator in Eq.~(\ref{eq:cfacts}) were 
computed with the code of~\cite{Czakon:2016dgf}.

\begin{figure}[t]
\centering
\includegraphics[scale=0.25,angle=270]{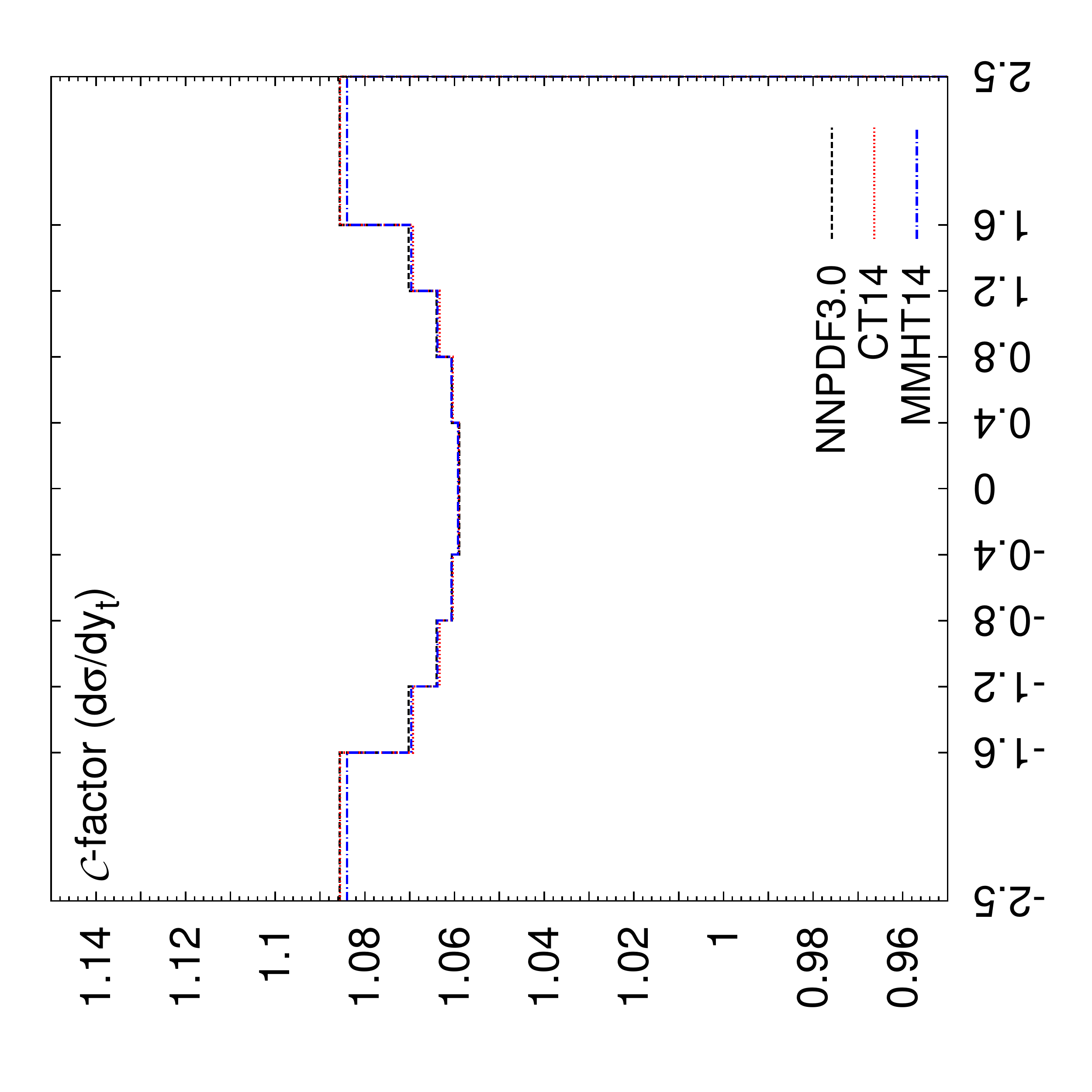}
\includegraphics[scale=0.25,angle=270]{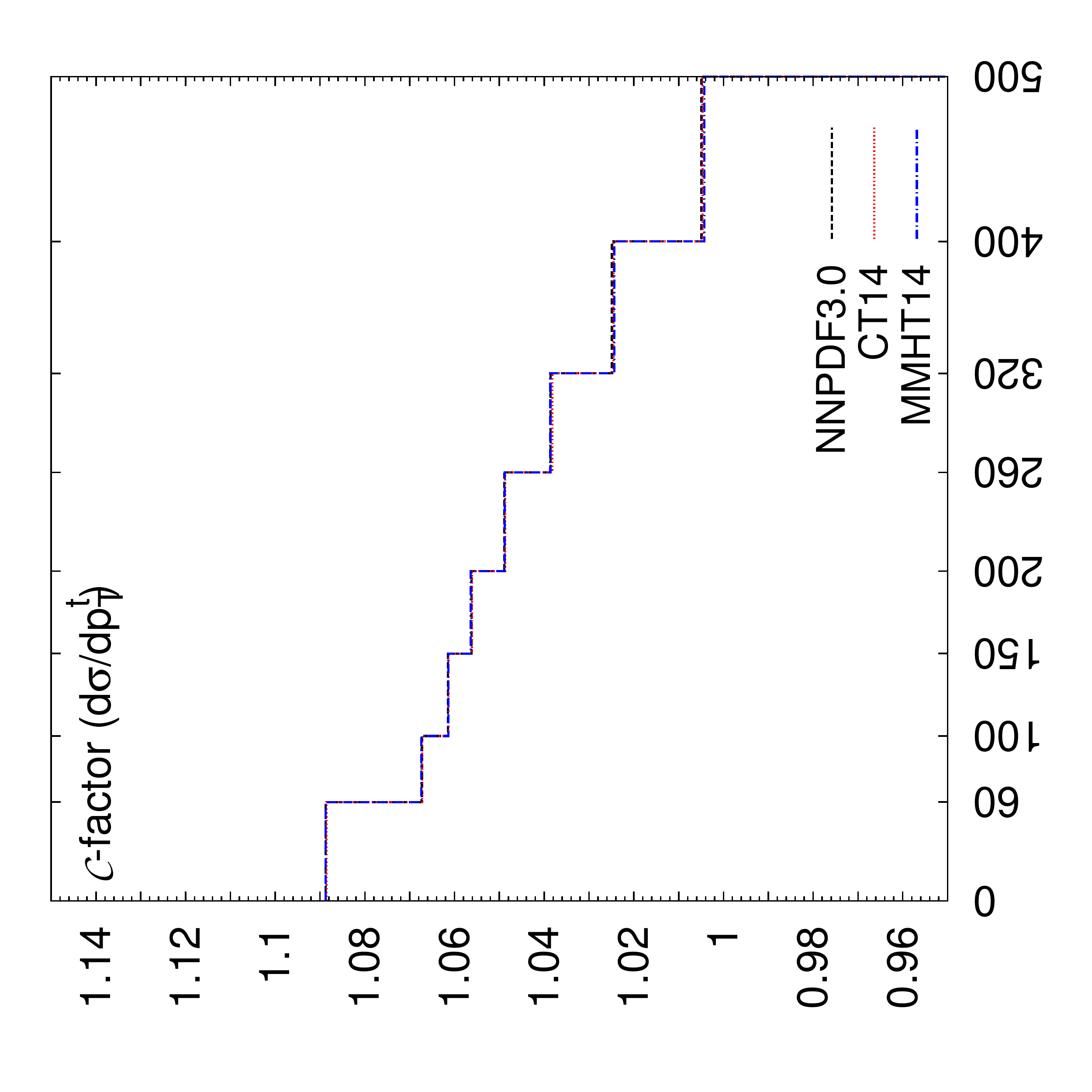}\\
\includegraphics[scale=0.25,angle=270]{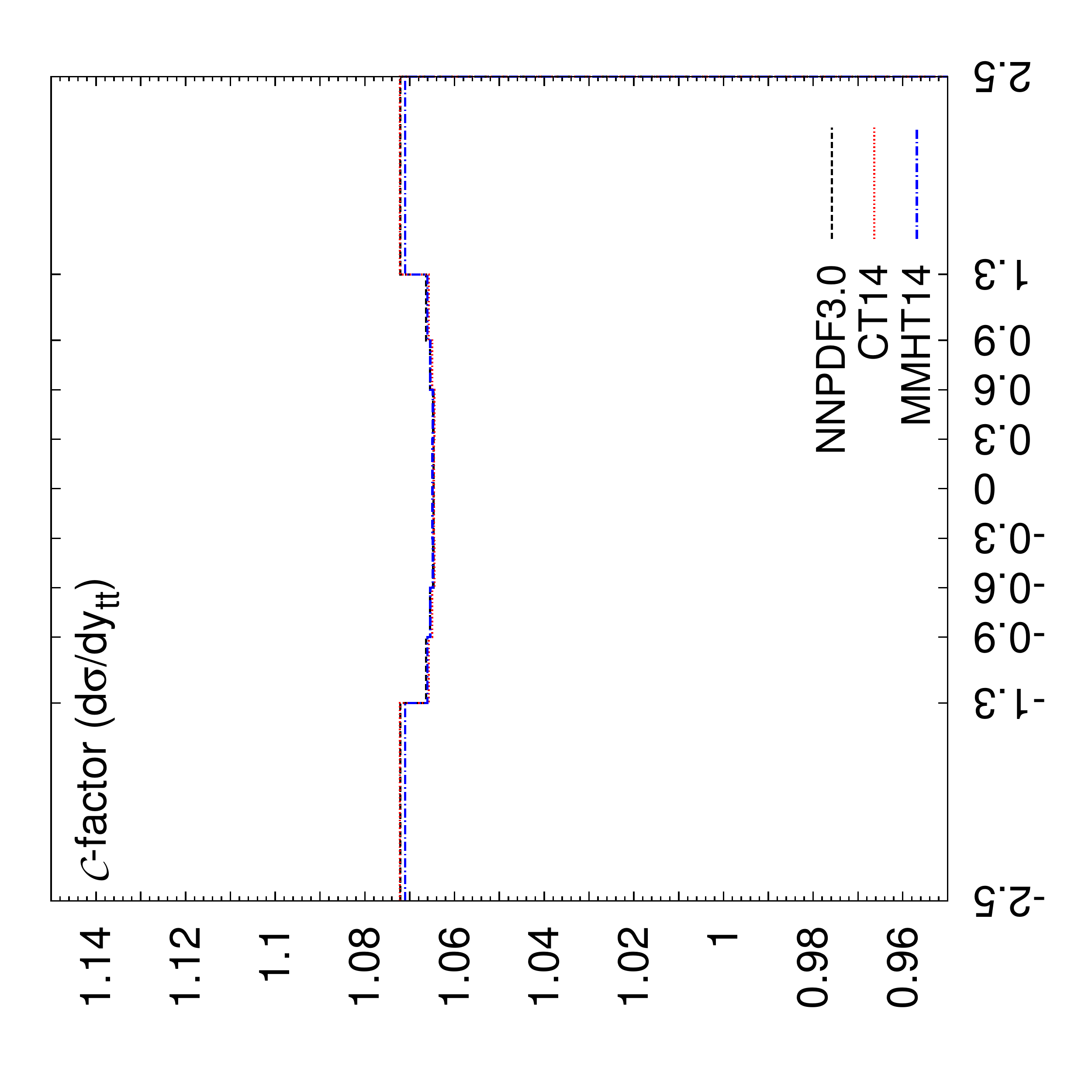}
\includegraphics[scale=0.25,angle=270]{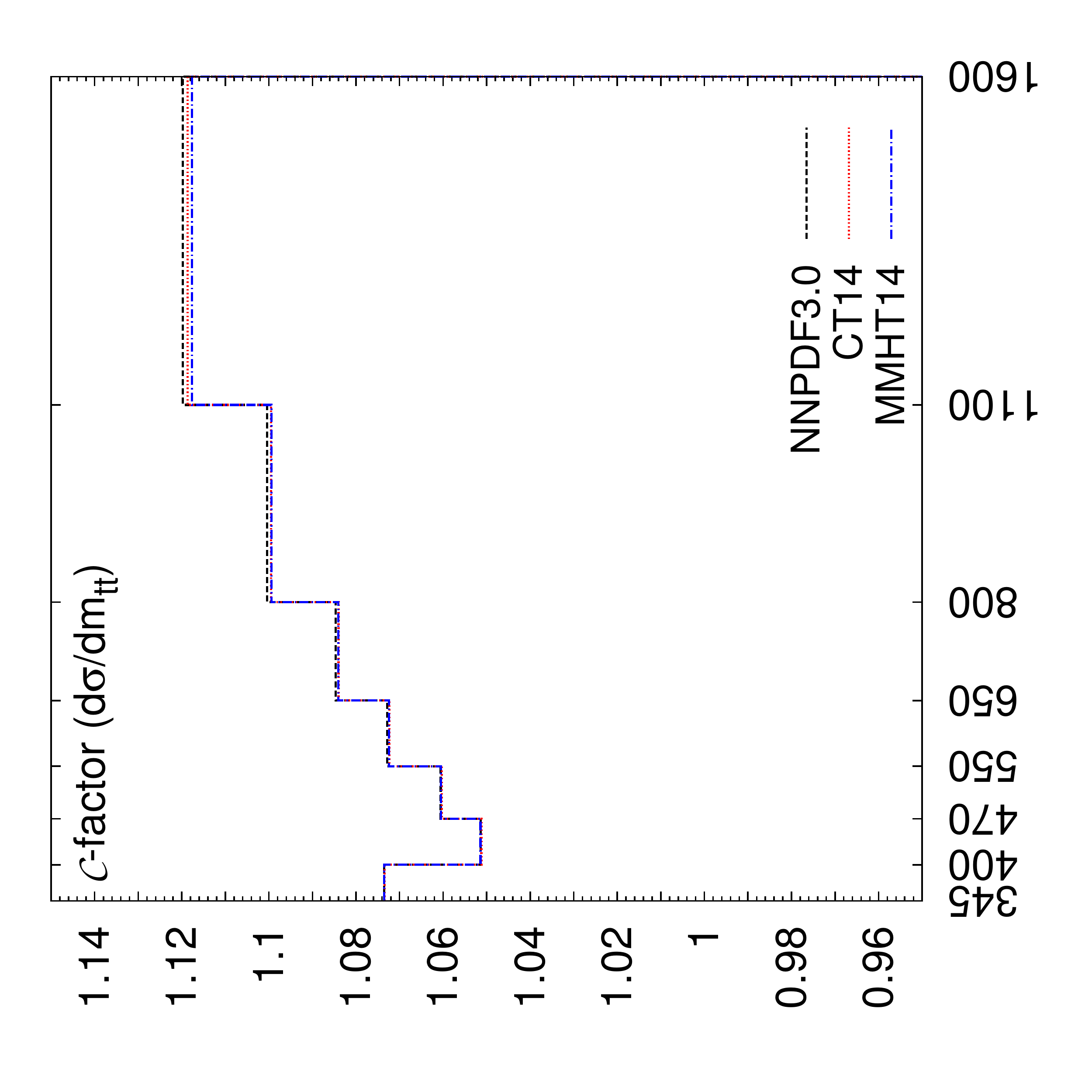}
\caption{\small The $\mathcal{C}$-factors Eq.~(\ref{eq:cfacts})
  for the four absolute differential distributions of
  Table~\ref{tab:unc}.
}
\label{fig:cfacts1}
\end{figure}

In Fig.~\ref{fig:cfacts1} we compare
the $\mathcal{C}$-factors computed with NNPDF3.0~\cite{Ball:2014uwa}, CT14~\cite{Dulat:2015mca} 
and MMHT2014~\cite{Harland-Lang:2014zoa}
for the absolute differential distributions in the following four variables:
the top quark rapidity $y_t$ and transverse momentum $p_T^t$,
and the top-pair rapidity $y_{t\bar{t}}$ 
and invariant mass $m_{t\bar{t}}$.
The binning here is the same
as that of the
ATLAS and CMS 8 TeV measurements listed in Table~\ref{tab:unc}.
 We find that
  the dependence of the $\mathcal{C}$-factors on the input
  PDF set is very small and can be safely neglected.
  In the case of the  $y_t$ and  $y_{t\bar{t}}$
  distributions, we find NNLO corrections 
  of between 6\% and 9\%, reasonably flat
  in the data region.
  For the $p_T^t$ distribution, the $\mathcal{C}$-factor decreases from 1.09
  at low transverse momentum to close to unity for $p_T^t\simeq 500$ GeV.
  For the
  invariant mass $m_{t\bar{t}}$,
  the $\mathcal{C}$-factor increases from 5\% at
  low masses to around 12\% above 1 TeV.

We note that, exactly as for the corresponding experimental measurements,
all NNLO distributions have been normalized with respect to the cross-section integrated 
over the considered kinematic range.
In other words, by construction, the integral of any normalized distribution over its kinematic range 
is unity.

As shown in Ref.~\cite{Czakon:2016dgf}, the integration of the differential distributions 
computed with the optimal dynamical scales Eqs.~(\ref{eq:scale1})--(\ref{eq:scale12}) 
returns a total cross-section which is about 2\% higher than the NNLO one 
from {\tt top++}~\cite{Czakon:2011xx}, and in close agreement with the 
NNLO+NNLL {\tt top++} result (recall that the total cross-section in {\tt top++} 
is computed with fixed scales $\mu_R=\mu_F=m_t$).
For this reason, 
when adding the inclusive cross-section data into PDF fits, it is more appropriate 
to compute the theory prediction with {\tt top++} at NNLO+NNLL.
Nonetheless, in the present work the total inclusive top-pair cross-section and 
corresponding $\mathcal{C}$-factors are  computed using {\tt top++} at NNLO.
As explained in Sect.~\ref{sec:global}, and given the exploratory nature of the 
present work, this choice is adequate since the overall impact of the total 
cross-sections on the global fits turns out to be small and this 
2\% difference is thus inconsequential for our study.

\begin{figure}[t]
\centering
\includegraphics[scale=0.39,clip=true]{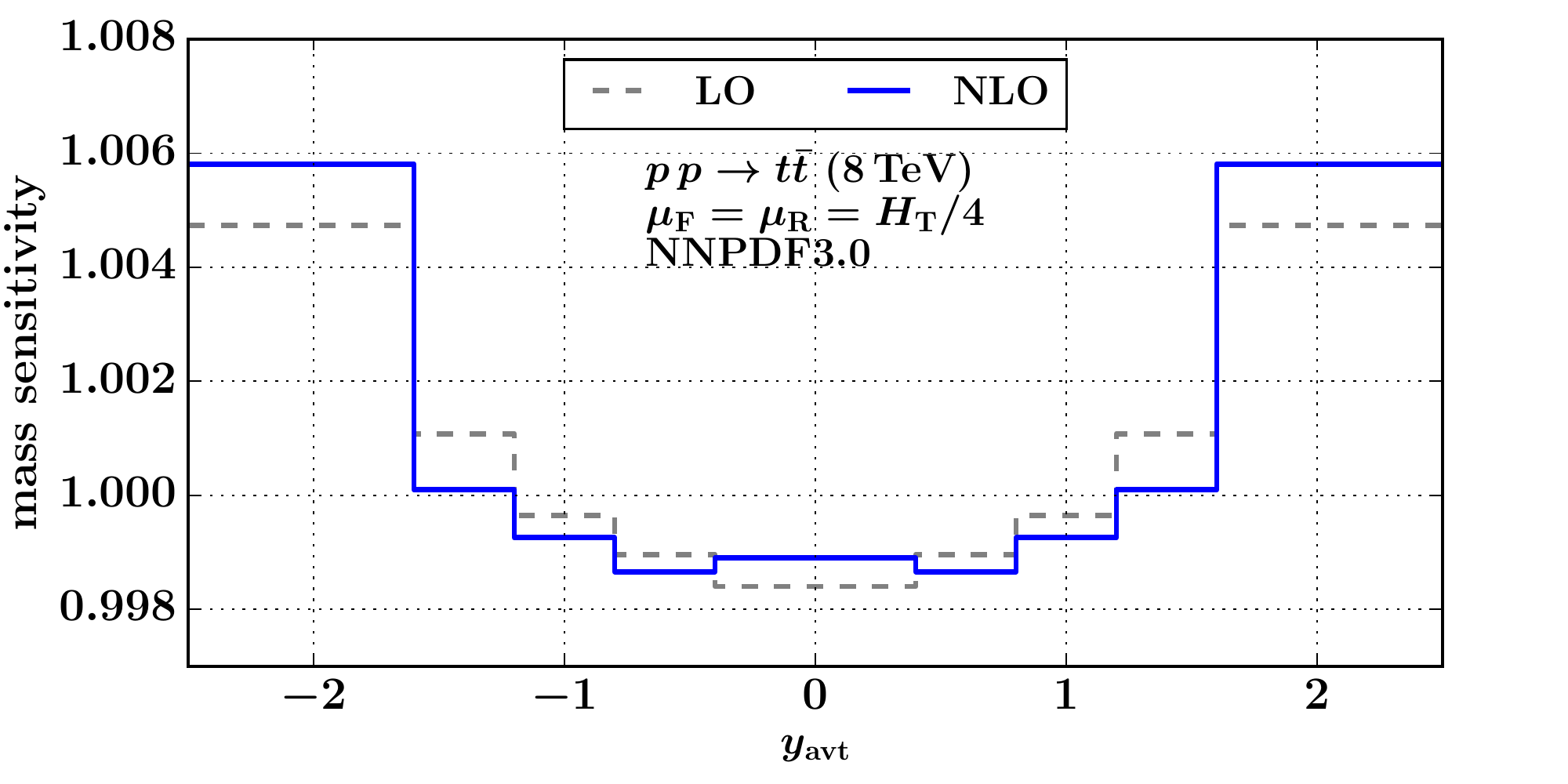}
\includegraphics[scale=0.39,clip=true]{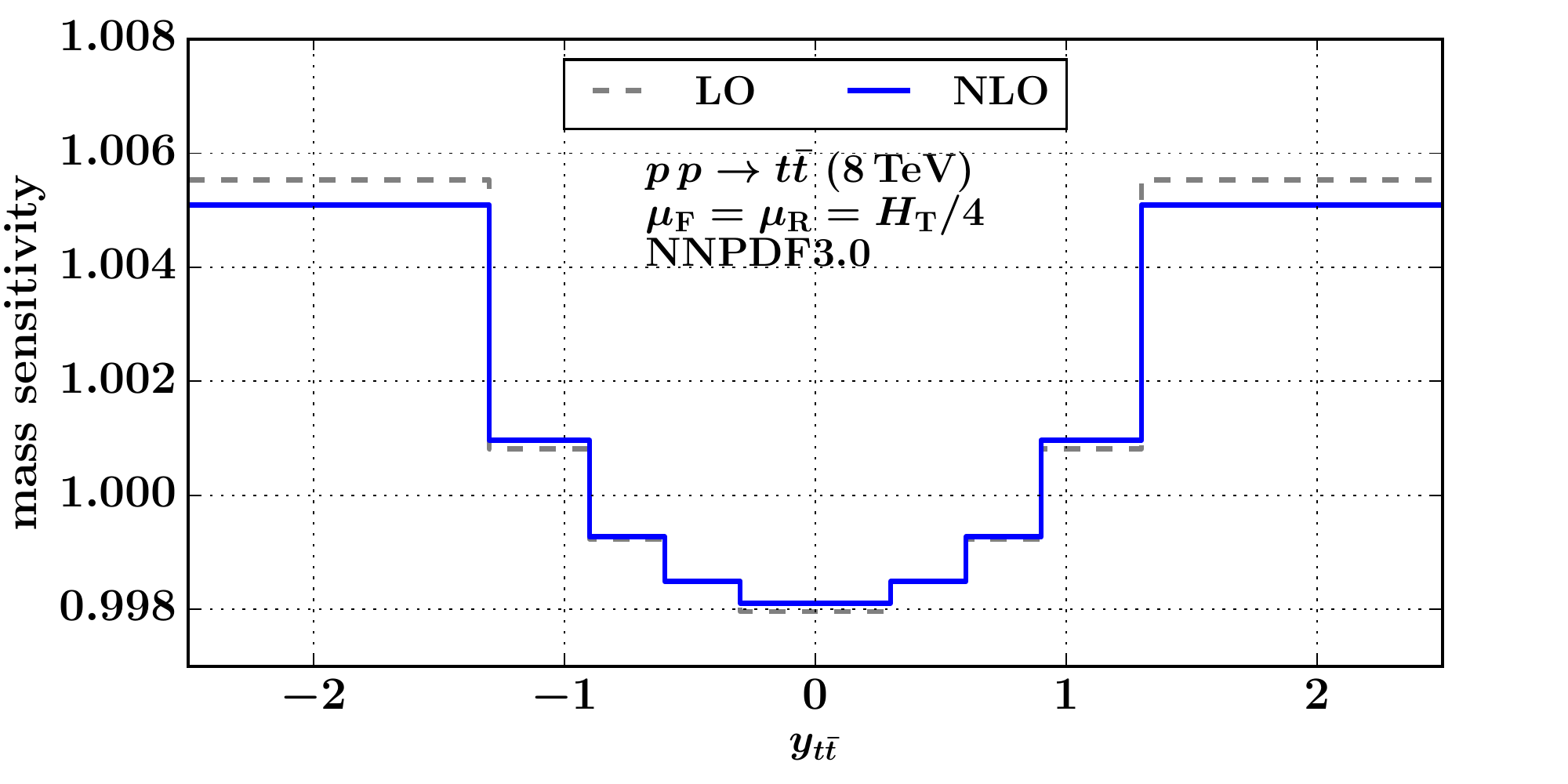}
\caption{\small The ratio between $m_t=172.3$ GeV and
  $m_t=173.3$ GeV (``mass sensitivity'') at LO and NLO
  for the normalized $y_t$ (left) and $y_{t\bar{t}}$ (right) distributions
  at 8 TeV, computed using NNPDF3.0.
}
\label{fig:mtsensitivity}
\end{figure}

The theoretical uncertainties due to the value of $m_t$ deserve special attention.
As mentioned above, in this work we use the PDG average of  $m_t=173.3$ GeV.
The significant spread among the individual measurements contributing to this average, however, suggests that in the future a shift in $m_t$ of up to $ \Delta m_{t}\simeq 1$ GeV, or even more, may be possible.
The sensitivity upon variations of $m_t$ of the four top-quark differential distributions considered here
has been studied in~\cite{Czakon:2016vfr}.
Shape modifications are pronounced in the $m_{t\bar{t}}$ and $p_T^t$ distributions, especially close to the threshold.
On the other hand, the $y_t$ and $y_{t\bar{t}}$ distributions exhibit a much reduced $m_t$ dependence.

To quantify this mass sensitivity, in Fig.~\ref{fig:mtsensitivity} we show the ratio between $m_t = 172.3$ GeV and the PDG average
$m_t = 173.3$ GeV for the LO and NLO normalized $y_t$ and $y_{t\bar{t}}$ distributions at 8 TeV.
We find that these two distributions are very stable upon a shift of $m_t$ by 1 GeV, varying at most by 0.6\%, which is much less than the experimental uncertainties or other sources of theory uncertainty such as PDFs and missing higher orders.
This robustness of the normalized $y_t$ and $y_{t\bar{t}}$ distributions with respect to $m_t$ variations is, therefore, an
important motivation in favour of using them as input to the PDF fits (see Sect.~\ref{sec:impact}).

\begin{figure}[t]
\centering
\includegraphics[scale=0.26,angle=270,clip=true]{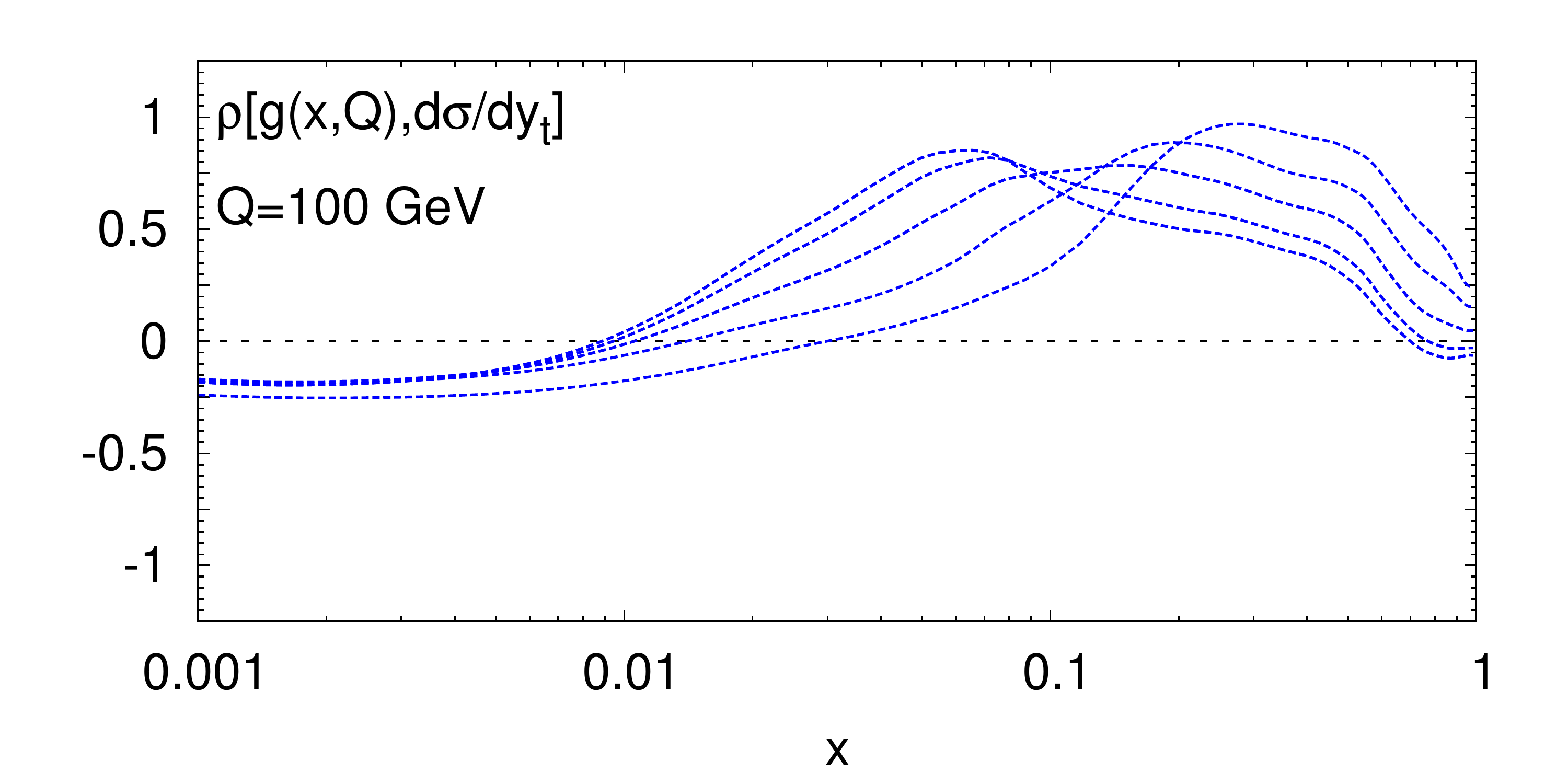}
\includegraphics[scale=0.26,angle=270,clip=true]{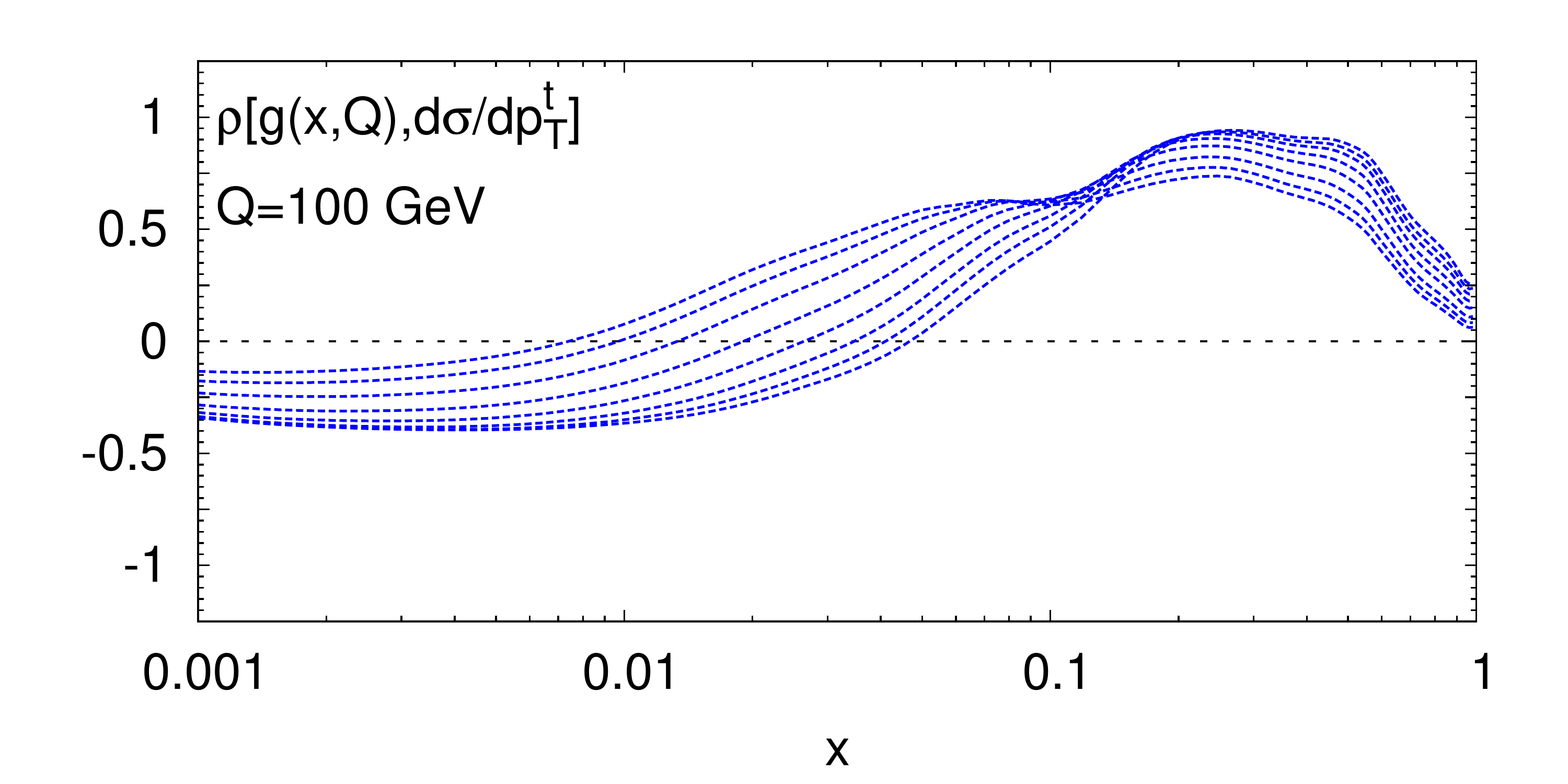}\\
\includegraphics[scale=0.26,angle=270,clip=true]{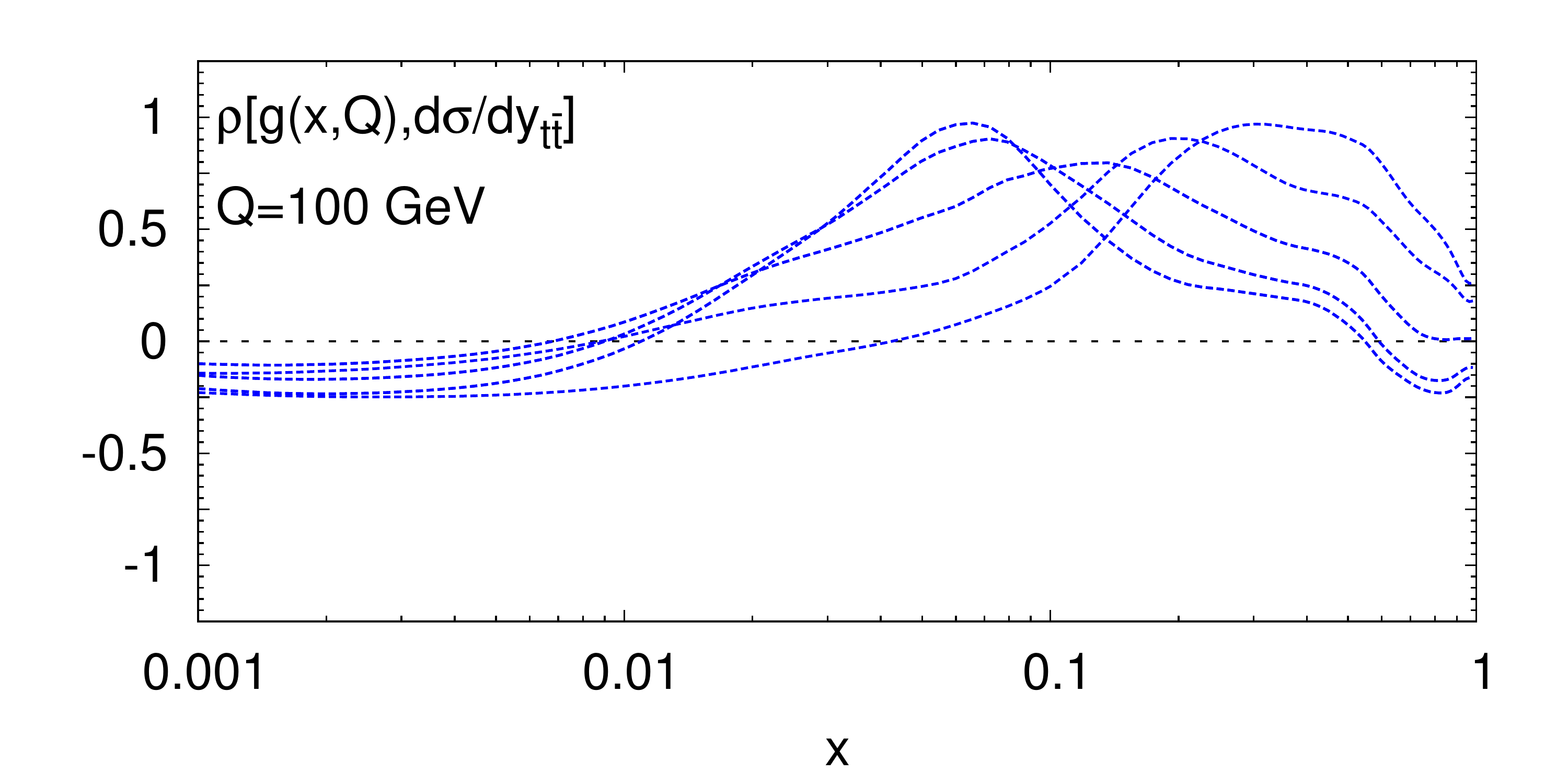}
\includegraphics[scale=0.26,angle=270,clip=true]{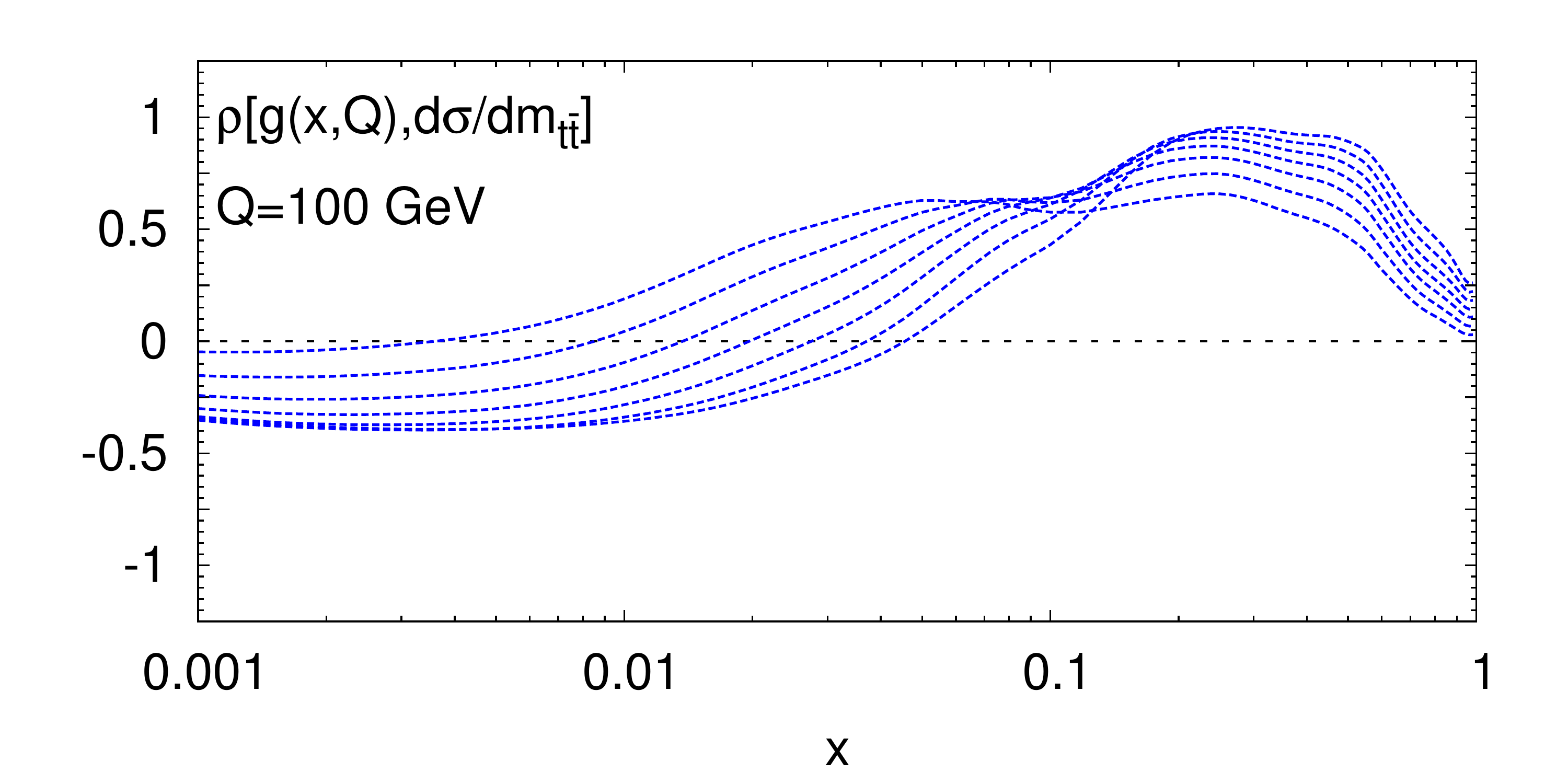}
\caption{\small The correlation coefficient $\rho$ between the gluon $g(x,Q^2)$,
  evaluated at $Q=100$ GeV, and each of the bins of the 
   $y_t$, $p_T^t$, $y_{t\bar{t}}$
  and $m_{t\bar{t}}$ top-quark
  differential distributions at the LHC 8 TeV.
}
\label{fig:correlations}
\end{figure}
\begin{figure}[t]
\centering
\includegraphics[scale=0.26,angle=270,clip=true]{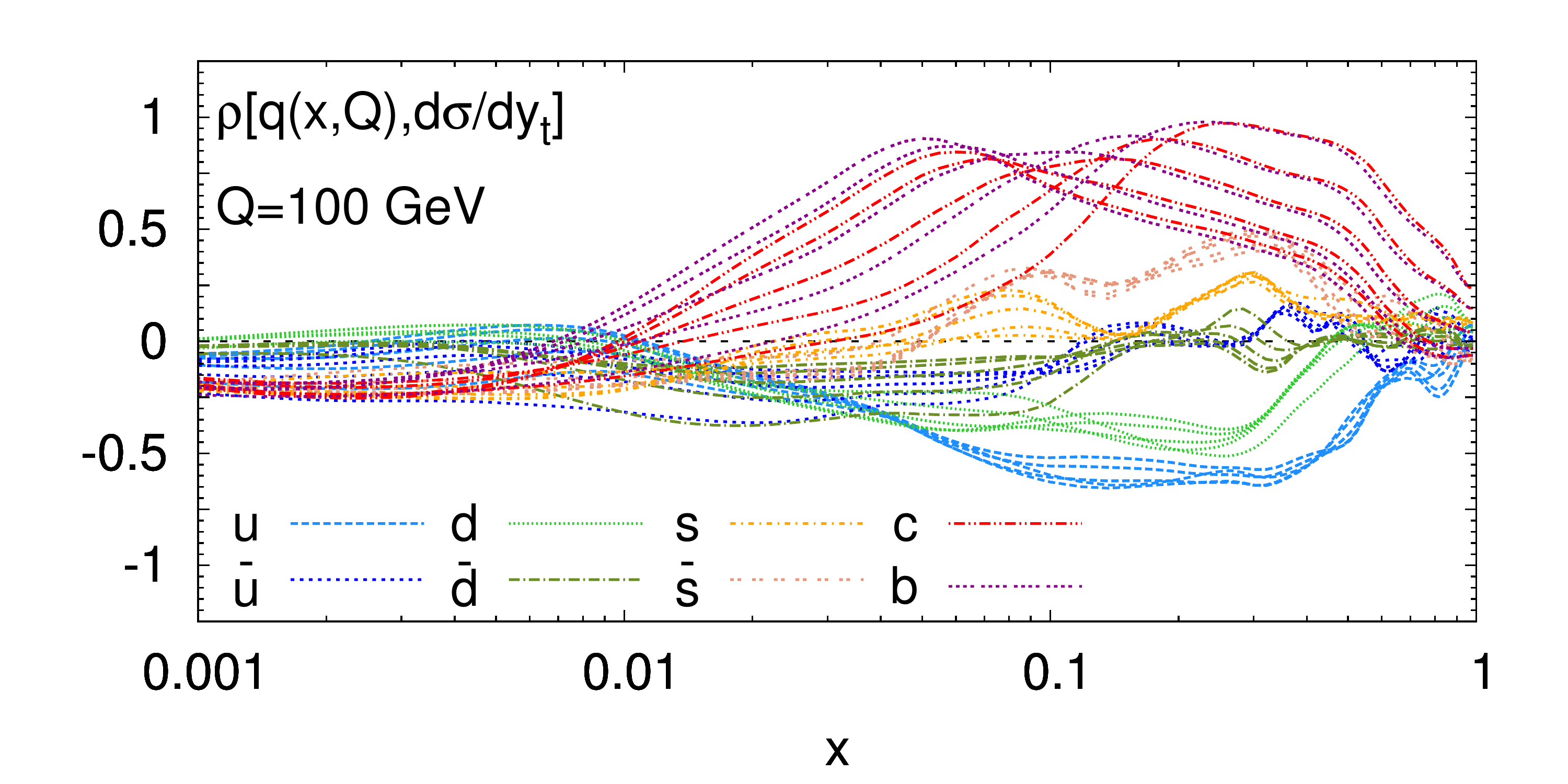}
\includegraphics[scale=0.26,angle=270,clip=true]{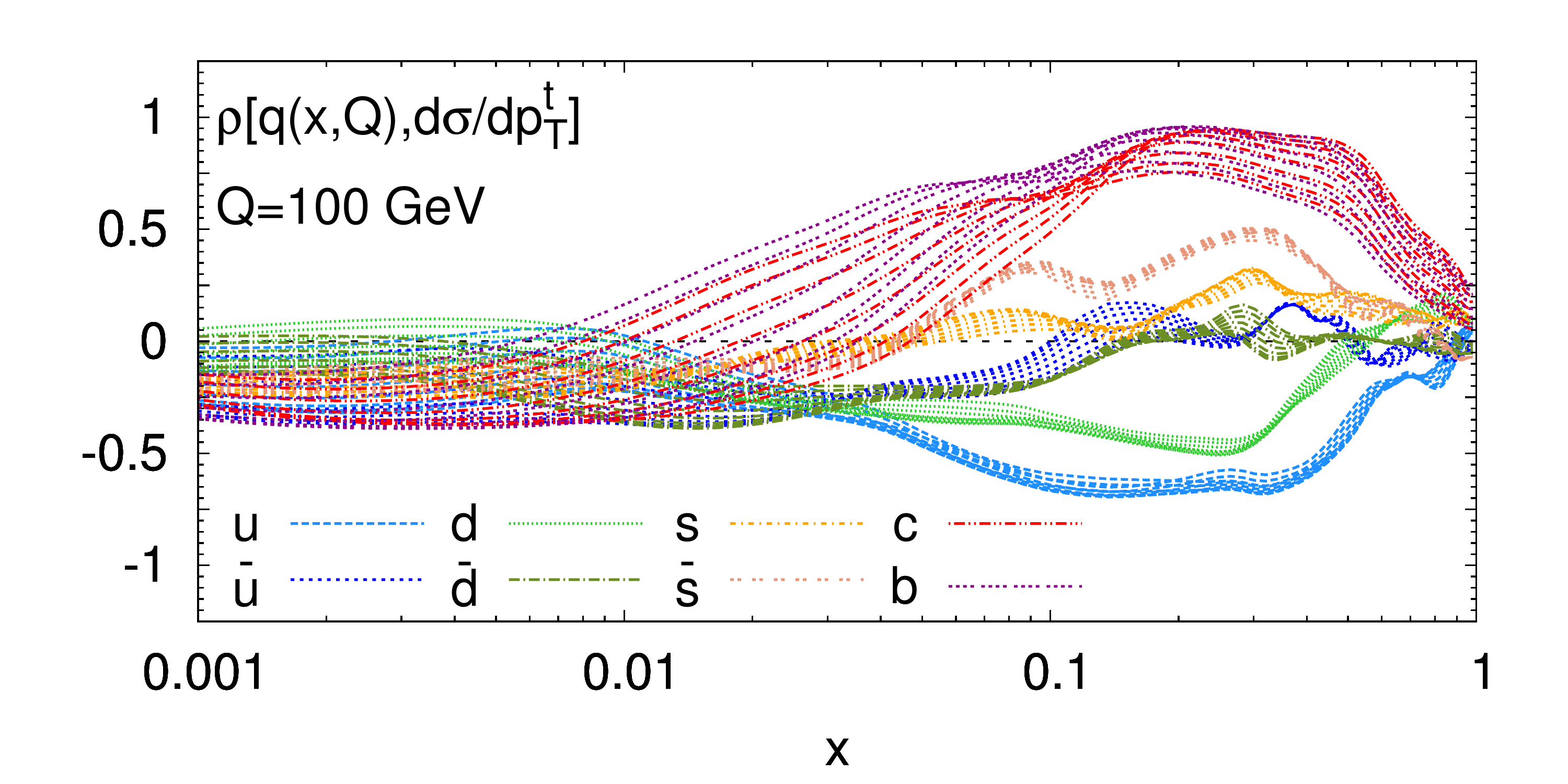}\\
\includegraphics[scale=0.26,angle=270,clip=true]{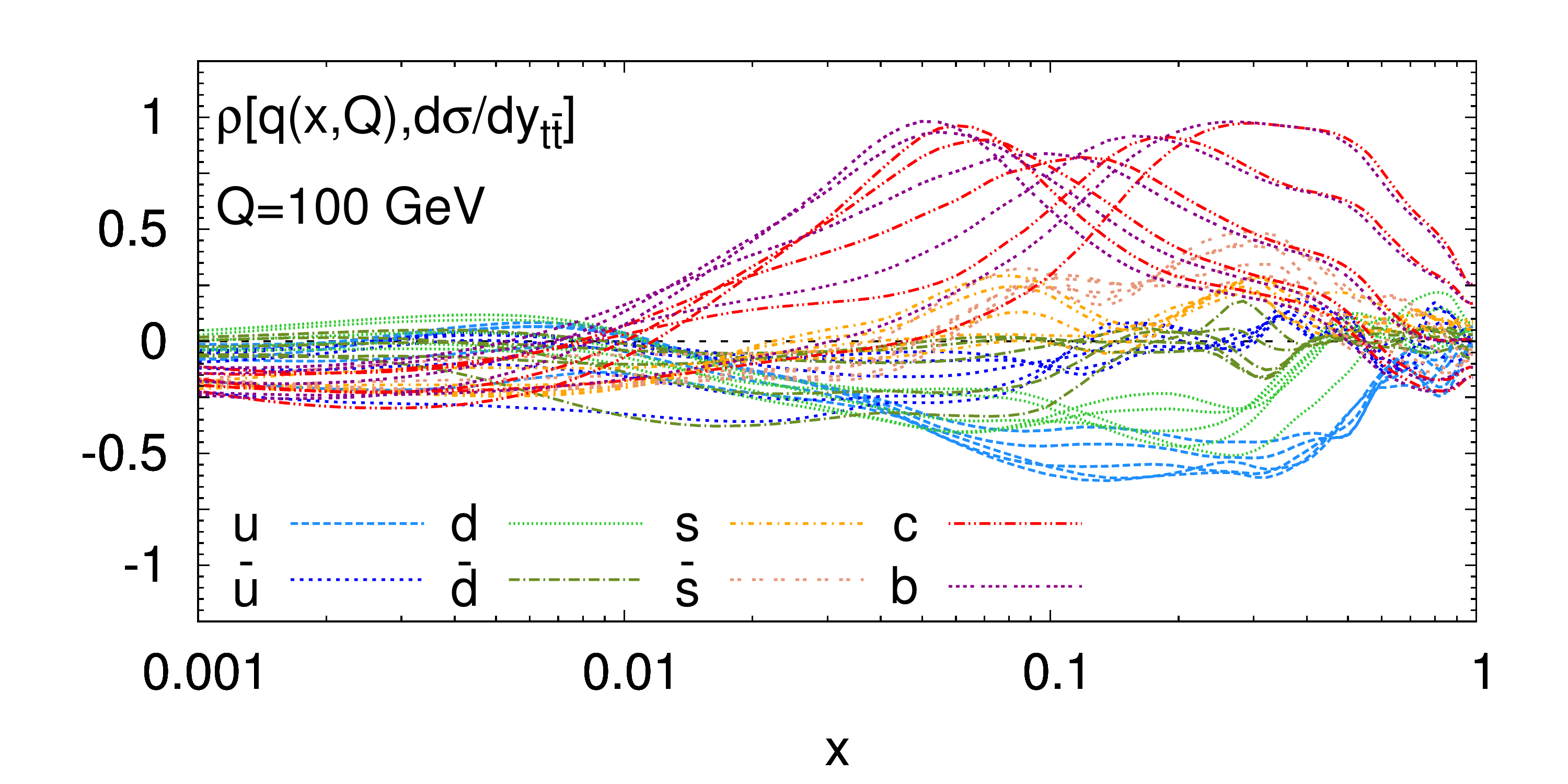}
\includegraphics[scale=0.26,angle=270,clip=true]{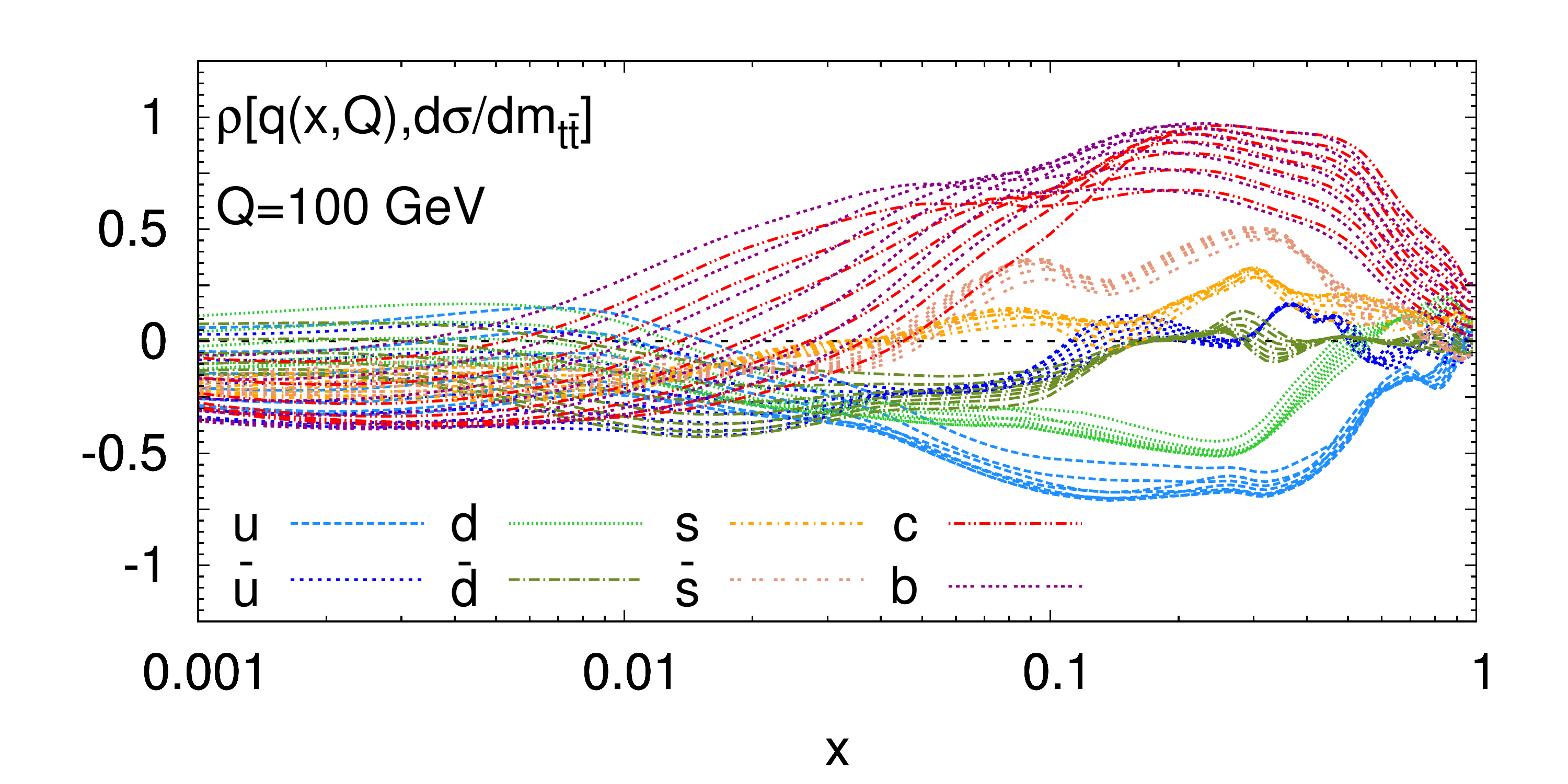}
\caption{\small Same as Fig.~\ref{fig:correlations}, but for quarks and 
antiquarks, $q(x,Q^2)$, $q=u,\bar{u},d,\bar{d},s,\bar{s},c,b$.}
\label{fig:correlations_quarks}
\end{figure}

The region of $x$ for which the LHC differential top data are sensitive to 
the various PDF flavours can be quantified by computing the correlation 
coefficient $\rho$ between them and each of the bins of a
given differential distribution~\cite{Demartin:2010er,Ball:2011mu}.
Large values of $|\rho|$ indicate regions in $x$ where the top-quark
data provide direct sensitivity to each PDF flavour.
These correlations are shown in Fig.~\ref{fig:correlations},
for the gluon $g(x,Q^2)$, and in Fig.~\ref{fig:correlations_quarks}, 
for quarks $q(x,Q)$, $q=u,\bar{u},d,\bar{d},s,\bar{s},c,b$.
PDFs are evaluated at $Q=100$ GeV from the NNPDF3.0 NNLO set.
In the case of the gluon, we find that already for $x\gsim 0.05$ the 
correlation coefficient can be larger than 0.5, while it peaks in the region 
between $x\simeq 0.08$ and $x\simeq 0.5$, depending on the specific bin and 
kinematical distribution. 
A similar trend is observed for the charm and bottom quarks, as a consequence 
of the fact that they are generated radiatively through the gluon splitting
in a quark-antiquark pair.
In the case of light quarks and antiquarks, moderate correlations are 
observed for $u$ and $d$, while correlations are almost negligible for 
$\bar{u}$, $\bar{d}$, $s$ and $\bar{s}$. 
As we will show in Sect.~\ref{sec:results}, top-quark data will mostly
constrain the gluon, and, as a consequence, the radiatively generated charm 
and bottom quarks, in the $x$ region where the correlation coefficient 
$|\rho|$ is larger, roughly $0.08\lesssim x \lesssim 0.5$.

\subsection{Comparison with the ATLAS and CMS differential distributions}

\begin{figure}[t]
\centering
\includegraphics[scale=0.31,angle=270,clip=true,trim=2cm 0cm 2cm 0cm]{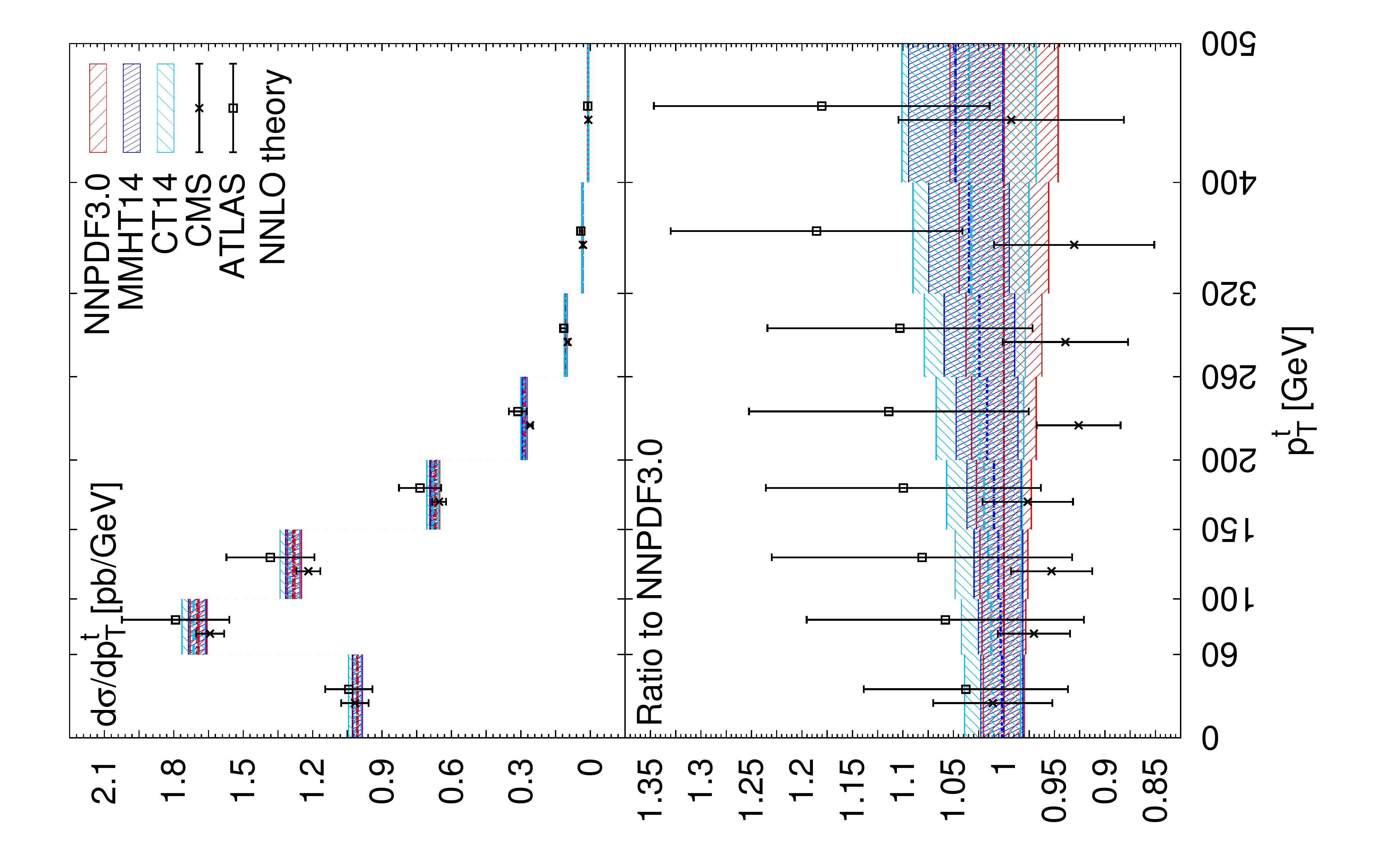}
\includegraphics[scale=0.31,angle=270,clip=true,trim=2cm 0cm 2cm 0cm]{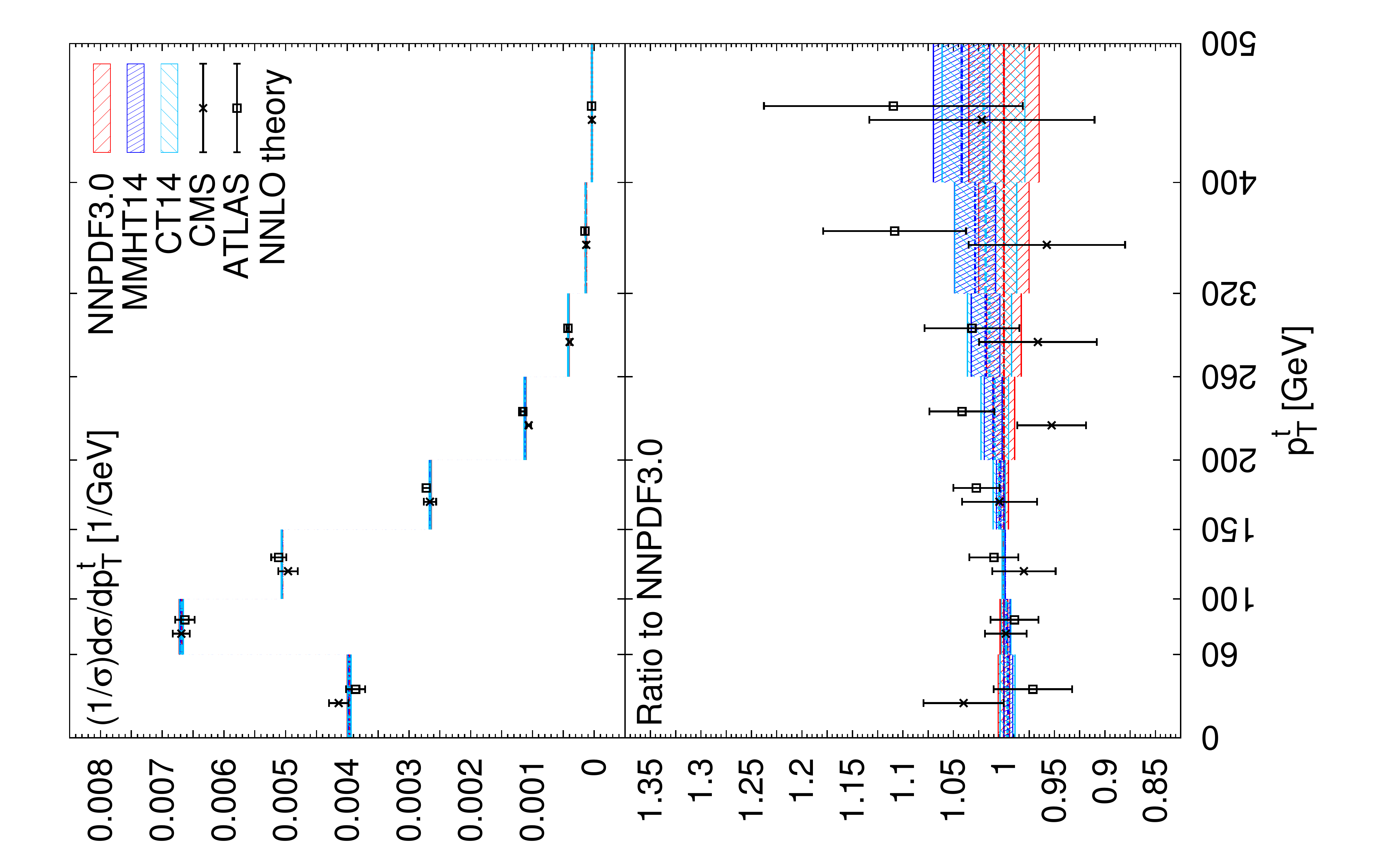}\\
\caption{\small Comparison between the NNLO predictions
  for the absolute (left) and normalized (right) $p_T^t$ differential 
  distributions in top-quark pair production and the
corresponding CMS and ATLAS measurements.
The theoretical predictions
have been computed with the NNPDF3.0, CT14 and
MMHT14 sets and include only
the 1--$\sigma$
PDF uncertainties,
while scale uncertainties are not shown.
In the lower panels, we display
the same results now as the ratio to the central NNPDF3.0 prediction.}
\label{fig:pTtnnlo}
\end{figure}

In order to assess the agreement between the data and the NNLO
theoretical predictions
based on our current knowledge of PDFs, we
perform now a systematic comparison 
of the calculations described in the previous section and
the ATLAS and CMS measurements.
This comparison is performed
at the level of both
absolute and normalized distributions,
allowing for an improved understanding of the differences and similarities 
between PDF sets.
This way, one can separate
differences induced by the shape of the gluon from those
induced by its normalisation.

The NNLO differential distributions with the binning
of the ATLAS and CMS measurements have been computed
using five different PDFs sets: NNPDF3.0,
CT14, MMHT2014, 
HERAPDF2.0~\cite{Abramowicz:2015mha} and
ABM12~\cite{Alekhin:2013nda}, in the last case with
the $n_f=5$ version.
For all these PDF sets, we consistently use the same
value of the strong coupling constant as in the NNLO matrix
elements.
This corresponds to  $\alpha_s(m_Z)=0.118$ for all sets except
for ABM12, for which PDFs are only available
 for their best-fit value of $\alpha_s(m_Z)=0.113$.

In Fig.~\ref{fig:pTtnnlo} we show the
NNLO predictions for the
absolute (left) and normalized (right) $p_T^t$ differential 
  distributions compared to the
corresponding CMS and ATLAS measurements.
The theory calculations are provided for
NNPDF3.0, CT14, and
MMHT14 and include only PDF uncertainties.
The data uncertainties correspond to the 
square root of the diagonal elements of the experimental covariance matrix.
At a qualitative level, we find that the theory calculations based on the three PDF sets used
in this comparison are in good agreement both among themselves and with the
data.
We also see that while at the level of normalized cross-sections
the experimental uncertainties are similar between ATLAS and CMS,
there are larger differences for absolute distributions.
Moreover, we note that the ATLAS and CMS measurements exhibit some
degree of tension.

\begin{figure}[t]
\centering
\includegraphics[scale=0.31,angle=270,clip=true,trim=2cm 0cm 2cm 0cm]{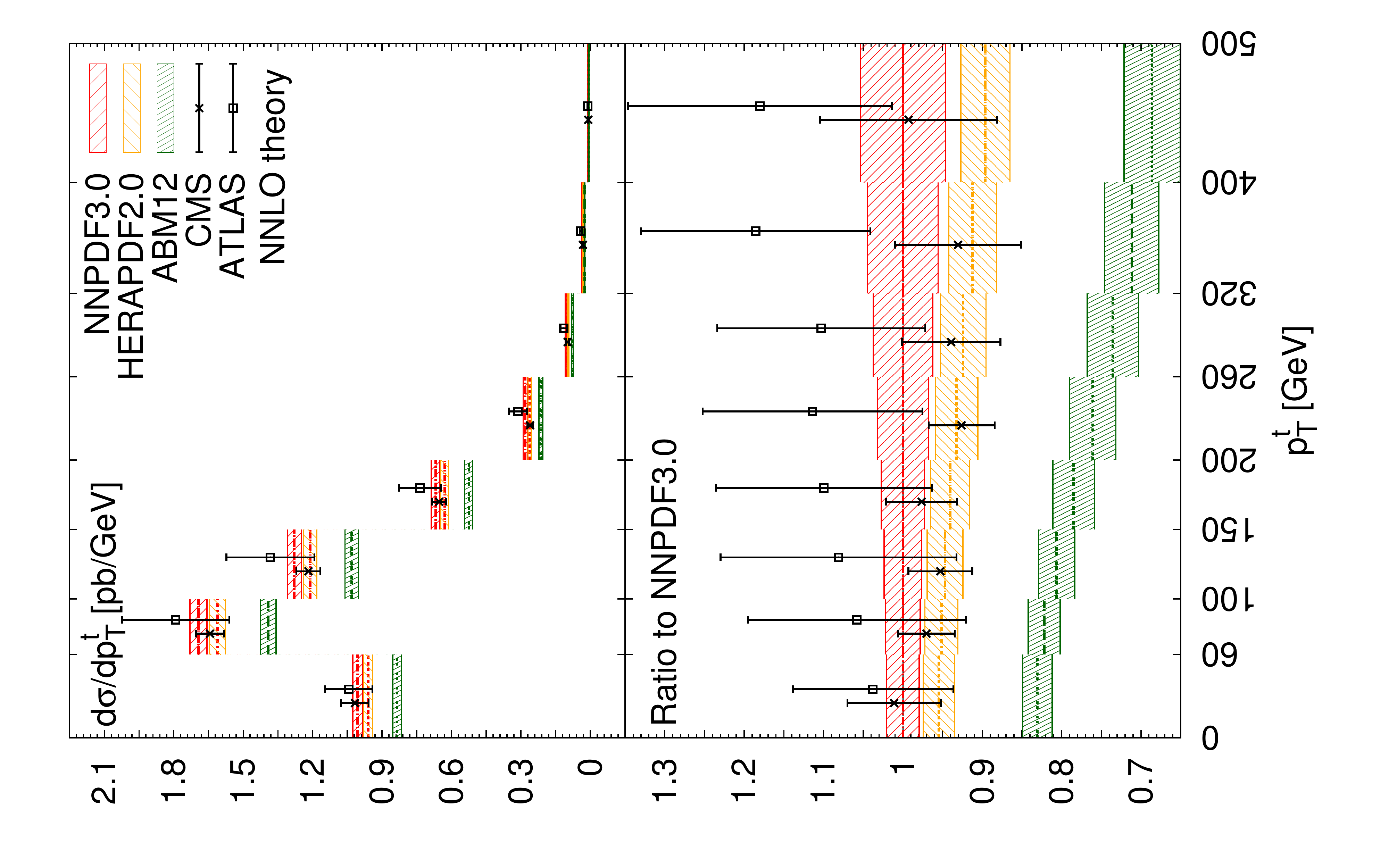}
\includegraphics[scale=0.31,angle=270,clip=true,trim=2cm 0cm 2cm 0cm]{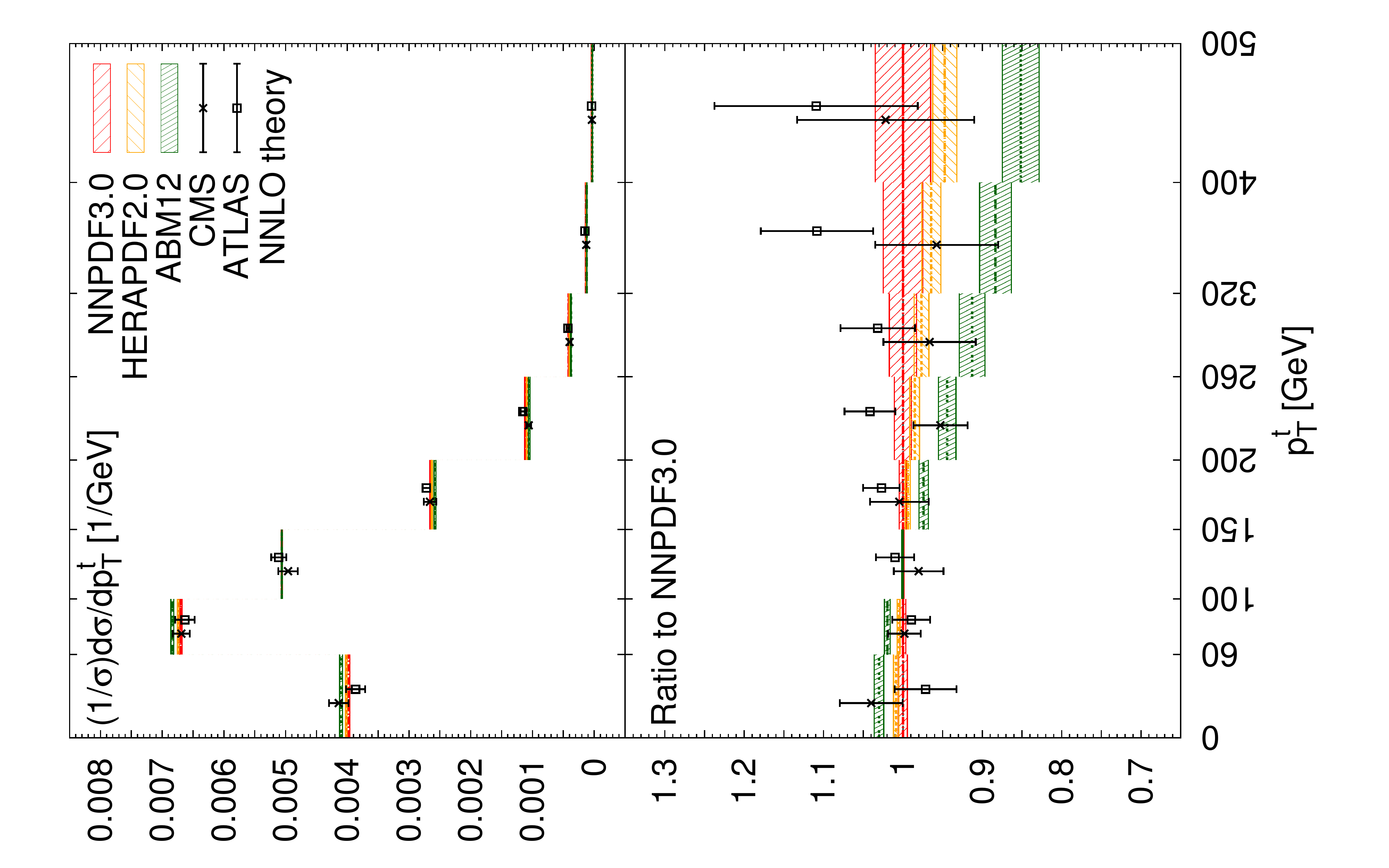}\\
\caption{\small Same as Fig.~\ref{fig:pTtnnlo} for the HERAPDF2.0
and ABM12 PDF sets.}
\label{fig:pTtnnlo1}
\end{figure}

Next, in Fig.~\ref{fig:pTtnnlo1} we show the same comparison but
now among NNPDF3.0, HERAPDF2.0 and ABM12.
In the case of HERAPDF2.0,
the PDF error band is the sum in quadrature of the 
statistical, model and parametrization uncertainties.
We note that while HERAPDF2.0 and NNPDF3.0 agree well, in particular for the normalized
distribution, this is not the case for ABM12, whose predictions
are substantially lower than those of the other PDF sets.
This effect is more pronounced for the absolute distributions,
and reflects intrinsic differences both in the gluon-gluon
luminosity and in the value of $\alpha_s(m_Z)$.
We will show that this trend reappears for other kinematical
distributions.
These differences between ABM12 and the other PDF sets
cannot be accommodated by a shift in the value of $m_t$ used.
As noted in Ref.~\cite{Czakon:2016vfr}, the sensitivity of the 
$p_T^t$ absolute differential distribution on the value of $m_t$ is
very non-uniform across the whole $p_T^t$ data range. 
In order for ABM12 to fit the data at the lowest $p_T^t$, one
should require an unreasonably small value of $m_t$, roughly around
$m_t=169$ GeV. 
However, even with such a shift of $m_t$, the large $p_T^t$ tail of the 
distribution will hardly move at all. 
Therefore, the shape of the ABM12 theoretical prediction will become even more 
different than that of the measured $p_T^t$ absolute differential distribution. 
This should remain true also for the normalised $p_T^t$ distribution, since its 
shape will shift similarly to the absolute one.

\begin{figure}[p]
\centering
\includegraphics[scale=0.31,angle=270,clip=true,trim=2cm 0cm 2cm 0cm]{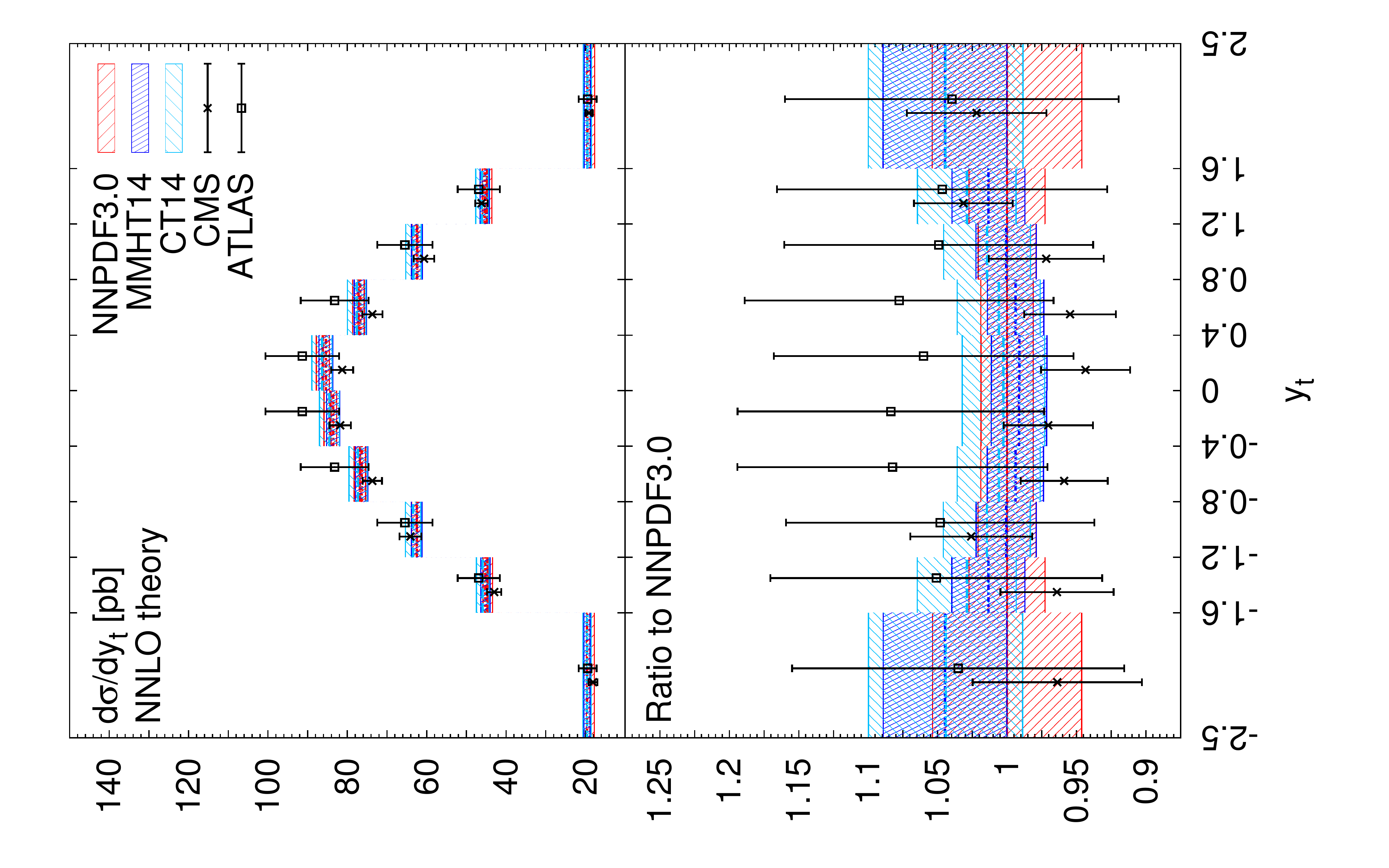}
\includegraphics[scale=0.31,angle=270,clip=true,trim=2cm 0cm 2cm 0cm]{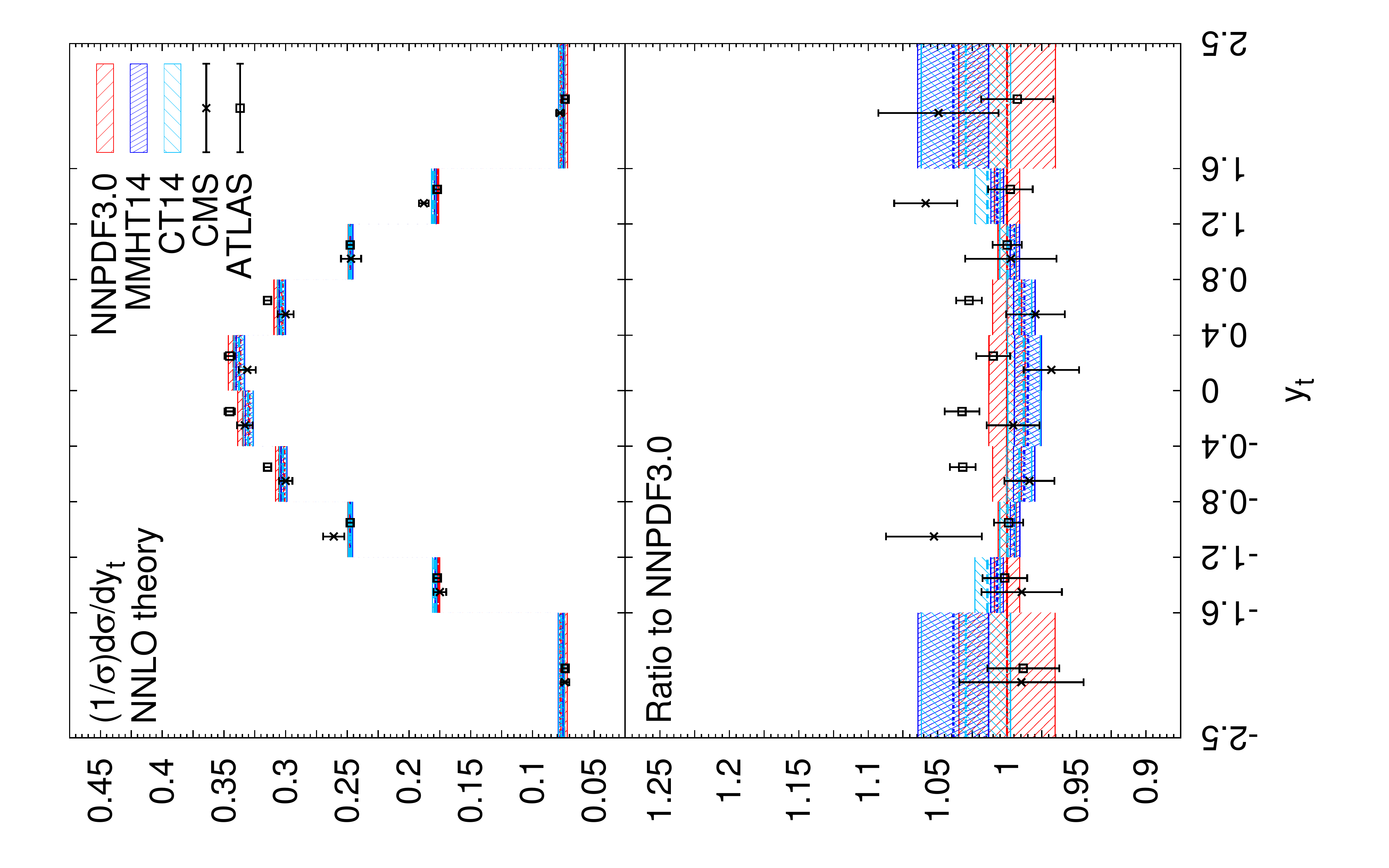}\\
\caption{\small Same as Fig.~\ref{fig:pTtnnlo}
for the top quark rapidity distribution $y_t$.
}
\label{fig:ytnnlo}
\end{figure}

In Fig.~\ref{fig:ytnnlo} we consider now the
top quark rapidity distribution, $y_t$.
Here too we find a good agreement among NNPDF3.0, CT14 and MMHT14, both
for the absolute and for the normalized distributions.
For forward rapidities, the PDF uncertainty
in NNPDF3.0 is larger than that of the other two PDF sets.
For this distribution, while CMS and ATLAS are consistent in the
absolute case, in the normalized case we again observe
some discrepancies between the two experiments in the
central region.
As we will show, this results in some difficulty
in being able to achieve a satisfactory fit of the distributions from 
both experiments simultaneously.

\begin{figure}[p]
\centering
\includegraphics[scale=0.31,angle=270,clip=true,trim=2cm 0cm 2cm 0cm]{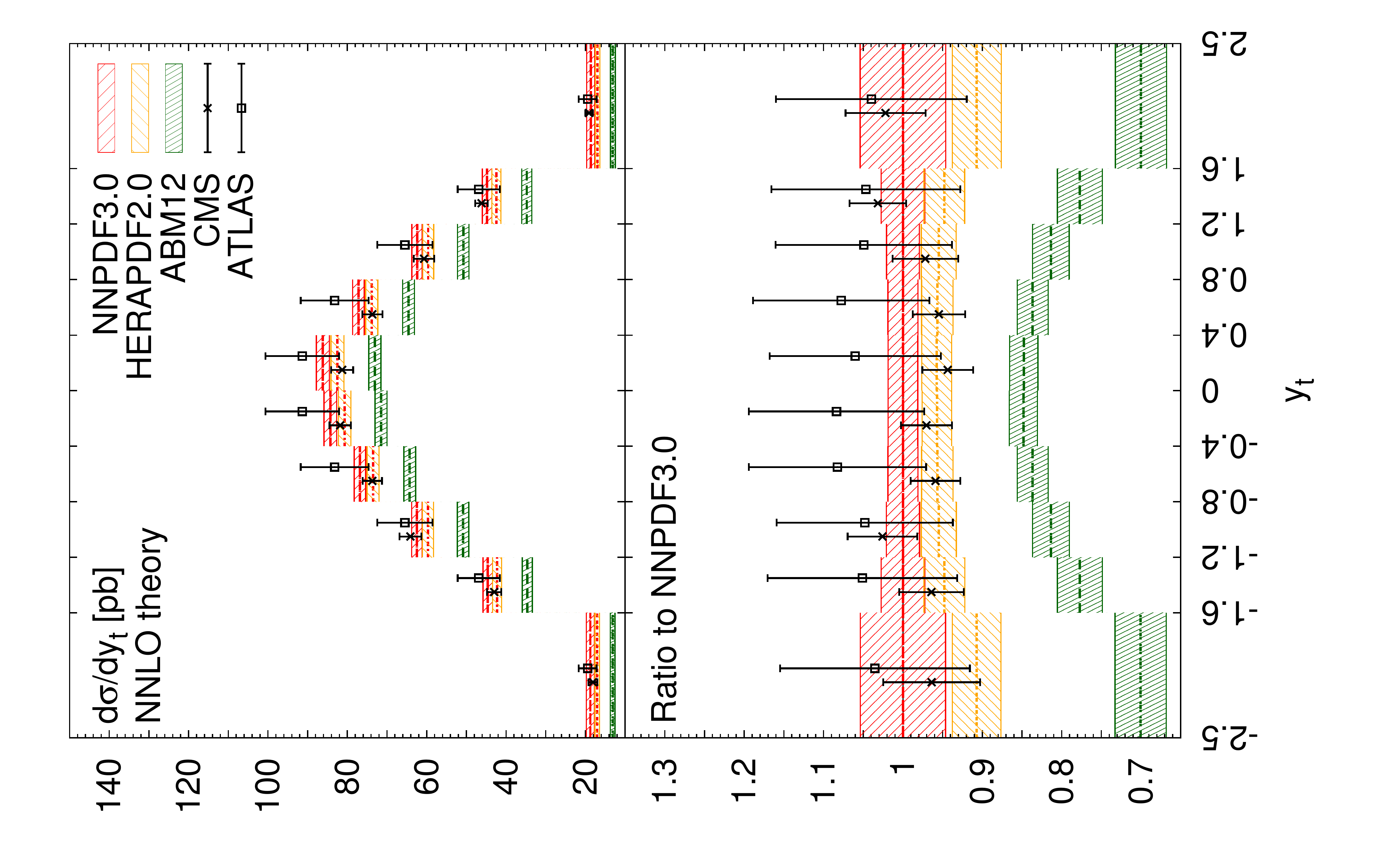}
\includegraphics[scale=0.31,angle=270,clip=true,trim=2cm 0cm 2cm 0cm]{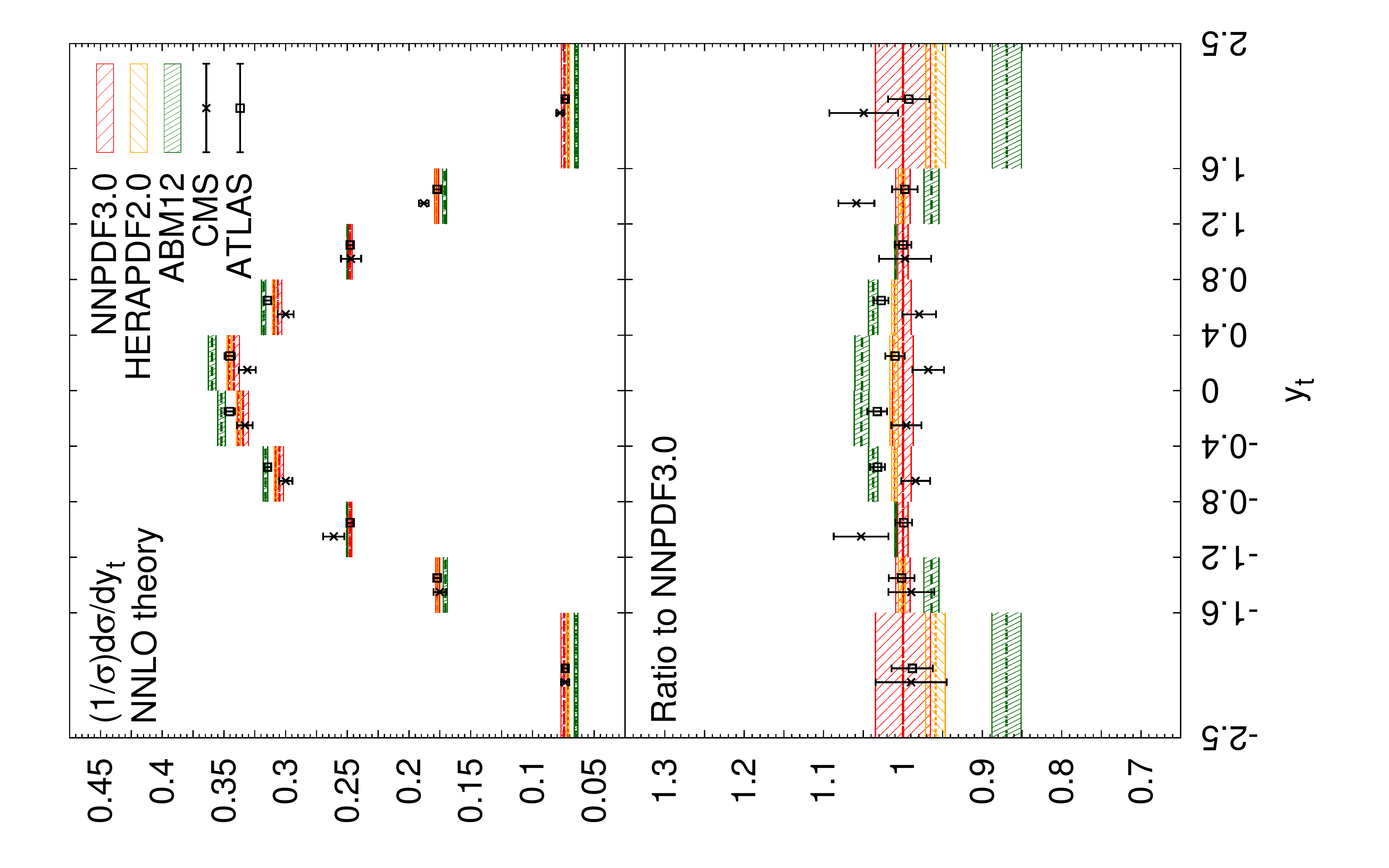}\\
\caption{\small Same as Fig.~\ref{fig:ytnnlo} now
  for NNPDF3.0, ABM12 and HERAPDF2.0.}
\label{fig:ytnnlo1}
\end{figure}

The corresponding comparisons between  theory predictions
and data for $y_t$, now among NNPDF3.0, ABM12 and HERAPDF2.0,
are shown in Fig.~\ref{fig:ytnnlo1}.
For the absolute distribution, HERAPDF2.0 is between 5\% and 10\% lower
than NNPDF3.0, with ABM12 lower by a larger amount, between 20\%
and 30\%.
These differences are reduced (but then the experimental
uncertainties are smaller as well) in the normalized case, where now
ABM12 is above NNPDF3.0 and HERAPDF2.0 in the central region
and undershoots them in the forward rapidity bins.
As we show below, these differences translate into a poor $\chi^2$
when the ABM12 predictions are compared with the experimental data.

\begin{figure}[p]
\centering
\includegraphics[scale=0.31,angle=270,clip=true,trim=2cm 0cm 2cm 0cm]{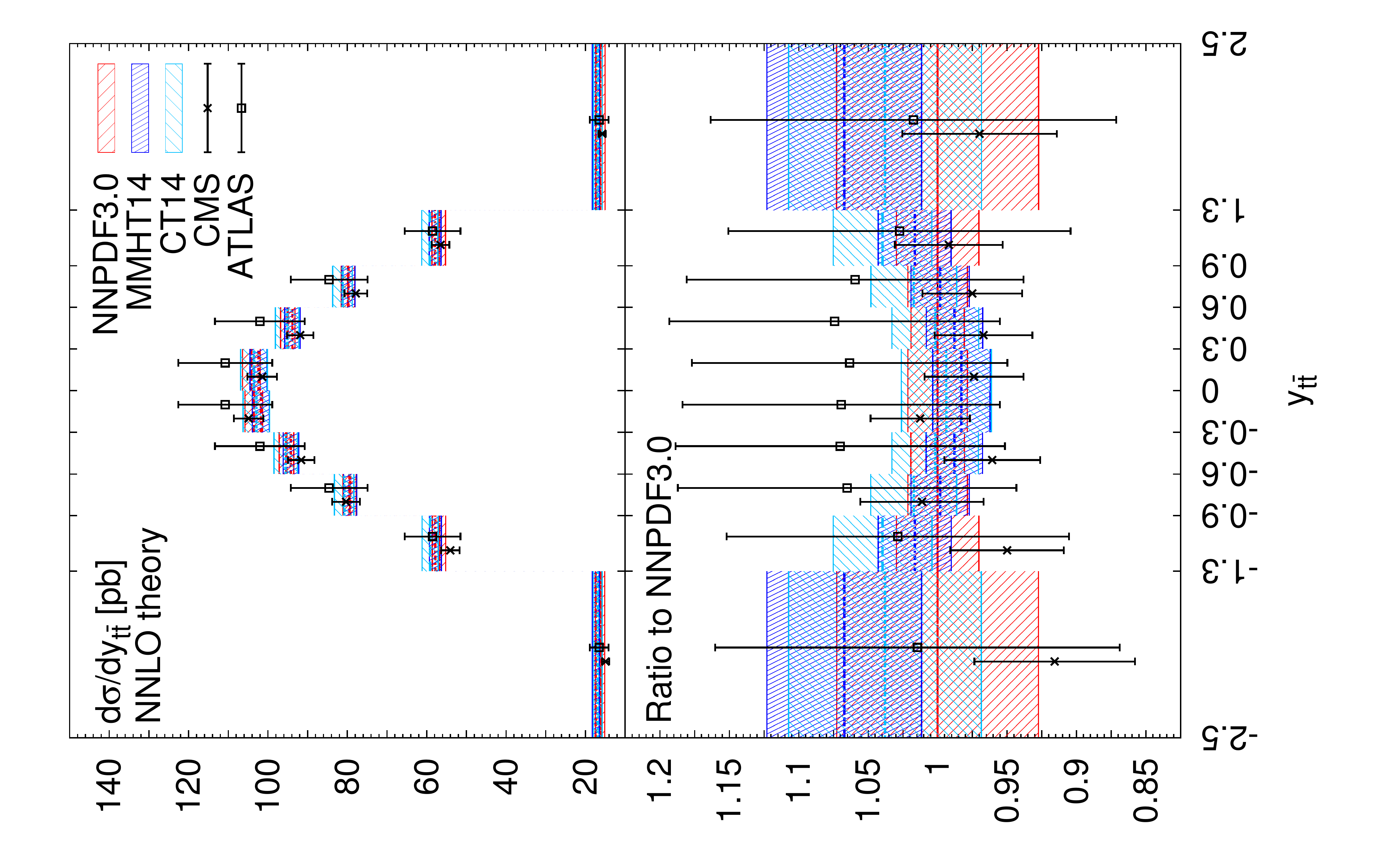}
\includegraphics[scale=0.31,angle=270,clip=true,trim=2cm 0cm 2cm 0cm]{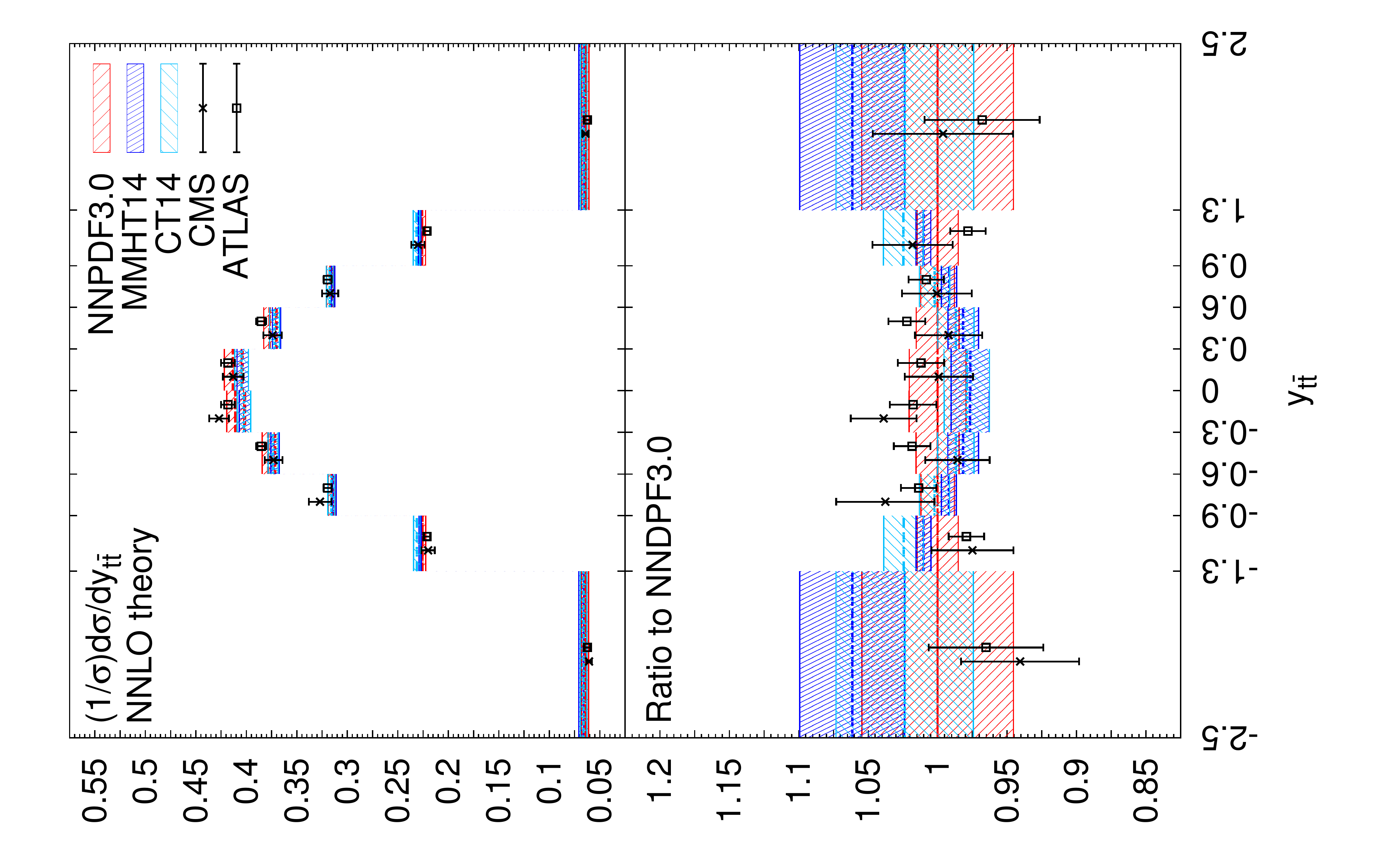}\\
\caption{\small Same as Fig.~\ref{fig:pTtnnlo} for the rapidity
distribution of the top-quark pair,  $y_{t\bar{t}}$.}
\label{fig:yttnnlo}
\end{figure}
\begin{figure}[p]
\centering
\includegraphics[scale=0.31,angle=270,clip=true,trim=2cm 0cm 2cm 0cm]{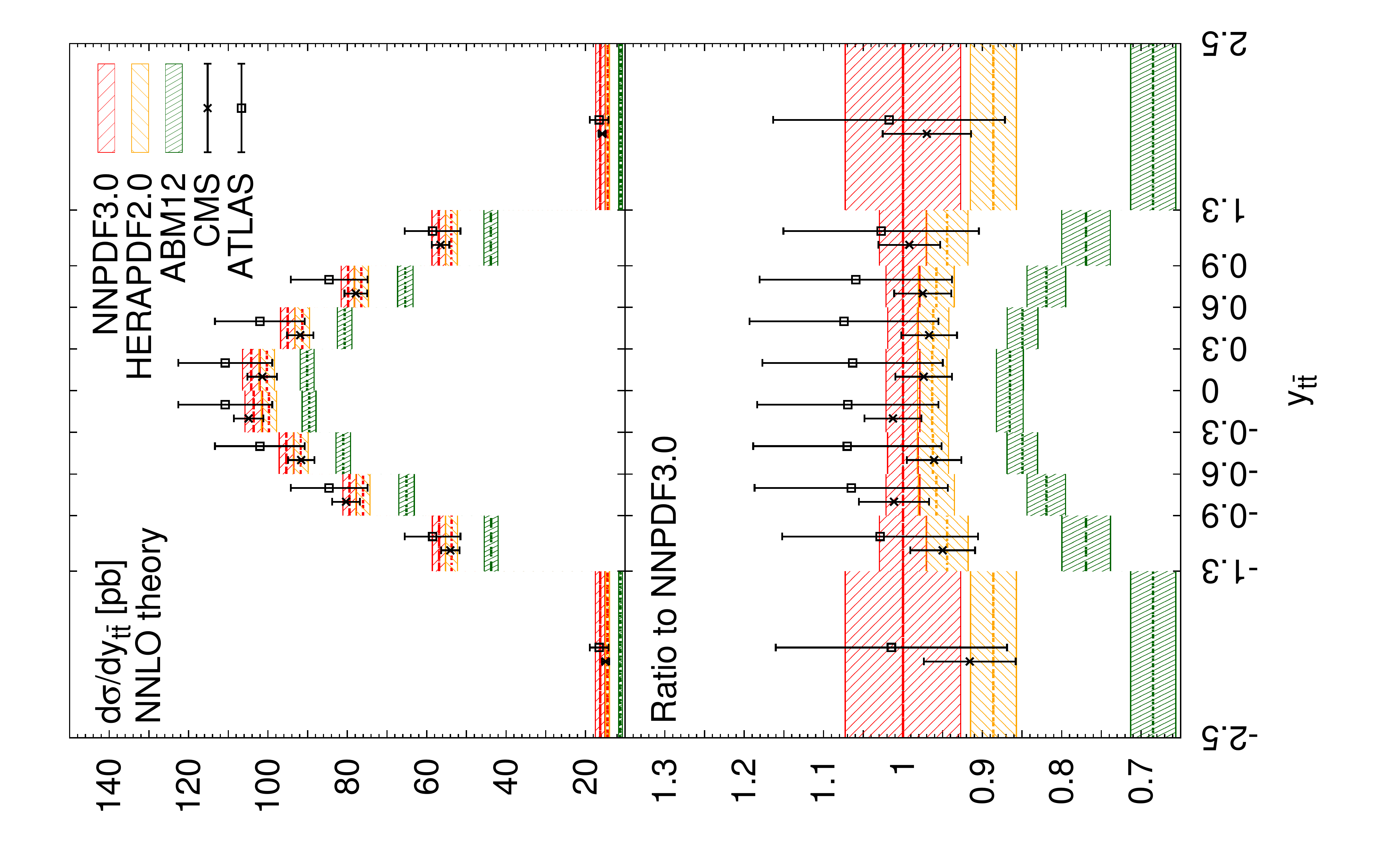}
\includegraphics[scale=0.31,angle=270,clip=true,trim=2cm 0cm 2cm 0cm]{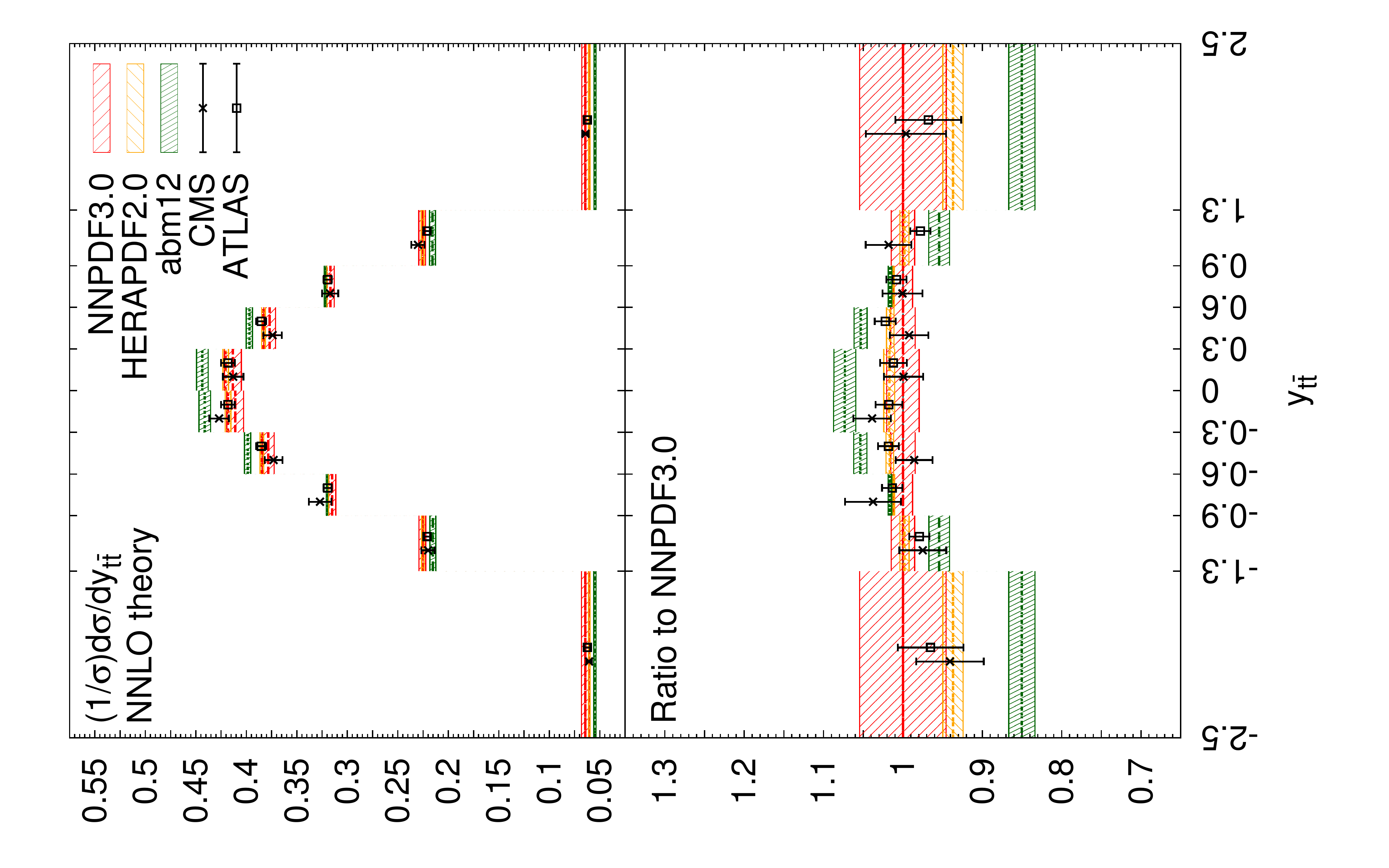}\\
\caption{\small Same as Fig.~\ref{fig:yttnnlo}
now
  for NNPDF3.0, ABM12 and HERAPDF2.0.}
\label{fig:ytttnnlo1}
\end{figure}

We now move to consider the comparison between data and theory
for the kinematical distributions constructed from the
top-quark pair kinematics, in particular the rapidity
$y_{t\bar{t}}$ and the invariant mass $m_{t\bar{t}}$ of the pair.
First of all, in Fig.~\ref{fig:yttnnlo} we compare the ATLAS and CMS
 $y_{t\bar{t}}$ measurements
with the corresponding NNLO predictions obtained using NNPDF3.0, 
CT14 and MMHT14.
Interestingly, unlike the cases of the $p_T^t$ and $y_t$ distributions,
the ATLAS and CMS  $y_{t\bar{t}}$ measurements are now in good agreement, 
both at the level of absolute and normalized
distributions, both in the central and forward regions.
As we will show, this has the important
consequence that $y_{t\bar{t}}$ is the only distribution
that can be satisfactorily described when ATLAS and CMS datasets
are included together in the same fit.
Both for the absolute and
the normalized distributions,
the theory predictions for $y_{t\bar{t}}$ with the
three PDF sets in Fig.~\ref{fig:yttnnlo} 
are consistent at the one-sigma level (in units of the
PDF uncertainty), and are also in reasonable agreement with the
experimental data.
As in the case of the $y_t$ distribution, for forward rapidities the
PDF uncertainties from NNPDF3.0 are larger than those of the other two
sets.

In Fig.~\ref{fig:ytttnnlo1} we show the corresponding comparison for the $y_{t\bar{t}}$
distributions, finding a similar trend as in the $y_t$ case in Fig.~\ref{fig:ytnnlo1}.
For the absolute distribution,
HERAPDF2.0 is somewhat lower than NNPDF3.0, with almost touching error bands
(this translates into a $\sqrt{2}$ sigma discrepancy between the two sets);
ABM12 is lower by an amount between 15\% and 30\% depending on the specific bin.
In the normalized distribution, ABM12 overshoots
the predictions of the other two PDF sets and the data for central rapidities and undershoots
them  in the forward region.

Finally we consider the differential distribution in the  invariant mass
of the top-antitop pair, $m_{t\bar{t}}$.
An accurate theoretical and experimental understanding of this
distribution is crucial in many searches for BSM physics,
 where new states couple to top quarks.
A prime example would be 
the case of heavy resonances that decay into a $t\bar{t}$ pair. Such decays 
would appear in the data as an excess in the invariant mass
distribution~\cite{Czakon:2016vfr,Frederix:2007gi,Hespel:2016qaf,Carena:2016npr}.

\begin{figure}[p]
\centering
\includegraphics[scale=0.31,angle=270,clip=true,trim=2cm 0cm 2cm 0cm]{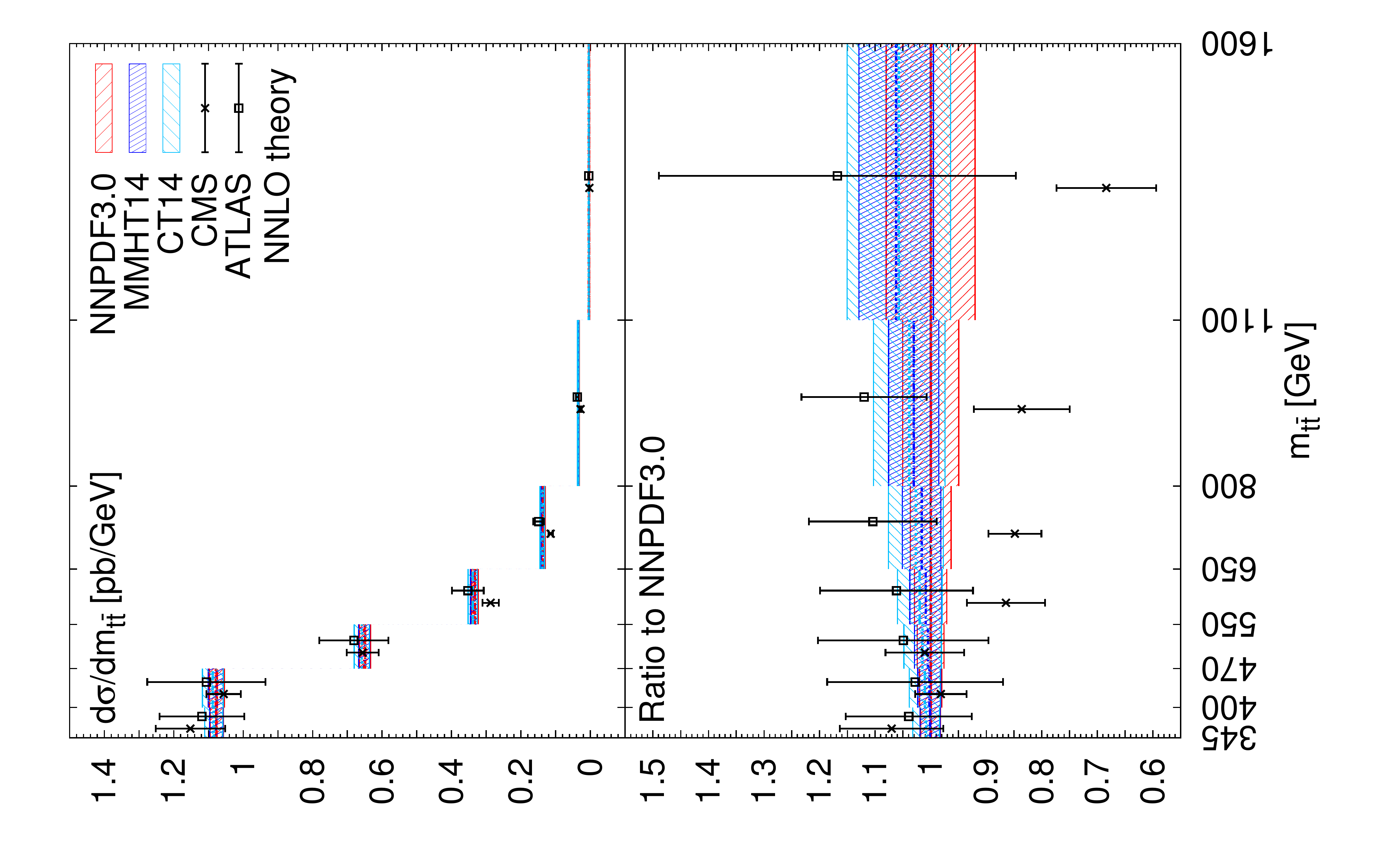}
\includegraphics[scale=0.31,angle=270,clip=true,trim=2cm 0cm 2cm 0cm]{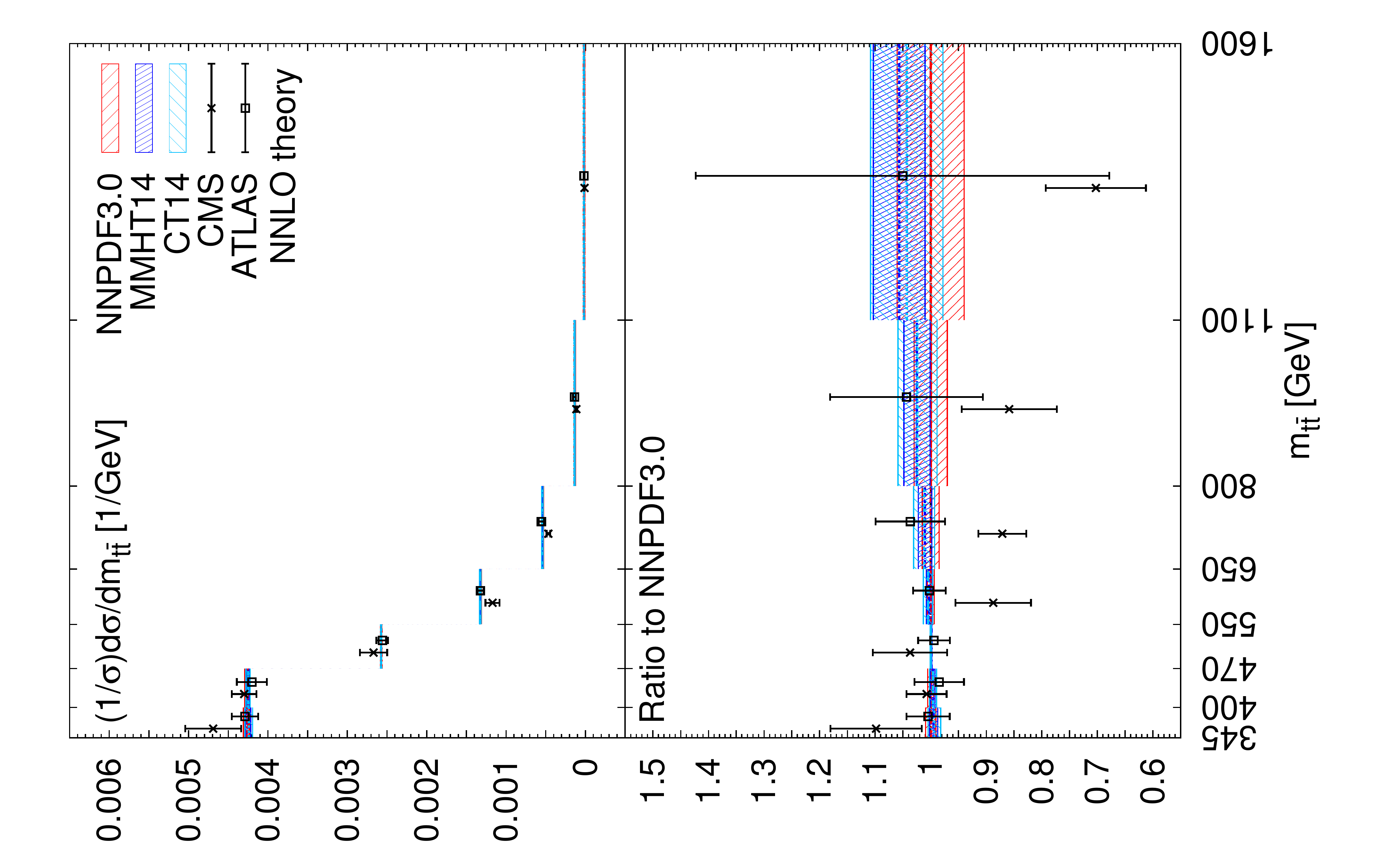}\\
\caption{\small Same as Fig.~\ref{fig:pTtnnlo} for the
  invariant mass distribution of the top-antitop pair, $m_{t\bar{t}}$.}
\label{fig:mttnnlo}
\end{figure}
\begin{figure}[p]
\centering
\includegraphics[scale=0.31,angle=270,clip=true,trim=2cm 0cm 2cm 0cm]{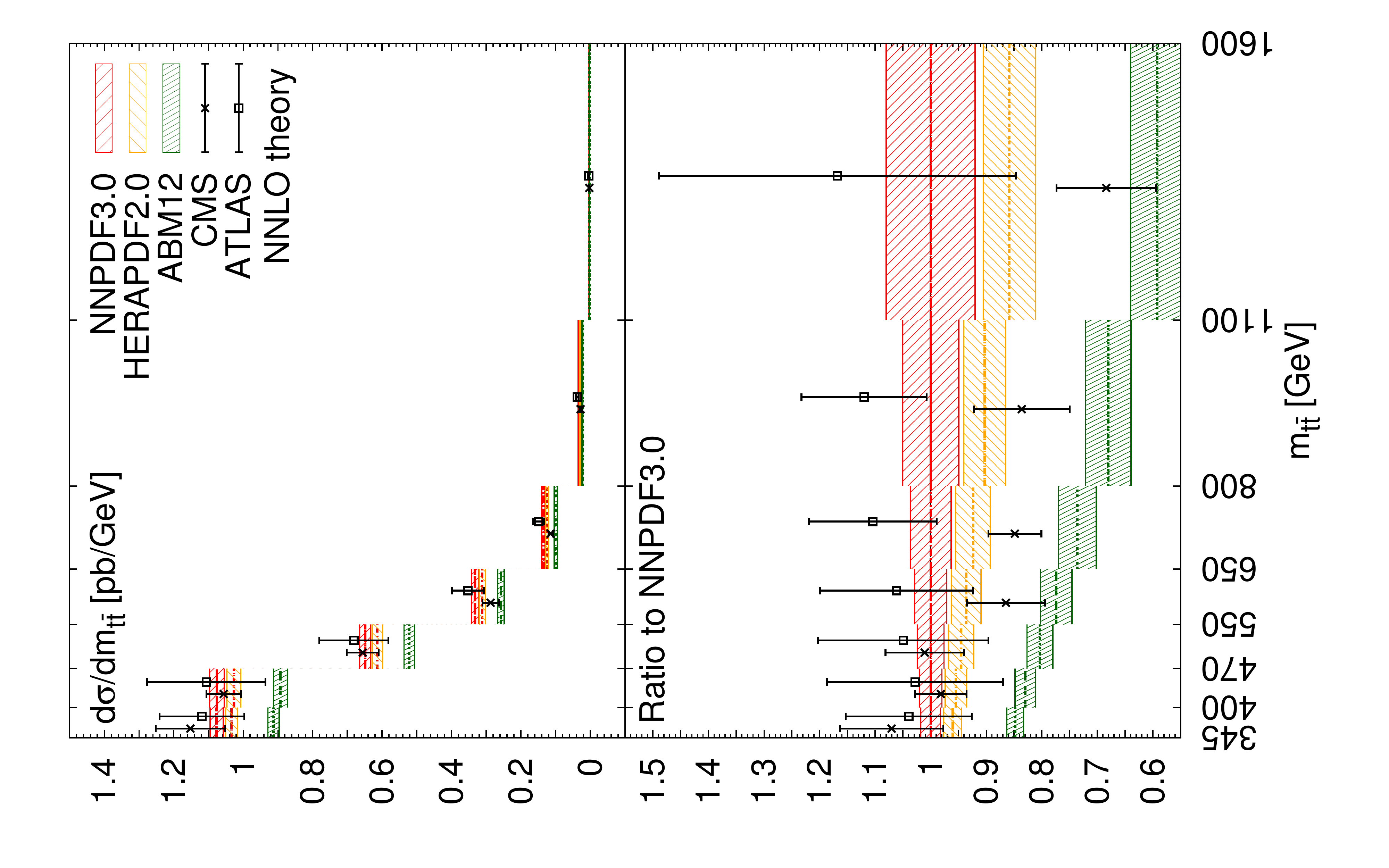}
\includegraphics[scale=0.31,angle=270,clip=true,trim=2cm 0cm 2cm 0cm]{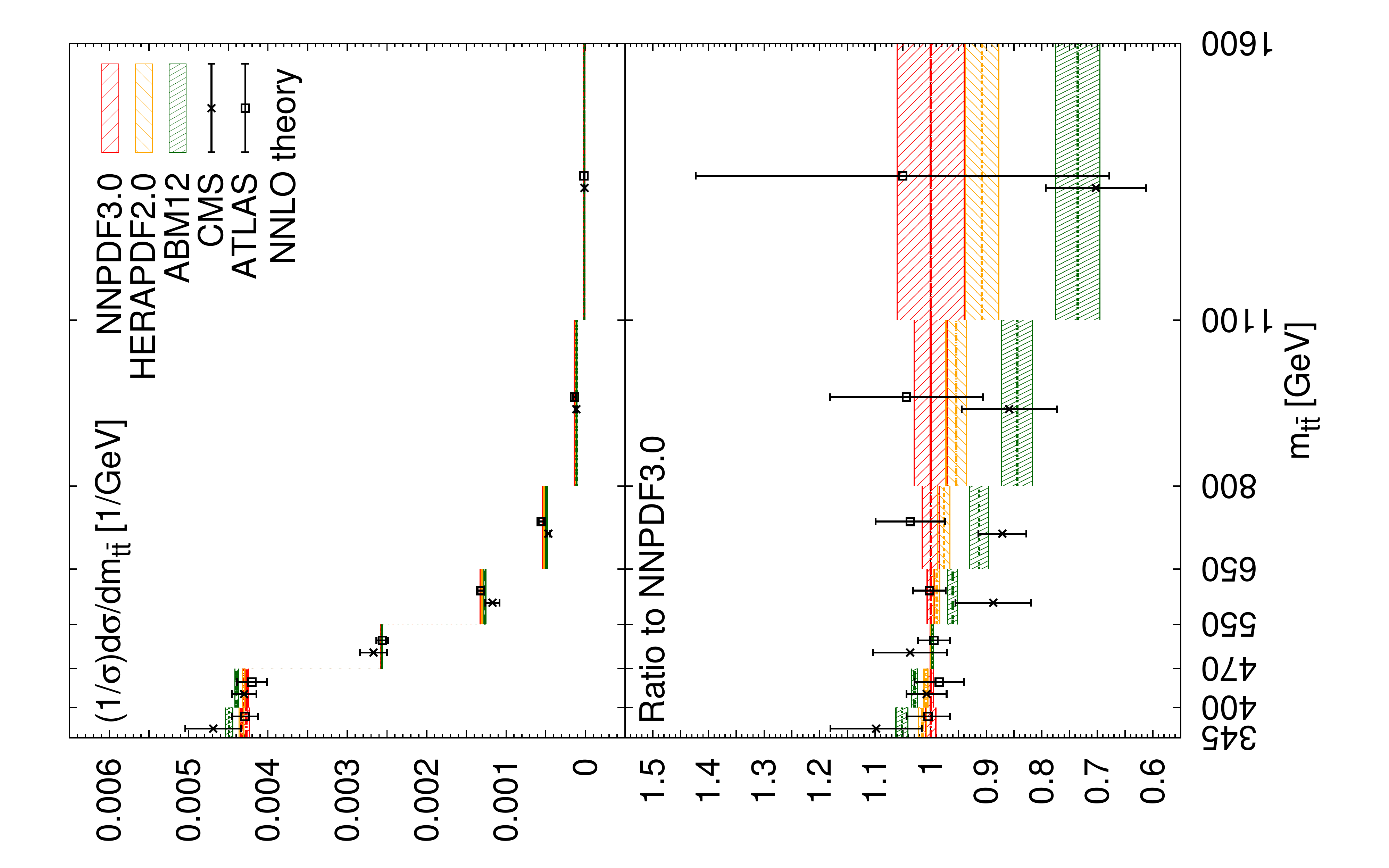}\\
\caption{\small
Same as Fig.~\ref{fig:mttnnlo},
now for NNPDF3.0, ABM12 and HERAPDF2.0.}
\label{fig:mttnnlo1}
\end{figure}

In Fig.~\ref{fig:mttnnlo} we show
the NNLO predictions for the invariant mass distribution of the
top-antitop pair, $m_{t\bar{t}}$, using NNPDF3.0, CT14 and MMHT14.
The first thing to note is the difference
between the ATLAS and CMS measurements, especially in the absolute distribution
and for intermediate values of $m_{t\bar{t}}$.
The difference in the size of the overall experimental uncertainties is also significant.
For instance, despite being based on the same
integrated luminosity, the ATLAS uncertainty 
in the highest $m_{t\bar{t}}$ bin is about four times larger than that of CMS.
We also find that the three PDF sets are in good agreement within
uncertainties, with NNPDF3.0 exhibiting a somewhat lower central value
and larger uncertainties at high $m_{t\bar{t}}$ as compared to the other two sets.
While the three PDF sets  agree qualitatively with the ATLAS measurements,
there seems to be some tension with the CMS data,
which exhibits lower central values in the intermediate 
and high $m_{t\bar{t}}$ regions and has smaller experimental uncertainties.

The corresponding comparison between NNPDF3.0, ABM12 and HERAPDF2.0 is shown
in Fig.~\ref{fig:mttnnlo1}, from which we observe common trends 
in the absolute and normalized distributions.
The HERAPDF2.0  prediction are lower than the NNPDF3.0 ones, with ABM12
being even lower, by up to 40\% (25\%) in the highest $m_{t\bar{t}}$ bin of
the absolute (normalized) distribution.
Given that the ATLAS and CMS measurements seem to be
pulling in opposite directions, the latter is favored by the ABM12 prediction,
while the former is in better agreement with NNPDF3.0 and HERAPDF2.0.

\begin{figure}[!t]
\centering
\includegraphics[scale=0.45,angle=270]{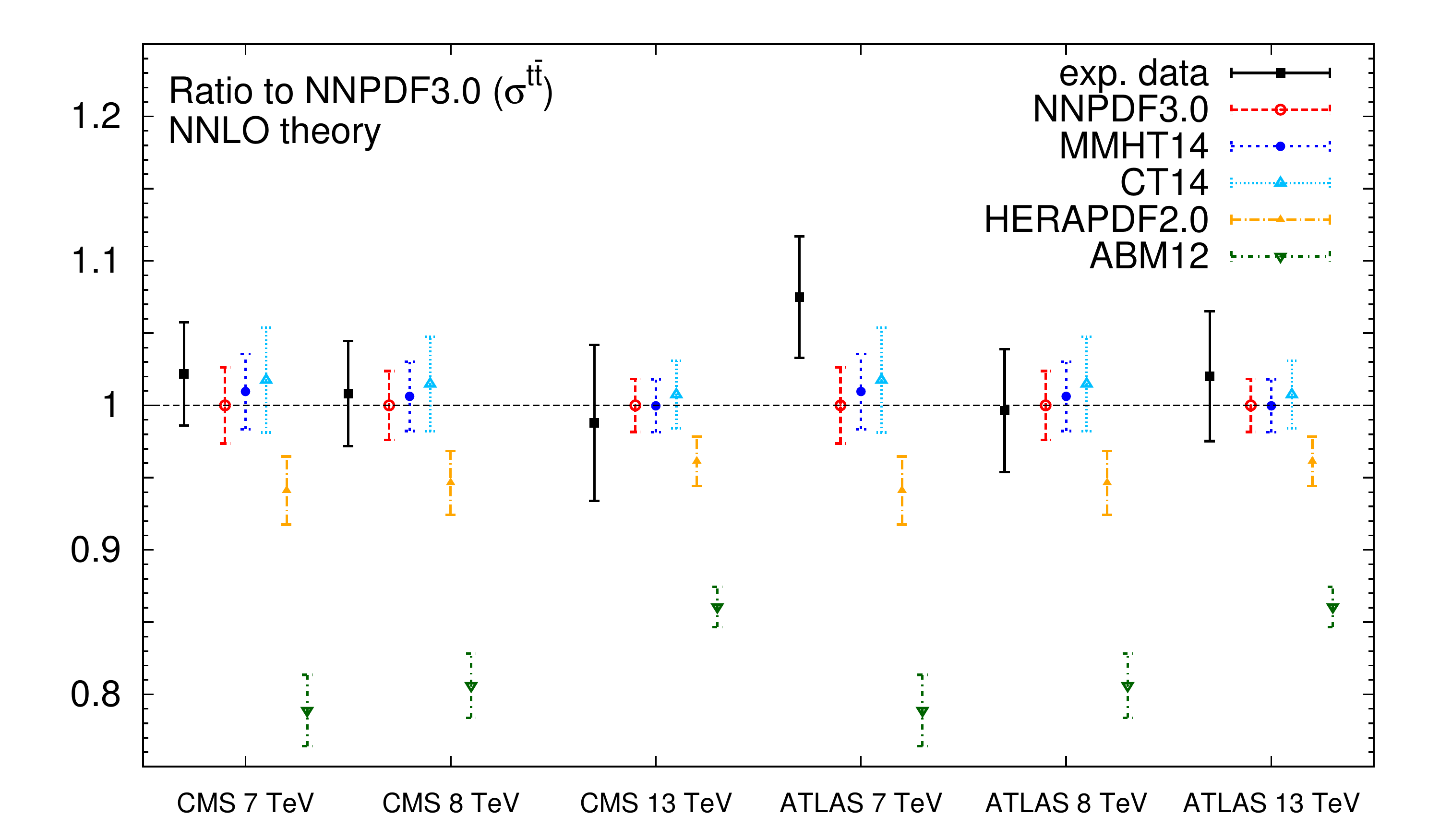}\\
\caption{\small The inclusive  cross-sections in top-quark pair
  production from ATLAS and CMS at different center-of-mass energies
  $\sqrt{s}$ (see Table~\ref{tab:uncTotalXsec}),
  compared to NNLO theory computed with the program {\tt top++} for the
  five PDF sets.
  Results
  are shown as ratios to the central NNPDF3.0 predictions.}
\label{fig:totxsec}
\end{figure}

Before moving to a more quantitative assessment of the agreement between
data and theory, we would like to compare the NNLO calculations with the experimental 
measurements of the total cross-section listed in Table.~\ref{tab:uncTotalXsec}.
This comparison is useful because inclusive data provide information
on the overall normalisation of the gluon for the cases where normalized distributions are fitted.
In Fig.~\ref{fig:totxsec} we show the inclusive  cross-sections
from ATLAS and CMS at different center-of-mass energies,
  compared to NNLO theory computed with {\tt top++} for the
  five PDF sets.
  Results
  are shown as ratios to the central NNPDF3.0 predictions.
The comparison follows the trend observed at the level of absolute 
differential distributions, with NNPDF3.0, MMHT14 and CT14 in good agreement both
among themselves and with the LHC measurements.
On the other hand, HERAPDF2.0 and ABM12 predict
cross-sections that are lower by about 6\% (4\%) and 20\% (15\%), respectively,
at 7 and 8 TeV (13 TeV) as compared to NNPDF3.0.

\subsection{Quantitative assessment of the agreement between theory and data}

Due to the presence of large correlated experimental uncertainties (both of statistical
and systematic origin),
it is not
possible to accurately assess the agreement between data and theory solely from the figures above.
An adequate measure of this agreement should fully take these correlations into account.
To this end we introduce a $\chi^2$ estimator, which
depends on the dataset, $\mathcal{D}$, and on the theoretical predictions
based on the PDFs $f$, $\mathcal{T}[f]$.
In this work, we use the following definition:
\begin{equation}
\chi^2\left\{\mathcal{T}[f], \mathcal{D} \right\} 
=\frac{1}{N_{\rm dat}}
\sum_{i,j}^{N_{\rm dat}}
\left(T_i[f] - D_i  \right)
C_{ij}^{-1}
\left(T_j[f] - D_j \right)
\,\mbox{.}
\label{eq:chi2}
\end{equation}
In this expression, $i$ and $j$ run over the experimental datapoints,
$D_i$ are the measured central values, $T_i$ 
are the corresponding NNLO theoretical predictions computed with a given set of 
PDFs
and $C_{ij}$ is the covariance matrix, constructed from the available 
information on experimental statistical and systematic uncertainties.

The definition of the covariance matrix in Eq.~(\ref{eq:chi2}) is
not unique (see, for example, the discussion
in Refs~\cite{Ball:2009qv,Ball:2012wy}).
In this section we will use the so-called {\it experimental}
definition:
\begin{equation}
C_{ij}^{\rm exp} 
\equiv 
\delta_{ij} \lp \sigma^{\rm stat}_{i}\rp ^2 
+ 
\left(
\sum_{\alpha =1}^{N_{\rm sysA}}\sigma_{i,\alpha}^{\rm sysA}\sigma_{j,\alpha}^{\rm sysA}
+
\sum_{\beta=1}^{N_{\rm sysM}}\sigma_{i,\beta}^{\rm sysM}
\sigma_{j,\beta}^{\rm sysM}
\right)
D_i D_j
\,\mbox{,}
\label{eq:covmat}
\end{equation}
where $\sigma^{\rm stat}_{i}$ is the uncorrelated uncertainty of the
data point $i$ (obtained by adding in 
quadrature statistical and uncorrelated systematic uncertainties), and
$\sigma_{i,\alpha}^{\rm sysA}$ ($\sigma_{i,\alpha}^{\rm sysM}$) are the $N_{\rm sysA}$
($N_{\rm sysM}$) correlated additive (multiplicative)
systematic
uncertainties.
 The total number of correlated uncertainties in this
 case is $N_{\rm sysM} + N_{\rm sysA}$.

\begin{table}[!t]
\centering
\footnotesize
\begin{tabular}{llccllcc}
\toprule
Dataset & PDF set & \multicolumn{2}{c}{$\chi^2$} &
Dataset & PDF set & \multicolumn{2}{c}{$\chi^2$} \\
\midrule
ATLAS $d\sigma/dp_T^t$ & & & &
CMS   $d\sigma/dp_T^t$ & & & \\
&  {  NNPDF3.0} & 0.84 & (0.66) & 
&  {  NNPDF3.0} & 1.24 & (0.91) \\
&  {  CT14}     & 0.76 & (0.42) & 
&  {  CT14}     & 1.67 & (1.77) \\
&  {  MMHT14}   & 0.63 & (0.44) & 
&  {  MMHT14}   & 1.54 & (1.47)  \\
&  {  HERA2.0}  & 1.13 & (1.69) & 
&  {  HERA2.0}  & 0.69 & (0.34)  \\
&  {  ABM12}    & 6.23 & (1.94) & 
&  {  ABM12}    & 12.5 & (3.00)  \\
ATLAS $d\sigma/dy_t$ & & & &  
CMS   $d\sigma/dy_t$ & & & \\
&  {  NNPDF3.0} & 0.73 & (0.28) & 
&  {  NNPDF3.0} & 3.04 & (1.05) \\
&  {  CT14}     & 1.28 & (0.20) & 
&  {  CT14}     & 2.23 & (1.47)  \\
&  {  MMHT14}   & 1.36 & (0.29) & 
&  {  MMHT14}   & 2.12 & (0.98)  \\
&  {  HERA2.0}  & 0.72 & (0.99) &
&  {  HERA2.0}  & 3.65 & (1.49)  \\
&  {  ABM12}    & 5.32 & (1.45) & 
&  {  ABM12}    & 22.1 & (9.78)  \\
ATLAS $d\sigma/dy_{t\bar{t}}$ & & & &   
CMS   $d\sigma/dy_{t\bar{t}}$ & & &  \\
&  {  NNPDF3.0} & 0.84 & (0.21) & 
&  {  NNPDF3.0} & 0.99 & (0.74) \\
&  {  CT14}     & 2.69 & (0.19) & 
&  {  CT14}     & 1.88 & (1.67) \\
&  {  MMHT14}   & 2.36 & (0.29) & 
&  {  MMHT14}   & 2.27 & (1.52) \\
&  {  HERA2.0}  & 0.53 & (0.74) & 
&  {  HERA2.0}  & 1.02 & (0.78)  \\
&  {  ABM12}    & 4.04 & (1.05) & 
&  {  ABM12}    & 18.0 & (5.48)  \\
ATLAS $d\sigma/dm_{t\bar{t}}$ & & & &  
CMS   $d\sigma/dm_{t\bar{t}}$ & & & \\
&  {  NNPDF3.0} & 0.77 & (0.38) & 
&  {  NNPDF3.0} & 5.73 & (4.36) \\
&  {  CT14}     & 0.61 & (0.19) & 
&  {  CT14}     & 7.28 & (6.06) \\
&  {  MMHT14}   & 0.58 & (0.24) & 
&  {  MMHT14}   & 7.32 & (5.74) \\
&  {  HERA2.0}  & 1.40 & (1.30) & 
&  {  HERA2.0}  & 3.32 & (1.49)  \\
&  {  ABM12}    & 5.72 & (3.81) & 
&  {  ABM12}    & 5.23 & (3.22)  \\
\bottomrule
\end{tabular}
\caption{\small
  The $\chi^2$ values
  for absolute distributions
  in top-quark pair production from ATLAS and CMS for different NNLO PDF sets.
  The first number is
   the $\chi^2$ from the full covariance matrix, Eq.~(\ref{eq:covmat}),
  while the value in parenthesis
  is obtained by adding in quadrature statistical
  and systematic errors.
  \label{tab:nnlochi2_1}
  }
\end{table}

The values of the $\chi^2$ computed using Eq.~(\ref{eq:covmat})
for each of the absolute and normalized differential distributions considered in this
work, and using the five NNLO PDF sets,
are summarized
in Tables~\ref{tab:nnlochi2_1} and~\ref{tab:nnlochi2_2}.
In order to facilitate the comparison 
with Figs.~\ref{fig:pTtnnlo}-\ref{fig:mttnnlo1}, 
we also indicate in parenthesis the $\chi^2$ values computed neglecting
bin-by-bin correlations.
As expected, the $\chi^2$ reduces substantially if experimental correlations
are not accounted for.

At the level of absolute distributions,
Table~\ref{tab:nnlochi2_1}, we find that
for NNPDF3.0 there is good agreement ($\chi^2\simeq 1$)
for all ATLAS distributions, while the agreement is poorer
for the CMS distributions except for $y_{t\bar{t}}$ and $p_T^t$.
A similar agreement between data and NNLO theory is found for HERAPDF2.0.
The same trend is also partly shared by CT14 and MMHT14,
though these two sets lead to a somewhat
worse description of the ATLAS and CMS $y_{t\bar{t}}$ distributions
as compared to NNPDF3.0 and HERAPDF2.0.
On the other hand, for ABM12 one finds a significantly worse $\chi^2$,
which reflects the fact that their predictions
tend to undershoot the LHC data, as observed in
Figs.~\ref{fig:pTtnnlo}-\ref{fig:mttnnlo}.
Concerning the top-quark transverse momentum $p_T^t$ absolute distributions, 
NNLO theory provides good description of both ATLAS and CMS data for
all PDF sets except for ABM12.

\begin{table}[!t]
\centering
\footnotesize
\begin{tabular}{llccllcc}
\toprule
Dataset & PDF set & \multicolumn{2}{c}{$\chi^2$} &
Dataset & PDF set & \multicolumn{2}{c}{$\chi^2$} \\
\midrule
ATLAS $(1/\sigma)d\sigma/dp_T^t$ & & & &  
CMS   $(1/\sigma)d\sigma/dp_T^t$ & & & \\
&  {  NNPDF3.0} & 3.13 & (0.94) & 
&  {  NNPDF3.0} & 2.03 & (0.51) \\
&  {  CT14}     & 2.33 & (0.62) & 
&  {  CT14}     & 2.88 & (0.70) \\
&  {  MMHT14}   & 2.23 & (0.54) & 
&  {  MMHT14}   & 3.15 & (0.77) \\
&  {  HERA2.0}  & 5.19 & (1.73) & 
&  {  HERA2.0}  & 1.12 & (0.33) \\
&  {  ABM12}    & 14.0 & (4.90) & 
&  {  ABM12}    & 2.80 & (0.80) \\
ATLAS $(1/\sigma)d\sigma/dy_t$ & & & & 
CMS   $(1/\sigma)d\sigma/dy_t$ & & & \\
&  {  NNPDF3.0} & 4.06 & (2.85) & 
&  {  NNPDF3.0} & 3.29 & (1.49) \\
&  {  CT14}     & 10.3 & (5.71) & 
&  {  CT14}     & 2.33 & (0.96) \\
&  {  MMHT14}   & 12.1 & (6.82) & 
&  {  MMHT14}   & 2.40 & (1.09) \\
&  {  HERA2.0}  & 1.76 & (1.62) & 
&  {  HERA2.0}  & 4.99 & (2.29) \\
&  {  ABM12}    & 15.5 & (7.09) & 
&  {  ABM12}    & 17.7 & (8.72) \\
ATLAS $(1/\sigma)d\sigma/dy_{t\bar{t}}$ & & & &  
CMS   $(1/\sigma)d\sigma/dy_{t\bar{t}}$ & & & \\
&  {  NNPDF3.0} & 3.59 & (1.48) & 
&  {  NNPDF3.0} & 1.17 & (0.75) \\
&  {  CT14}     & 12.7 & (5.26) & 
&  {  CT14}     & 2.53 & (1.51) \\
&  {  MMHT14}   & 15.6 & (5.49) & 
&  {  MMHT14}   & 3.33 & (2.10)  \\
&  {  HERA2.0}  & 1.20 & (0.60) & 
&  {  HERA2.0}  & 1.23 & (0.73)  \\
&  {  ABM12}    & 20.2 & (6.06) & 
&  {  ABM12}    & 8.26 & (4.52)  \\
ATLAS $(1/\sigma)d\sigma/dm_{t\bar{t}}$ & & & &  
CMS   $(1/\sigma)d\sigma/dm_{t\bar{t}}$ & & & \\
&  {  NNPDF3.0} & 1.57 & (0.10) & 
&  {  NNPDF3.0} & 10.6 & (3.87)  \\
&  {  CT14}     & 1.09 & (0.05) & 
&  {  CT14}     & 13.5 & (4.82)  \\
&  {  MMHT14}   & 1.01 & (0.05) & 
&  {  MMHT14}   & 13.5 & (4.93)  \\
&  {  HERA2.0}  & 4.36 & (0.30) & 
&  {  HERA2.0}  & 5.96 & (2.28)  \\
&  {  ABM12}    & 21.1 & (1.61) & 
&  {  ABM12}    & 1.24 & (0.47)  \\
\bottomrule
\end{tabular}
\caption{\small
  Same as Table~\ref{tab:nnlochi2_1} for the normalized
  differential distributions.
  \label{tab:nnlochi2_2}
  }
\end{table}

Moving to normalized distributions, Table~\ref{tab:nnlochi2_2}, one
finds  $\chi^2$ values that are in general higher than those from the absolute case.
In the case of the $p_T^t$ distribution,
the agreement between normalized data and theory is generally poor 
for all PDF sets and for both ATLAS and CMS, except
for HERAPDF2.0 in the former case.
For the normalized $y_{t}$ and $y_{t\bar{t}}$ distributions, HERAPDF2.0
provides a reasonable description except for the CMS $y_{t}$ distribution,
where one finds $\chi^2\simeq 5$.
None of the other NNLO sets achieves a satisfactory description
of these two distributions.

Concerning the normalized invariant mass $m_{t\bar{t}}$ distribution,
there is a stark difference between the comparisons of the ATLAS and the
CMS measurements with theory.
In the former case, NNPDF3.0, CT14 and MMHT14 lead to
a good $\chi^2$, while for the latter the same PDF sets lead to a much worse $\chi^2$.
For this distribution, HERAPDF2.0 provides a poor description of both ATLAS and CMS data,
while ABM12 can successfully describe the CMS data at the price
of a very poor $\chi^2$ to the ATLAS measurements.
Therefore, it seems not possible to achieve a simultaneous 
satisfactory description of both the ATLAS and CMS
normalized $m_{t\bar{t}}$ distributions.
As we will show in the next section, the same conclusions hold after the PDF fit.

A pattern arises from both Figs.~\ref{fig:pTtnnlo}--\ref{fig:mttnnlo1} and from the
$\chi^2$ comparisons in Tables~\ref{tab:nnlochi2_1}--\ref{tab:nnlochi2_2}:
a certain degree of tension is present between the
ATLAS and CMS measurements.
This tension is more marked in the normalized distributions, which are characterized
by smaller experimental uncertainties.
As we will demonstrate next, such tension
does not disappear when the top-quark
distributions are included in the global
PDF fit, though it is significantly alleviated when ATLAS and CMS
data are fitted separately.
Moreover, as we will show, it is possible to select a combination of
ATLAS and CMS data leading to significant constraining 
power on the large-$x$ gluon while at the same time providing a good $\chi^2$ 
description of the two experiments.

Before ending this discussion, let us mention that the
experimental covariance matrix defined in Eq.~(\ref{eq:covmat}), and used
in this section, is not suitable to be used in PDF fits, since
these would be affected by the D'Agostini bias~\cite{D'Agostini:1993uj}.
A more appropriate definition of the covariance matrix for PDF fits is 
provided by the $t_0$-{\it prescription}~\cite{Ball:2009qv},
\begin{equation}
C_{ij}^{t_0}
\equiv
\delta_{ij} \lp \sigma^{\rm stat}_{i}\rp ^2 
+ 
\left(
\sum_{\alpha =1}^{N_{\rm sysA}}\sigma_{i,\alpha}^{\rm sysA}\sigma_{j,\alpha}^{\rm sysA}
\right)
D_i D_j
+
\left(
\sum_{\alpha=1}^{N_{\rm sysM}}\sigma_{i,\beta}^{\rm sysM}
\sigma_{j,\beta}^{\rm sysM}
\right)
T_i^{(0)} T^{(0)}_j
\,\mbox{,}
\label{eq:covt0}
\end{equation}
in which a fixed theory prediction $\{T_i^{(0)}\}$ is used to define the 
contribution to the $\chi^2$
from the multiplicative systematic uncertainties.
The D'Agostini
bias, which would otherwise be introduced if the fit were performed
using 
the experimental definition Eq.~(\ref{eq:covmat}), is then avoided.
Therefore, all the PDF fits presented in the next section will be based
on Eq.~(\ref{eq:covt0}).

\section{PDF fits with top-quark pair differential distributions}
\label{sec:results}

We now present the main results
of this work, namely, NNLO PDF fits including top-quark pair 
differential distributions from ATLAS and CMS at $\sqrt{s}=8$ TeV.
We begin by describing the settings of the PDF fits,
based on the NNPDF framework, and
then present the results for various choices
of the baseline dataset (HERA-only or global)
and of the top-quark differential data (absolute or normalized,
and for different kinematic distributions).
With this procedure we
determine a suitable combination of top-quark
measurements to be used in PDF fits.
We then quantify the impact of the differential top data on the large-$x$ 
gluon and on the kinematical distributions not directly included in the fit.
Finally, we compare our results with the constraints
on the large-$x$ gluon provided by collider inclusive
jet measurements.

\subsection{Fit settings}
\label{sec:fitsettings}

The PDF fits presented in this work are based on a variant of the
NNPDF3.0 global analysis~\cite{Ball:2014uwa,Ball:2016neh}.
PDF evolution and deep-inelastic structure functions are
evaluated with the public code
{\tt APFEL}~\cite{Bertone:2013vaa,Bertone:2016lga},
with heavy quark structure functions computed in the FONLL-C general-mass 
scheme~\cite{Forte:2010ta} with pole masses and
with up to $n_f=5$ active quark flavours.
The charm PDF is generated perturbatively from light quarks and gluons. 
The value of the strong coupling is set to $\alpha_s(m_Z)=0.118$, 
consistently with the PDG average~\cite{Agashe:2014kda}.
For charm and bottom pole masses we use the values recommended
by the Higgs Cross-Section Working Group~\cite{deFlorian:2016spz},
namely 
$m_c=1.51$ GeV and $m_b=4.92$ GeV.
DGLAP evolution equations are solved up to NNLO using the 
truncated solution, and the input PDF parametrization
scale is taken to be $Q_0=1$ GeV.

For the fits presented here,
we have defined a baseline global dataset which includes:
fixed-target neutral-current DIS structure functions
from NMC~\cite{Arneodo:1996kd,Arneodo:1996qe}, BCDMS~\cite{bcdms1,bcdms2},
and SLAC~\cite{Whitlow:1991uw}; the legacy HERA
combinations for inclusive~\cite{Abramowicz:2015mha}
and charm~\cite{Abramowicz:1900rp} reduced cross-sections;
charged-current structure
functions from CHORUS inclusive neutrino DIS~\cite{Onengut:2005kv} and from
NuTeV dimuon production data~\cite{Goncharov:2001qe,MasonPhD};
fixed-target E605~\cite{Moreno:1990sf} and
E866~\cite{Webb:2003ps,Webb:2003bj,Towell:2001nh} DY production
data;
Tevatron collider data including
the
CDF~\cite{Aaltonen:2010zza} and D0~\cite{Abazov:2007jy} $Z$ rapidity
distributions; and LHC collider data including 
ATLAS~\cite{Aad:2011dm,Aad:2013iua,Aad:2011fp}, 
CMS~\cite{Chatrchyan:2012xt,Chatrchyan:2013mza,Chatrchyan:2013uja,CMSDY} 
and LHCb~\cite{Aaij:2012vn,Aaij:2012mda} vector boson production
measurements, adding up to a total of $N_{\rm dat}=3567$ data points.

This baseline global dataset is similar to that
of NNPDF3.0 with three important differences.
The first is in the HERA inclusive structure
functions, where the separate
HERA-II 
measurements from  H1 and
ZEUS~\cite{Aaron:2012qi,Collaboration:2010ry,Abramowicz:2012bx,
  Collaboration:2010xc}
have been replaced by the HERA legacy
combination~\cite{Abramowicz:2015mha}.
Secondly,  
inclusive top-quark production cross-sections
are excluded from the baseline, as we want to study
the impact of top data separately.
Finally, in order to ensure a consistent NNLO determination without 
approximations for the NNLO matrix elements,
we exclude jet production
measurements~\cite{Abulencia:2007ez,Aad:2011fc,Chatrchyan:2012bja,Aad:2013lpa}.
The impact of jet data as compared
to top data on the
large-$x$ gluon is discussed in Sect.~\ref{sec:jets}.

The influence of the differential top data on the gluon is assessed in two different scenarios.
In the first case, we start from a baseline PDF fit which 
includes only HERA deep-inelastic structure functions.
In the second case, we start from the NNPDF3.0-like baseline PDF fit 
described above.
Subsequently, for each fit, we include either the absolute or normalized
top-quark pair differential 
distributions, in the latter case supplemented with
the inclusive total cross-section data.
For completeness, we also perform a PDF fit
where only total cross-sections are included.
An overview of the datasets included in each fit is 
presented in Table~\ref{tab:listfits}.
We  emphasize again that including different
distributions from the same experiment would be double
counting, since the statistical correlations
among them are not available.

\begin{table}[t]
\footnotesize
\centering
\begin{tabular}{lcccccccccc}
\toprule
dataset & \multicolumn{10}{c}{Fit ID}\\
&  1 &  2  & 3 & 4  & 5  & 6  & 7  & 8  & 9  & 10 \\
\midrule
Baseline  &  {\bf y}  & {\bf y} & {\bf y} & {\bf y}& {\bf y} 
          &  {\bf y}  & {\bf y} & {\bf y}& {\bf y}& {\bf y} \\
\midrule
ATLAS\, $d\sigma/dp_T^t$ 
& n & n & {\bf y} & n & n & n & n & n & n & n \\
ATLAS\, $d\sigma/dy_t$  
& n & n & n & {\bf y} & n & n & n & n & n & n \\
ATLAS\, $d\sigma/dy_{t\bar{t}}$ 
& n & n & n & n & {\bf y} & n & n & n & n & n \\
ATLAS\, $d\sigma/dm_{t\bar{t}}$  
& n & n & n & n & n & {\bf y} & n & n & n & n \\
ATLAS\, $(1/\sigma)d\sigma/dp_T^t$  
& n & n & n & n & n & n & {\bf y} & n & n & n \\
ATLAS\, $(1/\sigma)d\sigma/dy_t$  
& n & n & n & n & n & n & n & {\bf y} & n & n \\
ATLAS\, $(1/\sigma)d\sigma/dy_{t\bar{t}}$  
& n & n & n & n & n & n & n & n & {\bf y} & n \\
ATLAS\, $(1/\sigma)d\sigma/dm_{t\bar{t}}$  
& n & n & n & n & n & n & n & n & n & {\bf y} \\
ATLAS\, $\sigma_{\rm t\bar{t}}$  
& n & {\bf y} & n & n & n & n & {\bf y} & {\bf y} & {\bf y} & {\bf y}\\
\midrule
CMS\, $d\sigma/dp_T^t$ 
& n & n & {\bf y} & n & n & n & n & n & n & n \\
CMS\, $d\sigma/dy_t$  
& n & n & n & {\bf y} & n & n & n & n & n & n \\
CMS\, $d\sigma/dy_{t\bar{t}}$ 
& n & n & n & n & {\bf y} & n & n & n & n & n \\
CMS\, $d\sigma/dm_{t\bar{t}}$  
& n & n & n & n & n & {\bf y} & n & n & n & n \\
CMS\, $(1/\sigma)d\sigma/dp_T^t$  
& n & n & n & n & n & n & {\bf y} & n & n & n \\
CMS\, $(1/\sigma)d\sigma/dy_t$  
& n & n & n & n & n & n & n & {\bf y} & n & n \\
CMS\, $(1/\sigma)d\sigma/dy_{t\bar{t}}$  
& n & n & n & n & n & n & n & n & {\bf y} & n \\
CMS\, $(1/\sigma)d\sigma/dm_{t\bar{t}}$  
& n & n & n & n & n & n & n & n & n & {\bf y} \\
CMS\, $\sigma_{\rm t\bar{t}}$  
& n & {\bf y} & n & n & n & n & {\bf y} & {\bf y} & {\bf y} & {\bf y}\\
\bottomrule
\end{tabular}
\caption{\small Overview of the fits presented in this work.
  The baseline dataset is composed by either the HERA
  structure functions or by the NNPDF3.0-like dataset (see text).
  For each fit, we indicate in boldface
  which top-quark pair measurement from ATLAS and
  CMS have been included. 
}
\label{tab:listfits}
\end{table}

\subsection{Results from the HERA-only fits}
\label{sec:heraonlyPDFfits}

We begin by discussing the results from the HERA-only fits
where the baseline dataset is composed exclusively of HERA deep-inelastic
structure function measurements.
The fit quality is assessed
by means of the $\chi^2$ computed
using the experimental
definition of the covariance matrix Eq.~(\ref{eq:covt0}).
These values are collected in Table~\ref{tab:chi2val1}
for each of the ten fits of Table~\ref{tab:listfits}.
The numbers in boldface
refer to the fits with the corresponding datasets included, 
whilst the rest of the entries have been obtained 
from the predictions of the resultant PDF fit in each column.
Note that fits of the normalized differential distributions
are supplemented by total cross-sections,
and that the first column is
the result of the HERA-only baseline fit.

From Table~\ref{tab:chi2val1}, we observe that in general it
is possible  to provide a satisfactory description
of most of the fitted differential distributions.
However in some cases the fit quality is somewhat poor: in the case of
 the CMS absolute (normalized) top rapidity
distributions $y_{t}$, we find that the value
of the $\chi^2$ is 1.75 (1.94), while the
corresponding ATLAS values are 1.06 (1.48).
The worst agreement between NNLO theory and data can be seen
in the top-quark pair invariant mass
$m_{t\bar{t}}$ normalized  distributions.
Here we find values of the $\chi^2$ as large as 6.26
and 3.03 for the ATLAS and CMS measurements,
respectively.
From Table~\ref{tab:chi2val1}, we also note that the quality of the description
of the HERA data does not deteriorates once top-quark pair differential 
distributions are added on top of it in the fit. The value of the $\chi^2$ 
per data point is remarkably stable among all fits and shows limited statistical
fluctuations.

\begin{table}[!t]
\footnotesize
\centering
\begin{tabular}{lcccccccccc}
\toprule
Dataset & \multicolumn{10}{c}{Fit ID}\\
          &  1 &  2  & 3 & 4  & 5  & 6  & 7  & 8  & 9  & 10 \\
\midrule   
HERA inclusive 
          & {\bf 1.18} & {\bf 1.18} & {\bf 1.19} & {\bf 1.19} & {\bf 1.19} 
          & {\bf 1.19} & {\bf 1.18} & {\bf 1.18} & {\bf 1.19} & {\bf 1.19} \\
HERA $F_2^c$
          & {\bf 1.07} & {\bf 1.05} & {\bf 1.05} & {\bf 1.05} & {\bf 1.06} 
          & {\bf 1.05} & {\bf 1.08} & {\bf 1.07} & {\bf 1.06} & {\bf 1.06} \\
\midrule
ATLAS\, $d\sigma/dp_T^t$ 
          & 2.30 & 2.48 & {\bf 0.73} & 3.16 & 3.46 
          & 2.04 & 1.34 & 3.28       & 4.88 & 2.89\\
ATLAS\, $d\sigma/dy_t$ 
          & 0.82 & 1.14 & 1.21 & {\bf 1.06} & 0.75 
          & 1.04 & 1.31 & 0.59 & 0.75       & 0.74 \\
ATLAS\, $d\sigma/dy_{t\bar{t}}$ 
          & 1.12 & 1.90 & 2.40 & 2.83 & {\bf 0.45} 
          & 4.43 & 1.96 & 1.88 & 0.40 & 1.49 \\
ATLAS\, $d\sigma/dm_{t\bar{t}}$  
          & 4.27        & 2.93 & 2.41 & 2.81 & 4.33 
          & {\bf 1.53}  & 2.70 & 2.88 & 4.37 & 5.09 \\
ATLAS\, $(1/\sigma)d\sigma/dp_T^t$  
          & 3.47 & 2.60       & 3.80 & 2.92 & 3.15 
          & 3.91 & {\bf 1.46} & 3.31 & 3.98 & 4.01 \\
ATLAS\, $(1/\sigma)d\sigma/dy_t$  
          & 1.21 & 6.07 & 3.32       & 5.95 & 1.34 
          & 2.24 & 4.27 & {\bf 1.48} & 1.58 & 1.61 \\
ATLAS\, $(1/\sigma)d\sigma/dy_{t\bar{t}}$  
          & 3.11 & 12.8 & 5.09 & 8.34       & 0.72 
          & 7.04 & 4.95 & 3.60 & {\bf 0.53} & 2.60 \\
ATLAS\, $(1/\sigma)d\sigma/dm_{t\bar{t}}$  
          & 8.14 & 3.07 & 6.53 & 4.94 & 5.42 
          & 20.5 & 6.44 & 5.61 & 4.40 & {\bf 3.03} \\
ATLAS\, $\sigma_{\rm t\bar{t}}$  
          & 3.88 & {\bf 0.35} & 3.38 & 0.63 & 1.58 
          & 1.29 & {\bf 0.87} & {\bf 0.37} & {\bf 0.42} & {\bf 0.66} \\
\midrule
CMS\, $d\sigma/dp_T^t$ 
          & 2.04 & 2.29 & {\bf 0.82} & 3.29 & 2.99 
          & 1.52 & 1.44 & 2.81 & 4.16 & 2.32 \\
CMS\, $d\sigma/dy_t$  
          & 3.38 & 2.48 & 2.91 & {\bf 1.75} & 3.51 
          & 3.47 & 2.32 & 3.03 & 3.48 & 4.81 \\
CMS\, $d\sigma/dy_{t\bar{t}}$ 
          & 1.00 & 1.58 & 2.29 & 1.68 & {\bf 1.08} 
          & 3.05 & 1.51 & 1.34 & 1.07 & 1.85 \\
CMS\, $d\sigma/dm_{t\bar{t}}$  
          & 3.96 & 5.85 & 4.81 & 4.70 & 4.23 
          & {\bf 1.73} & 4.46 & 4.23 & 4.71 & 3.74 \\
CMS\, $(1/\sigma)d\sigma/dp_T^t$  
          & 2.78 & 4.86 & 1.78 & 5.23 & 4.05 
          & 2.84 & {\bf 1.57} & 4.69 & 5.29 & 3.40 \\
CMS\, $(1/\sigma)d\sigma/dy_t$  
          & 5.73 & 3.15 & 4.10 & 2.35 & 5.04 
          & 4.88 & 3.13 & {\bf 1.94} & 4.60 & 6.71 \\
CMS\, $(1/\sigma)d\sigma/dy_{t\bar{t}}$  
          & 1.68 & 2.27 & 2.62 & 2.11 & 1.40 
          & 3.42 & 1.78 & 1.49 & {\bf 1.20} & 1.98 \\
CMS\, $(1/\sigma)d\sigma/dm_{t\bar{t}}$  
          & 5.30 & 10.3 & 7.83 & 8.24 & 7.06 
          & 2.71 & 7.45 & 7.41 & 8.06 & {\bf 6.26} \\
CMS\, $\sigma_{\rm t\bar{t}}$  
          & 6.95 & {\bf 1.04} & 6.17 & 1.59 & 3.24 
          & 2.75 & {\bf 1.02} & {\bf 1.09} & {\bf 1.17} & {\bf 1.64}\\
\midrule
TOTAL
& 1.18 & 1.18 & 1.17 & 1.19 & 1.18 
& 1.19 & 1.18 & 1.20 & 1.18 & 1.22\\
\bottomrule
\end{tabular}
\caption{\small The $\chi^2$ 
  from the HERA-only PDF fits for
  various combinations of top-quark data.
The numbers in boldface
indicate the datasets included in the fit,
while the other entries describe the quality of the 
predictions of the resultant PDF fit for 
the other distributions.
}
\label{tab:chi2val1}
\end{table}

Because the gluon PDF has little sensitivity to the HERA data in the region 
where instead it is sensitive to the LHC top differential data (roughly
$0.08\lesssim x \lesssim 0.5$, see Fig.~\ref{fig:correlations}),
the poor agreement between data and
theory for some distributions
cannot be attributed to a tension with one of the other
input datasets in the fit. 
%
The disagreement therefore appears to be the result of 
a genuine tension between the ATLAS and CMS measurements.
As we will show below, this effect is 
only exacerbated in the global fits, where there are additional
constraints on the gluon from other experiments.
Further evidence for an inconsistency is provided by examining fits
where ATLAS and CMS data are included separately. In such a case
the description of the data by NNLO theory is substantially improved
(see appendix~\ref{sec:compatibility}).
It is also interesting to note that the inclusion 
of the total cross-section data in the fit does not necessarily imply a good 
description of the differential distributions.
This highlights the fact that constraints on the large-$x$ gluon 
stemming from inclusive cross-sections are only a subset of those
 obtained when fitting the fully differential distributions.

\begin{figure}[t]
\centering
\includegraphics[scale=0.26,angle=270]{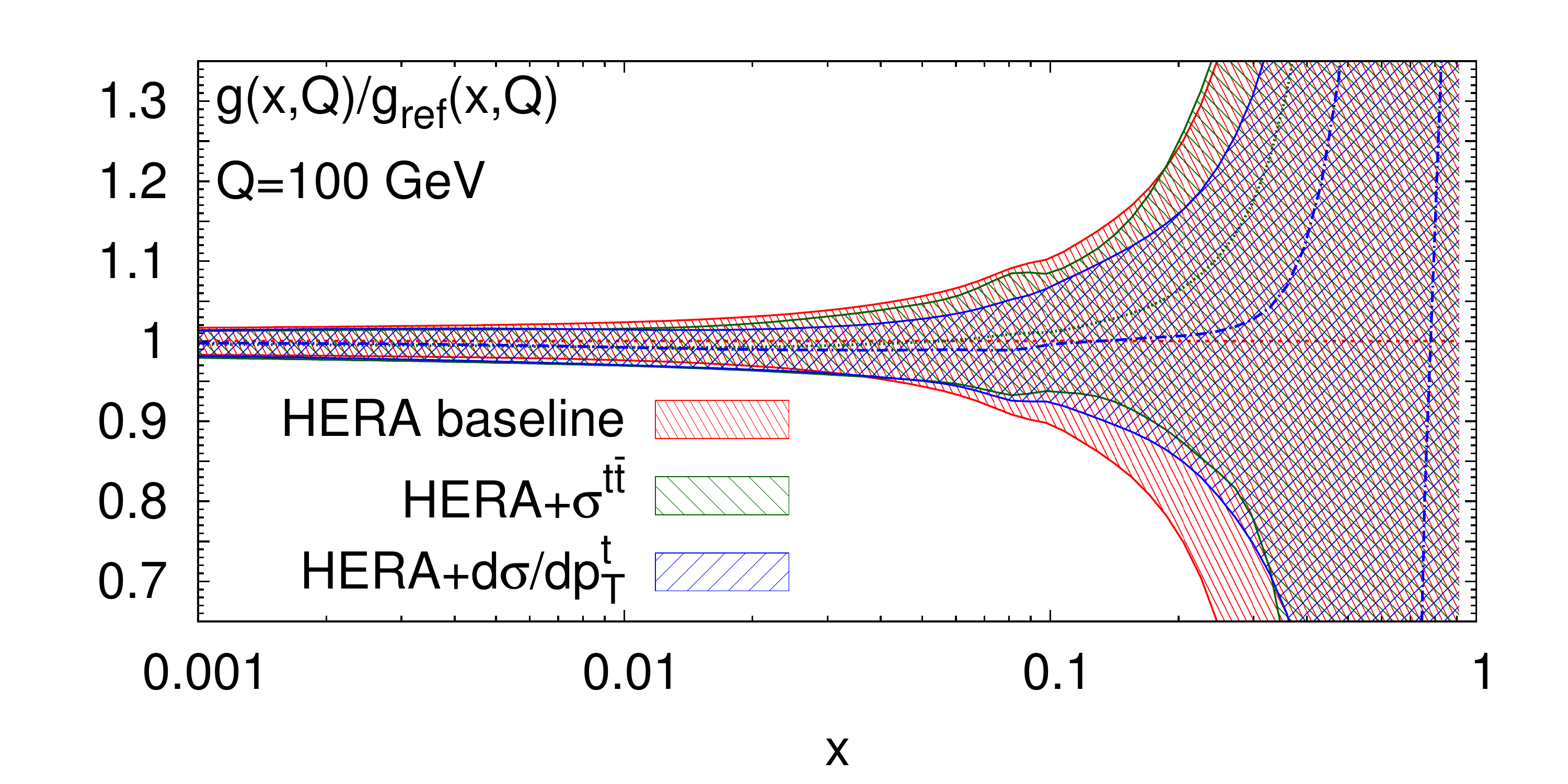}
\includegraphics[scale=0.26,angle=270]{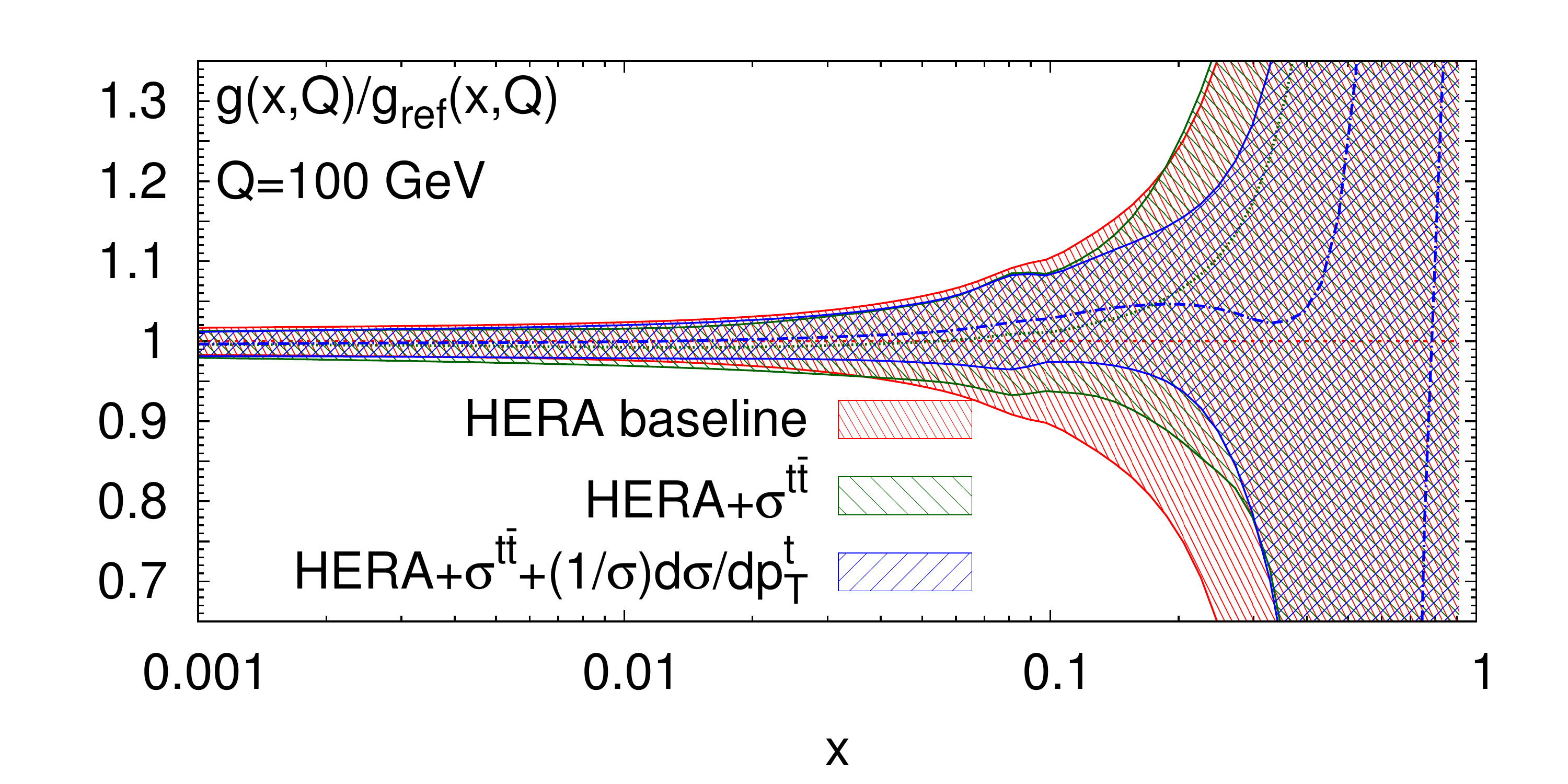}\\
\includegraphics[scale=0.26,angle=270]{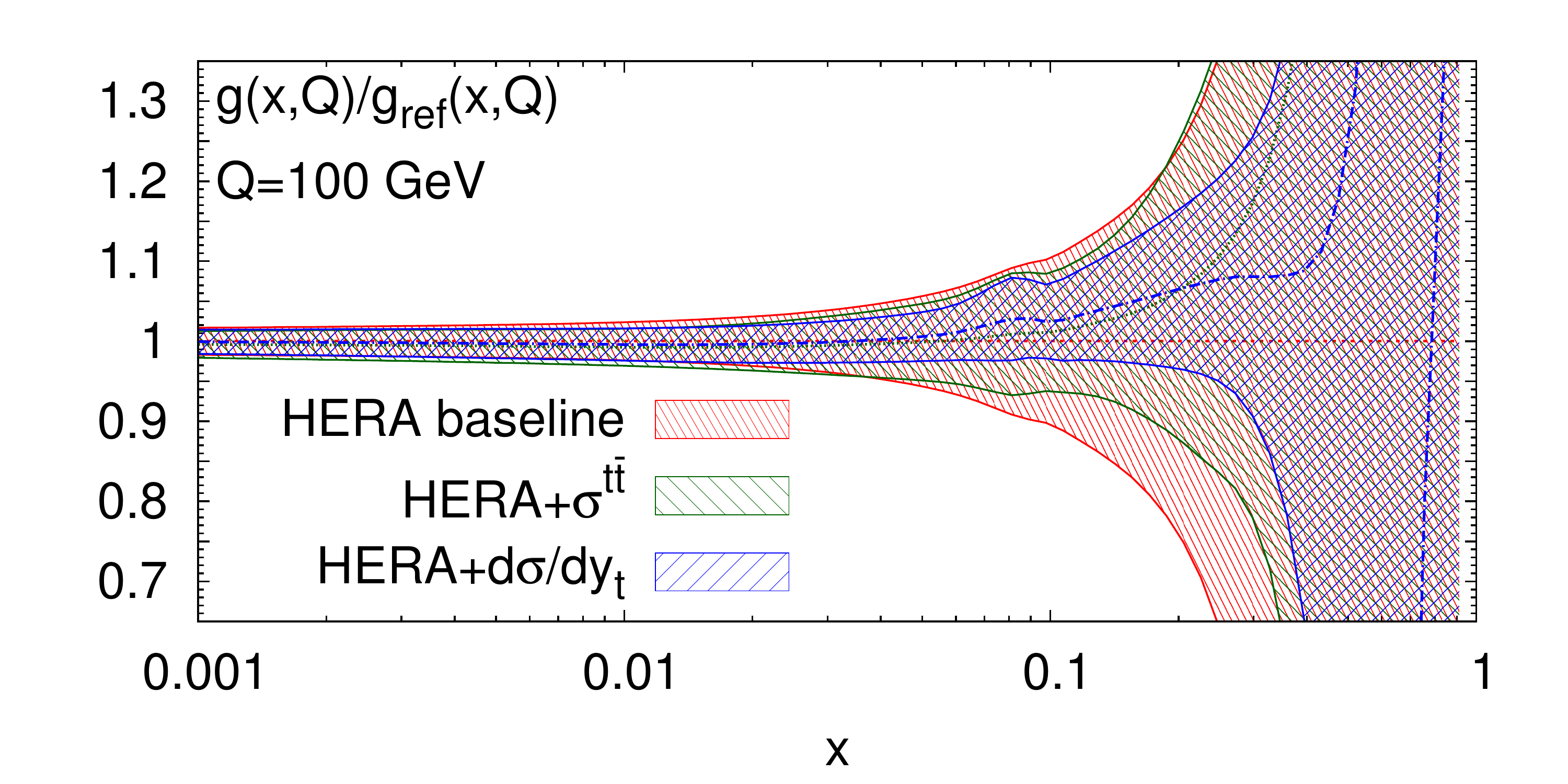}
\includegraphics[scale=0.26,angle=270]{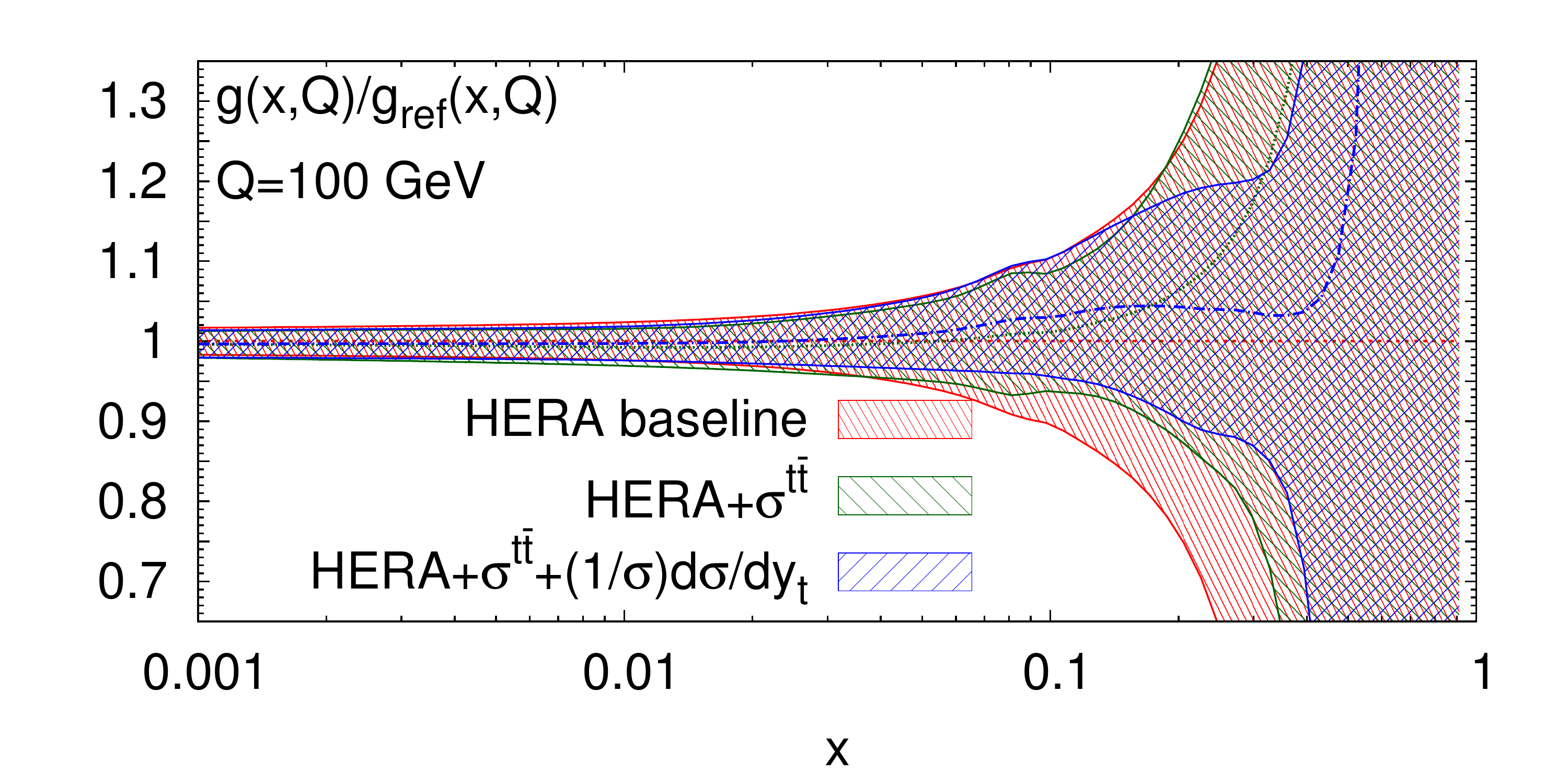}\\
\includegraphics[scale=0.26,angle=270]{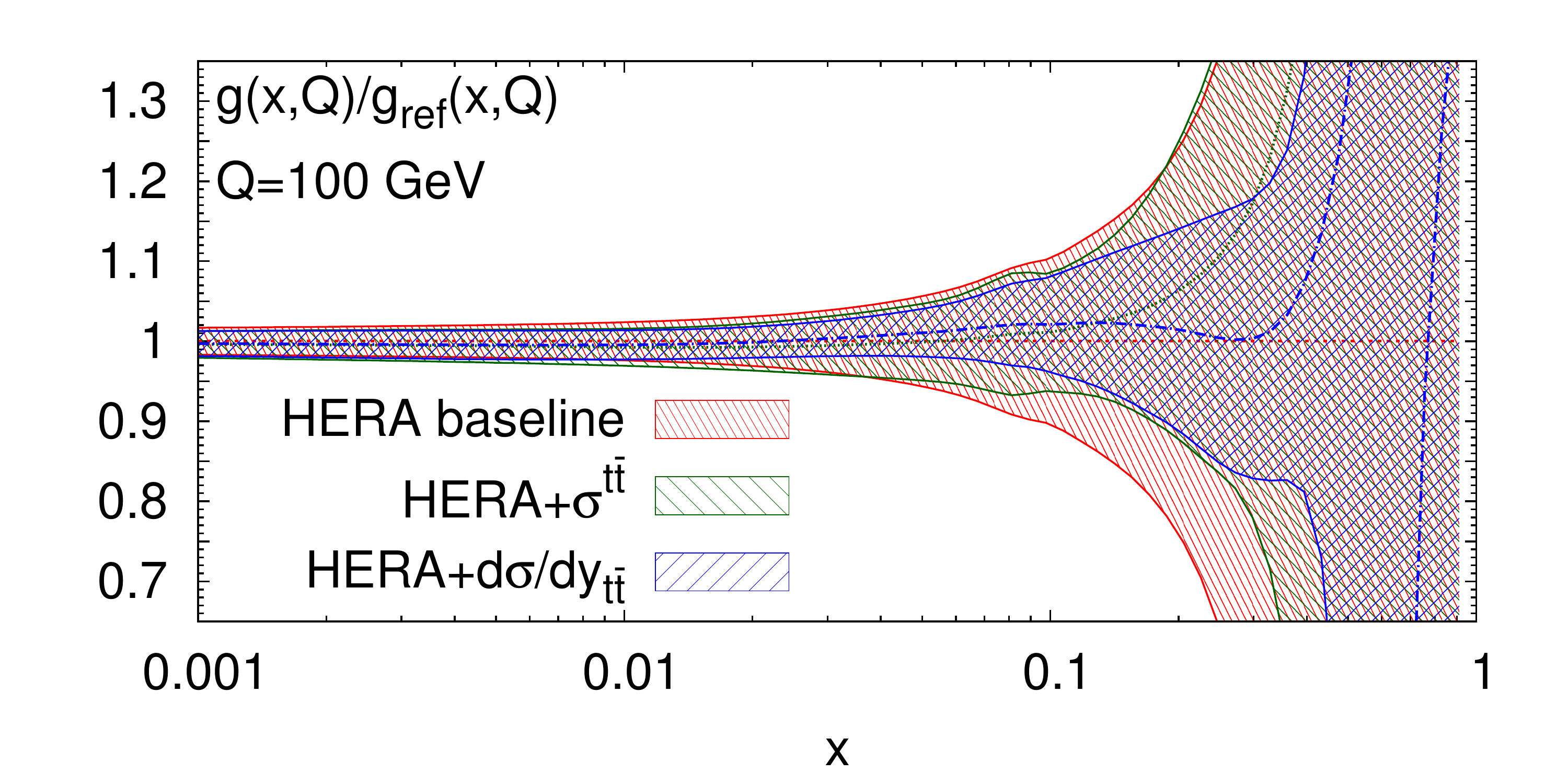}
\includegraphics[scale=0.26,angle=270]{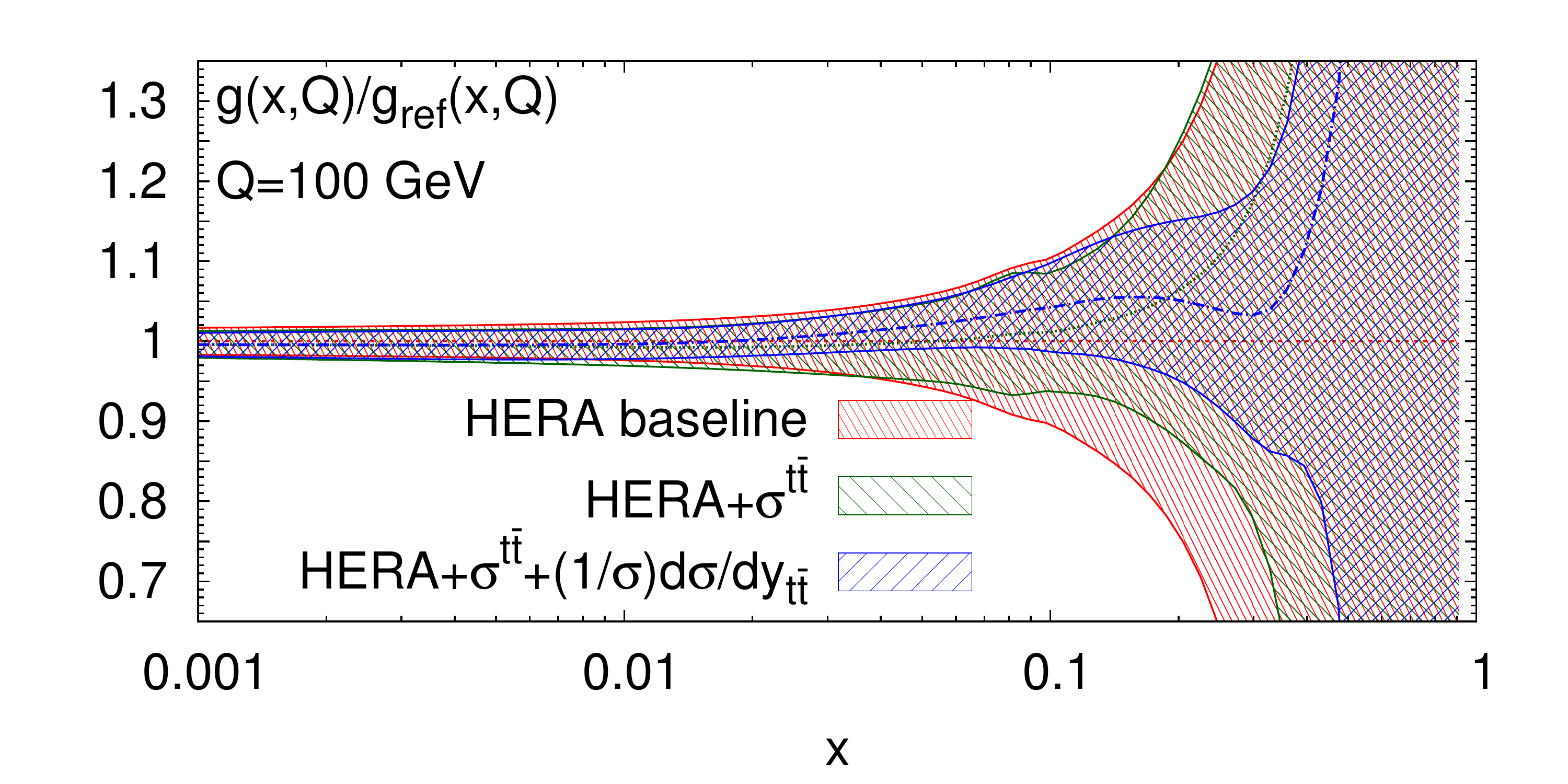}\\
\includegraphics[scale=0.26,angle=270]{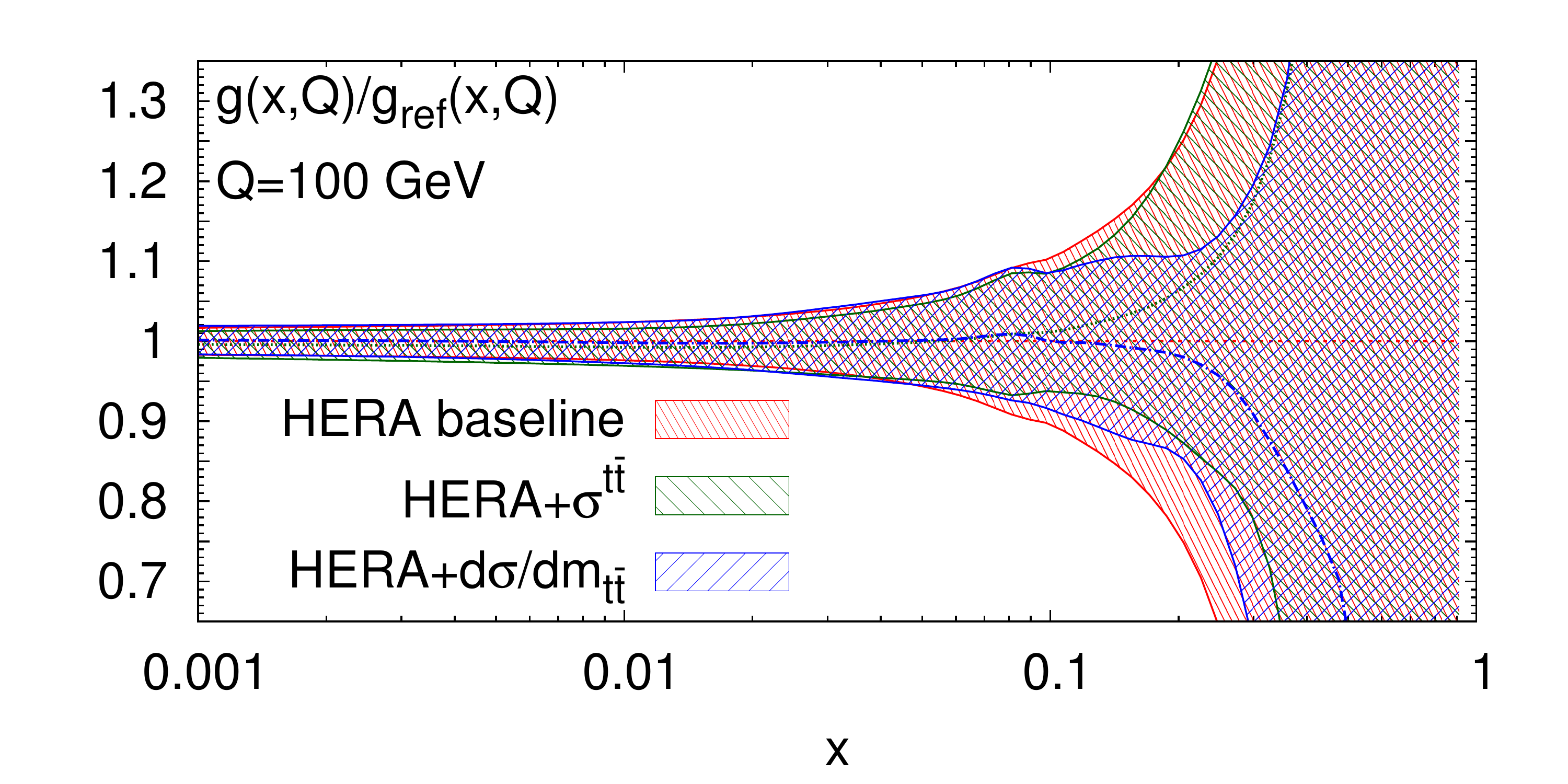}
\includegraphics[scale=0.26,angle=270]{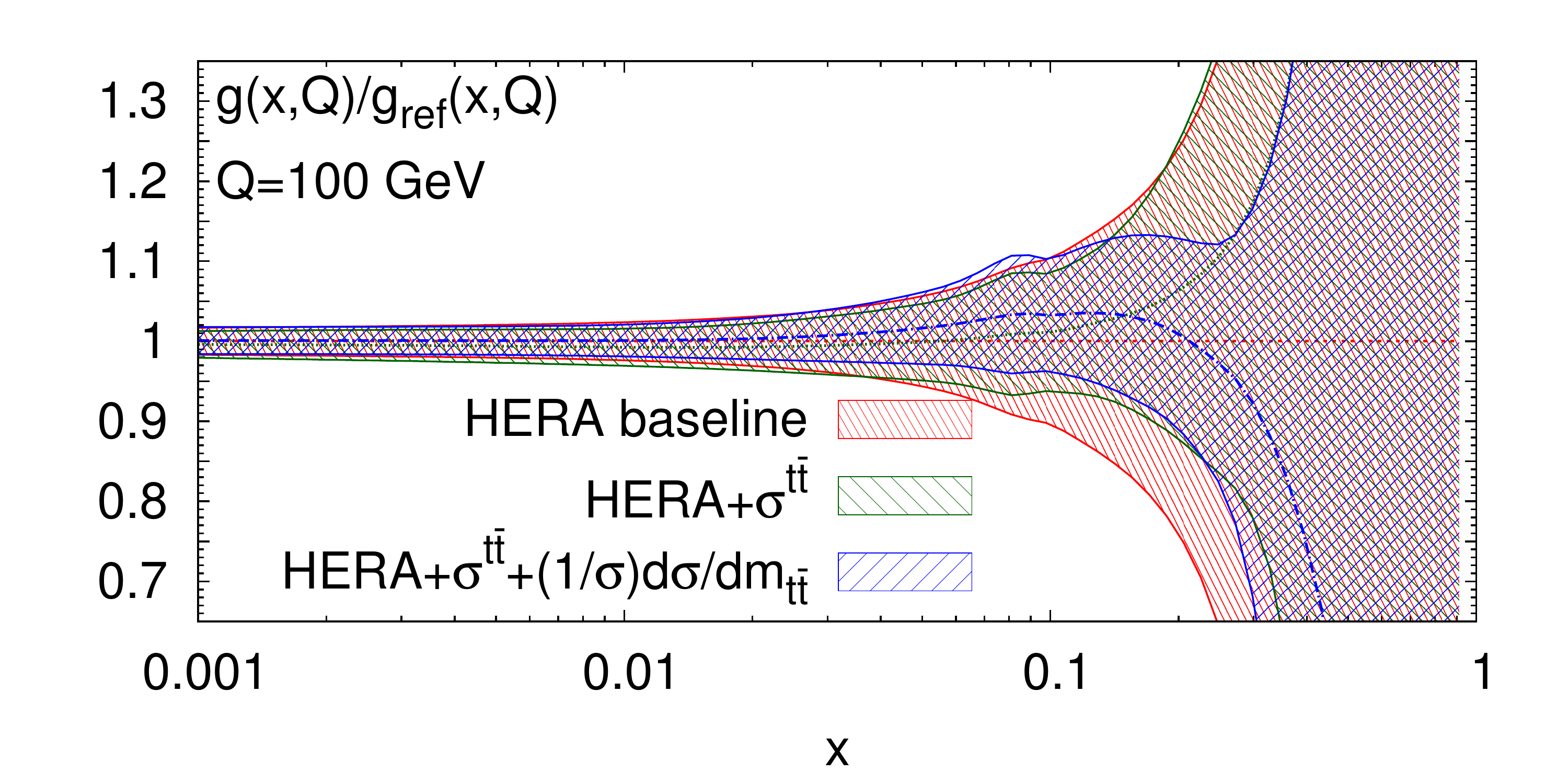}\\
\caption{\small
  The gluon PDF from the HERA-only fits corresponding to the same
  input combinations of LHC top-quark data,
both at the level of absolute (left) and
   of normalized
   distributions (right).
  \label{fig:gluonfit}
  }
\end{figure}

\begin{figure}
\centering
\includegraphics[scale=0.26,angle=270]{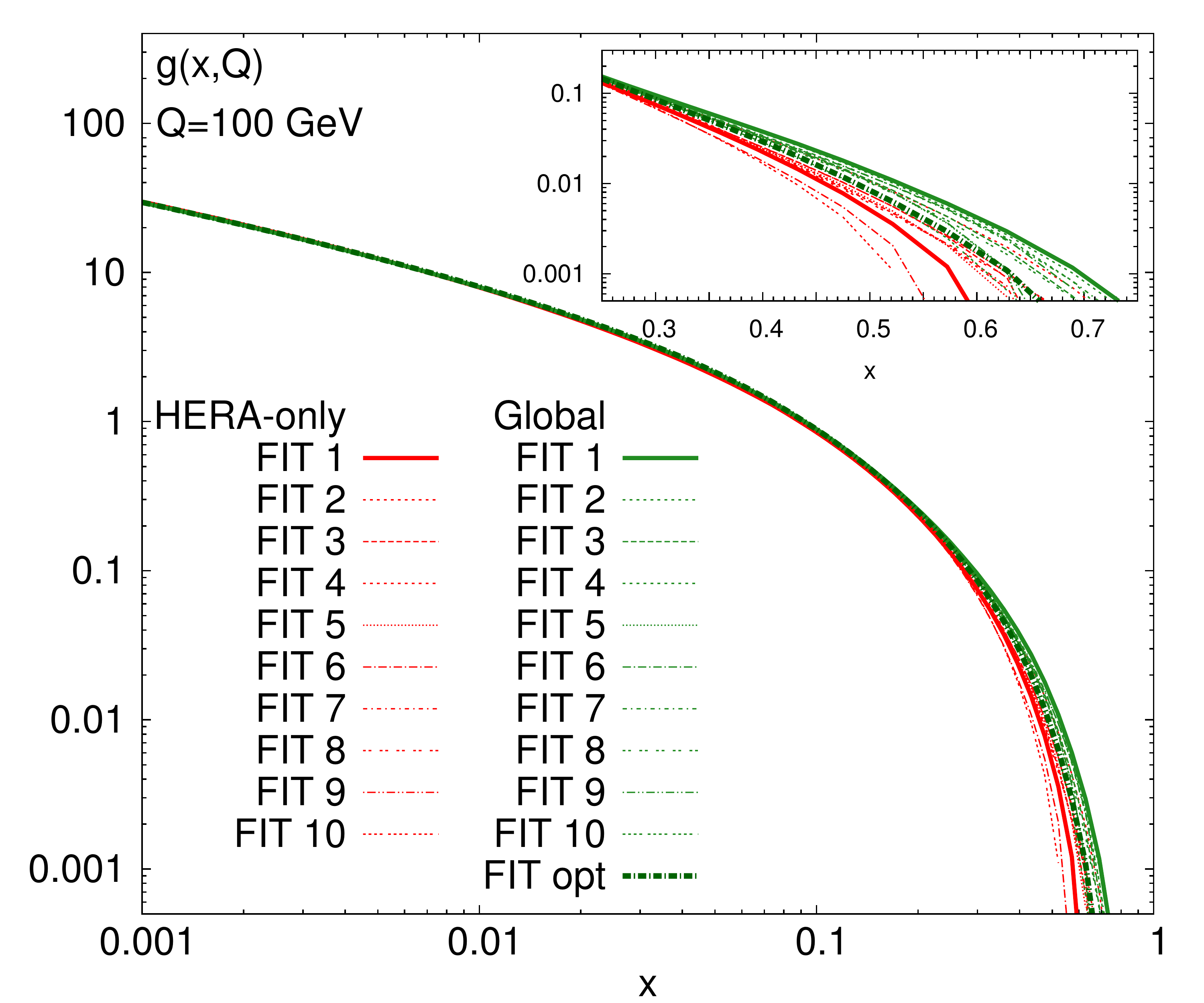}
\includegraphics[scale=0.26,angle=270]{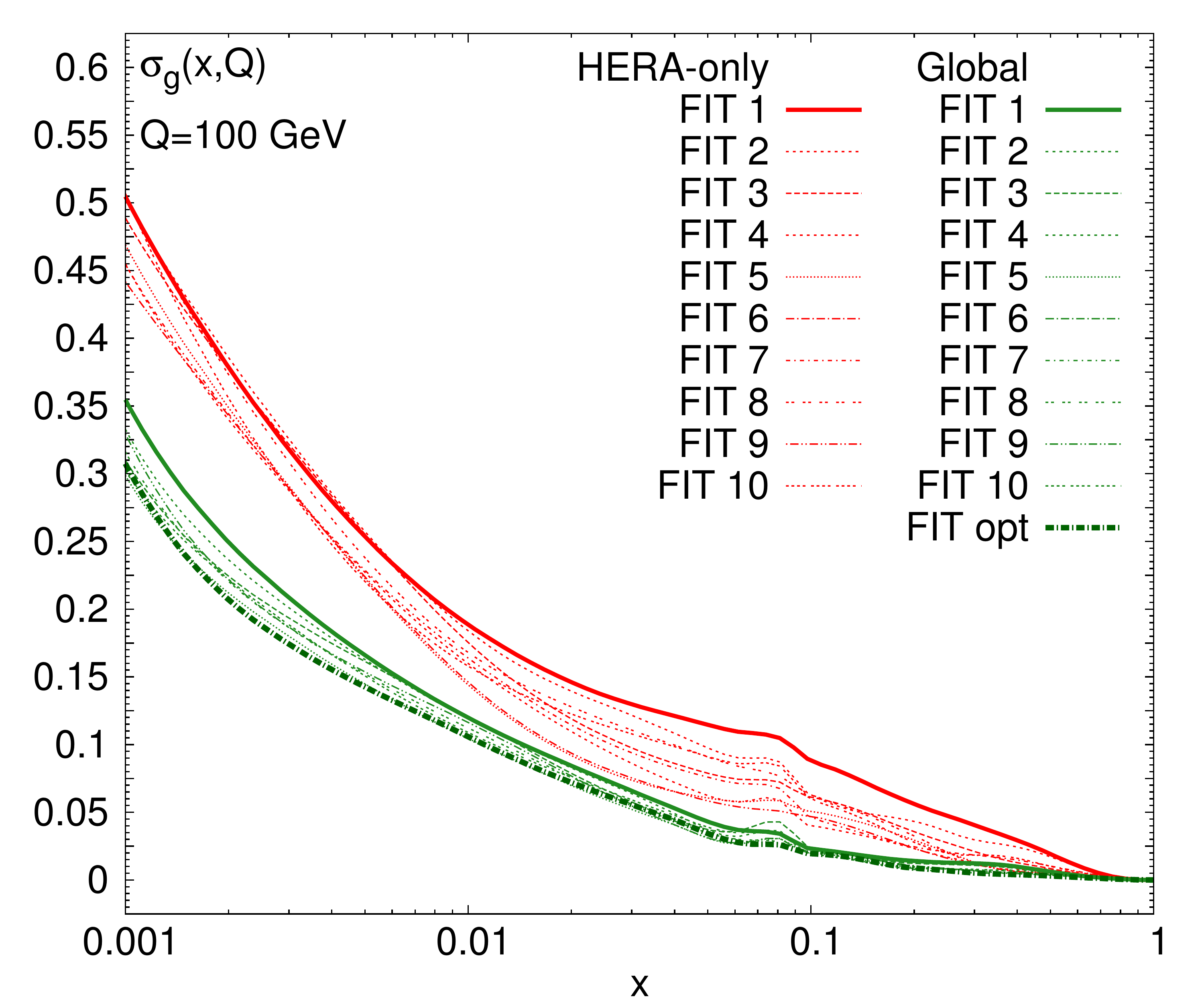}\\
\caption{\small The central value (left) and the one-sigma uncertainty (right) of the 
gluon PDF at $Q=100$ GeV for all the HERA-only and the global fits, including
our optimal fit to the optimal combination of top-quark pair differential
distributions (FIT opt), 
collected in Tables~\ref{tab:chi2val1}-\ref{tab:chi2val2}.
The inset in the left plot focuses on the gluon central values
in the large-$x$ region.}
\label{fig:gluonexplicit}
\end{figure}

In Fig.~\ref{fig:gluonfit} we compare the gluon
from the HERA-only baseline fit with those obtained
through fits to the various
combinations of ATLAS and CMS top quark differential cross-sections.
The comparison is performed at the scale
$Q=100$ GeV, and the results are shown
normalized to the central value of the  HERA-only baseline.
For completeness, we also show the results of the fit where
only the total cross-sections $\sigma^{t\bar{t}}$ is included.
In Fig.~\ref{fig:gluonexplicit}, we also display the central value and the 
one-sigma uncertainty of the gluon PDF at $Q=100$ GeV for all the HERA-only 
fits collected in Table~\ref{tab:chi2val1}.
We observe that the
various distributions demonstrate
a fair degree of consistency in their impact on the gluon.
For most of the considered kinematic distributions,
both normalized and absolute, we find that top-quark data prefers a harder
gluon at large $x$ as compared to the HERA-only fit.
This trend can also be observed for the fits including
only total cross-sections.
An exception arises in the fits with the differential $m_{t\bar{t}}$
distributions, which are however those for which the fit
quality is worst.

We also observe that the three types of fits (HERA-only, HERA with
total cross-sections, and HERA with differential
distributions) turn out to be fully consistent within
the respective PDF uncertainties.
Moreover, the reduction of
the PDF uncertainty in the
large-$x$ gluon appears to be similar
for both absolute and normalized distributions.
The PDF uncertainty is reduced
for $x\gsim 0.05$, which is the kinematic range accessed
by differential top-quark pair production data
(see Fig.~\ref{fig:correlations}).
In these HERA-only fits, the considered four kinematic 
distributions exhibit comparable constraining power.

While the HERA-only fits provide a clean testing ground to validate
the implementation of top-quark differential distributions
in a PDF fit, it is important to investigate the impact
of these datasets in a global analysis,
which then could be used for LHC phenomenology.
This is done in the next section.

\subsection{Results from the global fits}
\label{sec:global}

We now present the results of the NNLO fits in which the top-quark
data has been added to a baseline fit based on the global dataset.
As in the case of the HERA-only fits in Table~\ref{tab:chi2val1},
first of all we collect the values of the $\chi^2$ 
in Table~\ref{tab:chi2val2}. 
Like in the HERA-only case,
the description of the global baseline dataset does not deteriorate when 
any top-quark pair differential distribution is added on top of it in the 
fit. A slight worsening of the $\chi^2$ per data point is  observed for 
fixed-target DY and ATLAS and CMS vector boson production datasets, though it 
does not seem to be statistically significant.
For some top-quark pair distributions, a good agreement
between NNLO theory
and data is found after the fit.
These include the ATLAS absolute and normalized $y_t$ and
absolute $y_{t\bar{t}}$ distributions and the CMS
absolute and normalized $y_{t\bar{t}}$ ones.

On the other hand, we also find that for some distributions
the values of the $\chi^2$ worsen in the global fits as
compared to the HERA-only fits.
In the case of the $p_T^t$ distribution, the $\chi^2$ values for
the ATLAS and CMS absolute (normalized) distributions
are 1.99 and 2.60 (2.96 and 3.56) respectively, to be compared
with 0.73 and 0.82 (1.46 and 1.57) in the HERA-only fits.
This increase in the $\chi^2$ is also pronounced for the $m_{t\bar{t}}$
distribution, where in the global fits the
values of the $\chi^2$ for
the ATLAS and CMS data are 4.02 and 5.11 (2.98 and 7.27) 
for the absolute (normalized) distributions, while
we found 1.53 and 1.73 (3.03 and 6.26) in the HERA-only fits.
For the normalized $m_{t\bar{t}}$ distributions instead, the
$\chi^2$ is equally poor in the global and in the HERA-only fits.
When the ATLAS and CMS data are included separately in the fit,
the $\chi^2$ values exhibit a significant reduction, though they do not turn 
out to be as good as in the corresponding HERA-only case
(see appendix~\ref{sec:compatibility}). This behaviour might be related to a
residual tension between some top-quark pair distributions and other 
experiments included in the global fit, as we will discuss further in 
appendix~\ref{sec:reduceddatasets}.

\begin{table}[!t]
\footnotesize
\centering
\begin{tabular}{lccccccccccc}
\toprule
Dataset & \multicolumn{11}{c}{Fit ID}\\
&  1 &  2  & 3 & 4  & 5  & 6  & 7  & 8  & 9  & 10 & opt\\
\midrule
NMC
& {\bf 1.39} & {\bf 1.37} & {\bf 1.37} & {\bf 1.38} & {\bf 1.37} 
& {\bf 1.38} & {\bf 1.38} & {\bf 1.40} & {\bf 1.39} & {\bf 1.39} & {\bf 1.39}\\
\midrule
SLAC
& {\bf 0.67} & {\bf 0.70} & {\bf 0.66} & {\bf 0.67} & {\bf 0.66} 
& {\bf 0.70} & {\bf 0.67} & {\bf 0.65} & {\bf 0.65} & {\bf 0.65} & {\bf 0.65}\\
\midrule
BCDMS
& {\bf 1.25} & {\bf 1.25} & {\bf 1.26} & {\bf 1.24} & {\bf 1.25} 
& {\bf 1.24} & {\bf 1.25} & {\bf 1.25} & {\bf 1.25} & {\bf 1.25} & {\bf 1.25}\\
\midrule
CHORUS
& {\bf 1.10} & {\bf 1.09} & {\bf 1.11} & {\bf 1.11} & {\bf 1.10} 
& {\bf 1.09} & {\bf 1.09} & {\bf 1.09} & {\bf 1.10} & {\bf 1.10} & {\bf 1.10}\\
\midrule
NuTeV 
& {\bf 0.76} & {\bf 0.71} & {\bf 0.69} & {\bf 0.68} & {\bf 0.67} 
& {\bf 0.74} & {\bf 0.71} & {\bf 0.70} & {\bf 0.66} & {\bf 0.67} & {\bf 0.67}\\
\midrule
HERA inlusive
& {\bf 1.21} & {\bf 1.20} & {\bf 1.21} & {\bf 1.21} & {\bf 1.21} 
& {\bf 1.21} & {\bf 1.21} & {\bf 1.22} & {\bf 1.21} & {\bf 1.21} & {\bf 1.22}\\
\midrule
HERA $F_2^c$
& {\bf 1.18} & {\bf 1.20} & {\bf 1.19} & {\bf 1.18} & {\bf 1.21} 
& {\bf 1.20} & {\bf 1.20} & {\bf 1.18} & {\bf 1.19} & {\bf 1.20} & {\bf 1.18}\\
\midrule
E866
& {\bf 1.29} & {\bf 1.28} & {\bf 1.31} & {\bf 1.32} & {\bf 1.31} 
& {\bf 1.31} & {\bf 1.29} & {\bf 1.31} & {\bf 1.30} & {\bf 1.31} & {\bf 1.33}\\
\midrule
E605
& {\bf 0.80} & {\bf 0.82} & {\bf 0.82} & {\bf 0.83} & {\bf 0.83} 
& {\bf 0.82} & {\bf 0.82} & {\bf 0.82} & {\bf 0.83} & {\bf 0.83} & {\bf 0.83}\\
\midrule
CDF $Z$ rapidity
& {\bf 1.42} & {\bf 1.44} & {\bf 1.43} & {\bf 1.42} & {\bf 1.43} 
& {\bf 1.45} & {\bf 1.45} & {\bf 1.42} & {\bf 1.43} & {\bf 1.46} & {\bf 1.40}\\
\midrule
D0 $Z$ rapidity
& {\bf 0.59} & {\bf 0.59} & {\bf 0.60} & {\bf 0.59} & {\bf 0.59} 
& {\bf 0.60} & {\bf 0.59} & {\bf 0.59} & {\bf 0.59} & {\bf 0.60} & {\bf 0.60}\\
\midrule
LHCb $W$, $Z$ rapidity
& {\bf 1.09} & {\bf 1.13} & {\bf 1.12} & {\bf 1.10} & {\bf 1.12} 
& {\bf 1.13} & {\bf 1.12} & {\bf 1.13} & {\bf 1.12} & {\bf 1.12} & {\bf 1.11}\\
\midrule
ATLAS $W$, $Z$ 2010
& {\bf 1.10} & {\bf 1.10} & {\bf 1.18} & {\bf 1.14} & {\bf 1.17} 
& {\bf 1.16} & {\bf 1.17} & {\bf 1.15} & {\bf 1.18} & {\bf 1.20} & {\bf 1.15}\\
ATLAS high-mass DY
& {\bf 1.36} & {\bf 1.34} & {\bf 1.32} & {\bf 1.33} & {\bf 1.30} 
& {\bf 1.33} & {\bf 1.30} & {\bf 1.32} & {\bf 1.29} & {\bf 1.30} & {\bf 1.31}\\
\midrule
ATLAS\, $d\sigma/dp_T^t$ 
& 2.37 & 2.30 & {\bf 1.99} & 2.36 & 2.24 
& 2.23 & 2.09 & 2.18 & 2.34 & 2.24 & 2.19\\
ATLAS\, $d\sigma/dy_t$  
& 0.93 & 0.80 & 0.74 & {\bf 1.09} & 0.76 
& 0.76 & 0.86 & 0.69 & 0.76 & 0.66 & 0.64\\
ATLAS\, $d\sigma/dy_{t\bar{t}}$ 
& 2.44 & 2.03 & 1.96 & 2.59 & {\bf 1.32} 
& 2.32 & 2.11 & 1.74 & 1.26 & 1.80 & 1.84\\
ATLAS\, $d\sigma/dm_{t\bar{t}}$  
& 4.27 & 4.47 & 4.68 & 4.14 & 4.92 
& {\bf 4.02} & 4.34 & 4.79 & 4.98 & 4.99 & 5.01\\
ATLAS\, $(1/\sigma)d\sigma/dp_T^t$  
& 2.93 & 3.97 & 3.29 & 4.36 & 5.22 
& 4.35 & {\bf 2.96} & 4.26 & 4.92 & 5.68 & 2.49\\
ATLAS\, $(1/\sigma)d\sigma/dy_t$  
& 5.00 & 3.17 & 2.47 & 6.36 & 1.55 
& 2.93 & 3.94 & {\bf 1.68} & 1.45 & 1.10 & {\bf 1.16}\\
ATLAS\, $(1/\sigma)d\sigma/dy_{t\bar{t}}$  
& 9.69 & 5.59 & 5.89 & 8.95 & 2.68 
& 5.73 & 6.73 & 3.57 & {\bf 2.17} & 3.73 & 3.81\\
ATLAS\, $(1/\sigma)d\sigma/dm_{t\bar{t}}$  
& 2.30 & 2.80 & 3.31 & 2.67 & 3.96 
& 4.21 & 3.09 & 3.68 & 3.77 & {\bf 2.98} & 4.55\\
ATLAS\, $\sigma_{\rm t\bar{t}}$  
& 0.12 & {\bf 0.10} & 0.21 & 0.10 & 0.10 
& 0.12 & {\bf 0.36} & {\bf 0.29} & {\bf 0.26} & {\bf 0.10} & {\bf 0.78}\\
\midrule
CMS $W$ electron asy
& {\bf 0.58} & {\bf 0.57} & {\bf 0.55} & {\bf 0.59} & {\bf 0.57} 
& {\bf 0.58} & {\bf 0.56} & {\bf 0.58} & {\bf 0.57} & {\bf 0.56} & {\bf 0.58}\\
CMS $W$ muon asy
& {\bf 1.66} & {\bf 1.69} & {\bf 1.67} & {\bf 1.75} & {\bf 1.70} 
& {\bf 1.65} & {\bf 1.65} & {\bf 1.64} & {\bf 1.70} & {\bf 1.65} & {\bf 1.68}\\
CMS 2D DY 2011
& {\bf 1.62} & {\bf 1.61} & {\bf 1.63} & {\bf 1.62} & {\bf 1.63} 
& {\bf 1.64} & {\bf 1.63} & {\bf 1.64} & {\bf 1.63} & {\bf 1.64} & {\bf 1.63}\\
\midrule
CMS\, $d\sigma/dp_T^t$ 
& 3.50 & 3.46 & {\bf 2.60} & 3.50 & 3.03 
& 3.00 & 2.85 & 3.11 & 3.24 & 2.92 & 2.91\\
CMS\, $d\sigma/dy_t$  
& 3.48 & 3.71 & 4.05 & {\bf 2.66} & 4.18 
& 3.49 & 3.38 & 4.23 & 4.43 & 4.99 & 4.98\\
CMS\, $d\sigma/dy_{t\bar{t}}$ 
& 1.36 & 1.13 & 1.00 & 1.32 & {\bf 0.89} 
& 0.86 & 1.00 & 1.01 & 1.04 & 1.24 & 1.07\\
CMS\, $d\sigma/dm_{t\bar{t}}$  
& 7.07 & 6.27 & 5.79 & 6.33 & 5.09 
& {\bf 5.11} & 6.00 & 5.37 & 5.21 & 4.31 & 4.77\\
CMS\, $(1/\sigma)d\sigma/dp_T^t$  
& 4.31 & 4.00 & 3.39 & 4.28 & 3.65 
& 3.59 & {\bf 3.56} & 3.57 & 3.73 & 3.48 & 3.33\\
CMS\, $(1/\sigma)d\sigma/dy_t$  
& 3.66 & 4.10 & 4.45 & 3.10 & 4.98 
& 4.06 & 3.65 & {\bf 4.76} & 5.13 & 6.09 & 5.78\\
CMS\, $(1/\sigma)d\sigma/dy_{t\bar{t}}$  
& 1.59 & 1.20 & 1.06 & 1.73 & 0.94 
& 1.01 & 1.20 & 0.99 & {\bf 1.05} & 1.32 & {\bf 1.05}\\
CMS\, $(1/\sigma)d\sigma/dm_{t\bar{t}}$  
& 12.0 & 10.8 & 9.81 & 11.1 & 8.72 
& 8.72 & 10.3 & 9.15 & 8.97 & {\bf 7.27} & 8.05 \\
CMS\, $\sigma_{\rm t\bar{t}}$  
& 0.10 & {\bf 0.05} & 0.26 & 0.19 & 0.32 
& 0.21 & {\bf 0.11} & {\bf 0.10} & {\bf 0.15} & {\bf 0.35} & {\bf 0.50}\\
\midrule
TOTAL
& 1.20 & 1.19 & 1.20 & 1.20 & 1.19
& 1.21 & 1.20 & 1.21 & 1.20 & 1.21 & 1.20\\
\bottomrule
\end{tabular}
\caption{\small Same as Table~\ref{tab:chi2val1} for the global 
fits, including (last column) our optimal fit to the optimal combination of 
top-quark pair differential distributions.}
\label{tab:chi2val2}
\end{table}

\begin{figure}[!t]
\centering
\includegraphics[scale=0.26,angle=270]{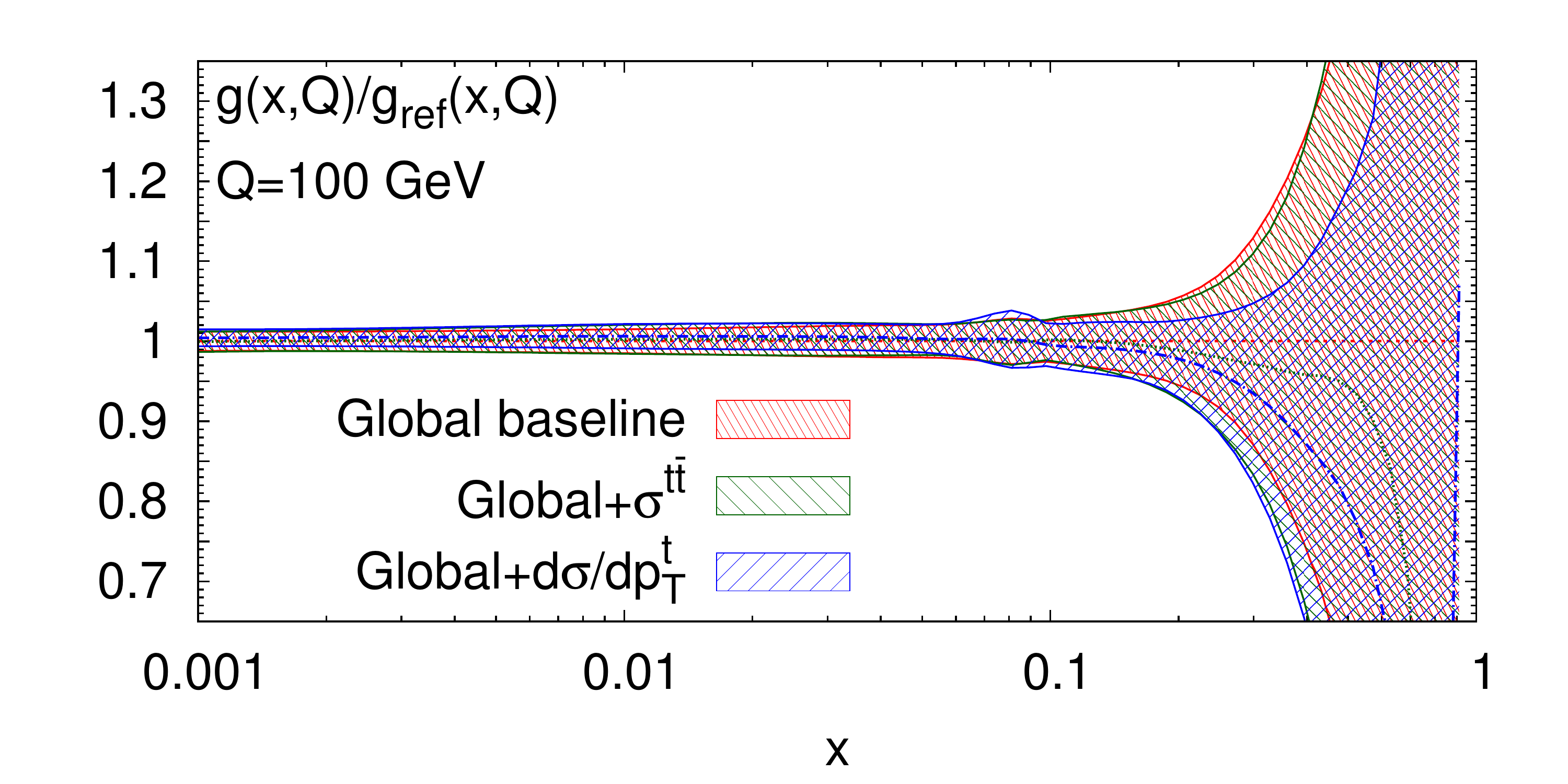}
\includegraphics[scale=0.26,angle=270]{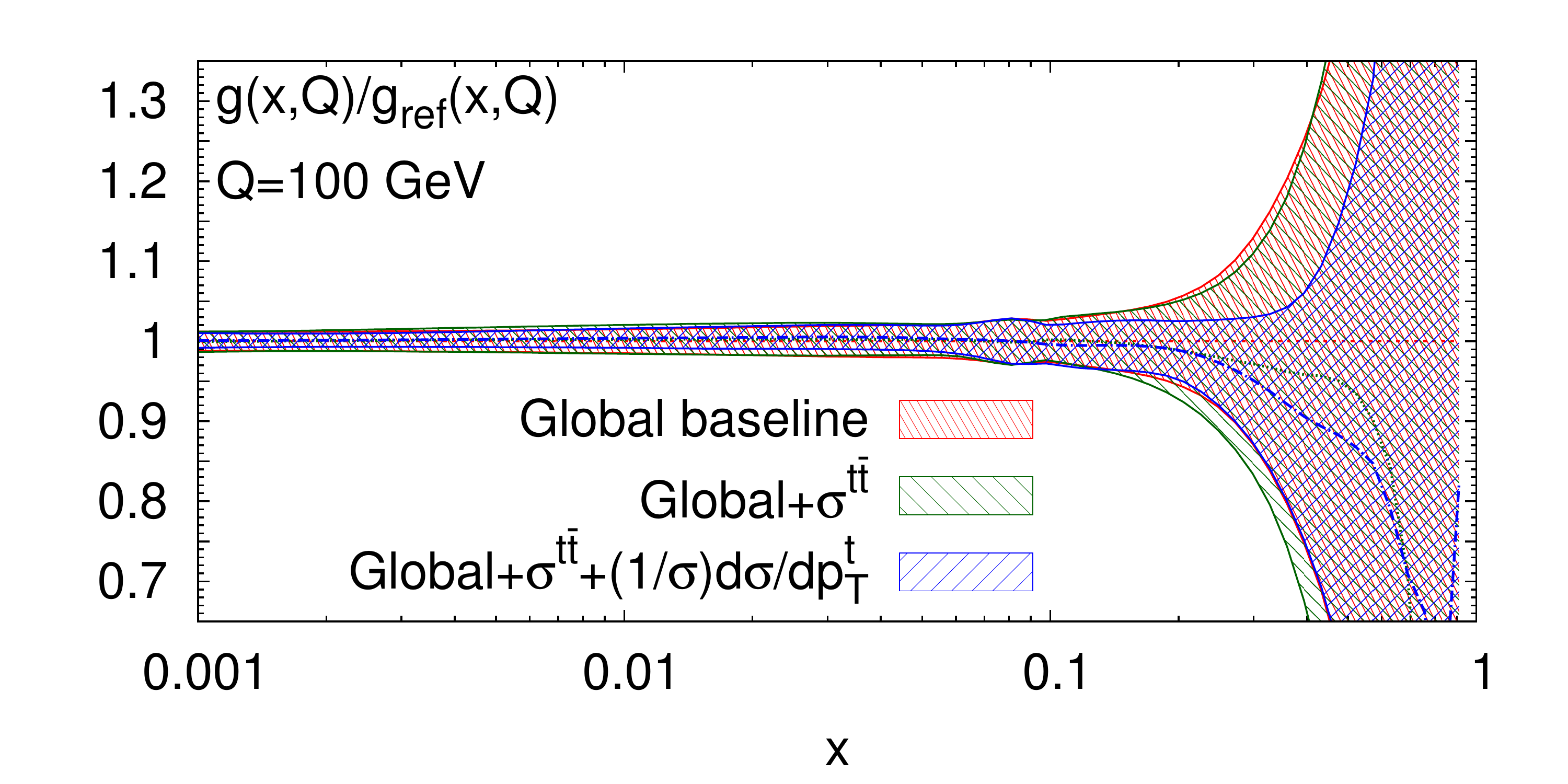}\\
\includegraphics[scale=0.26,angle=270]{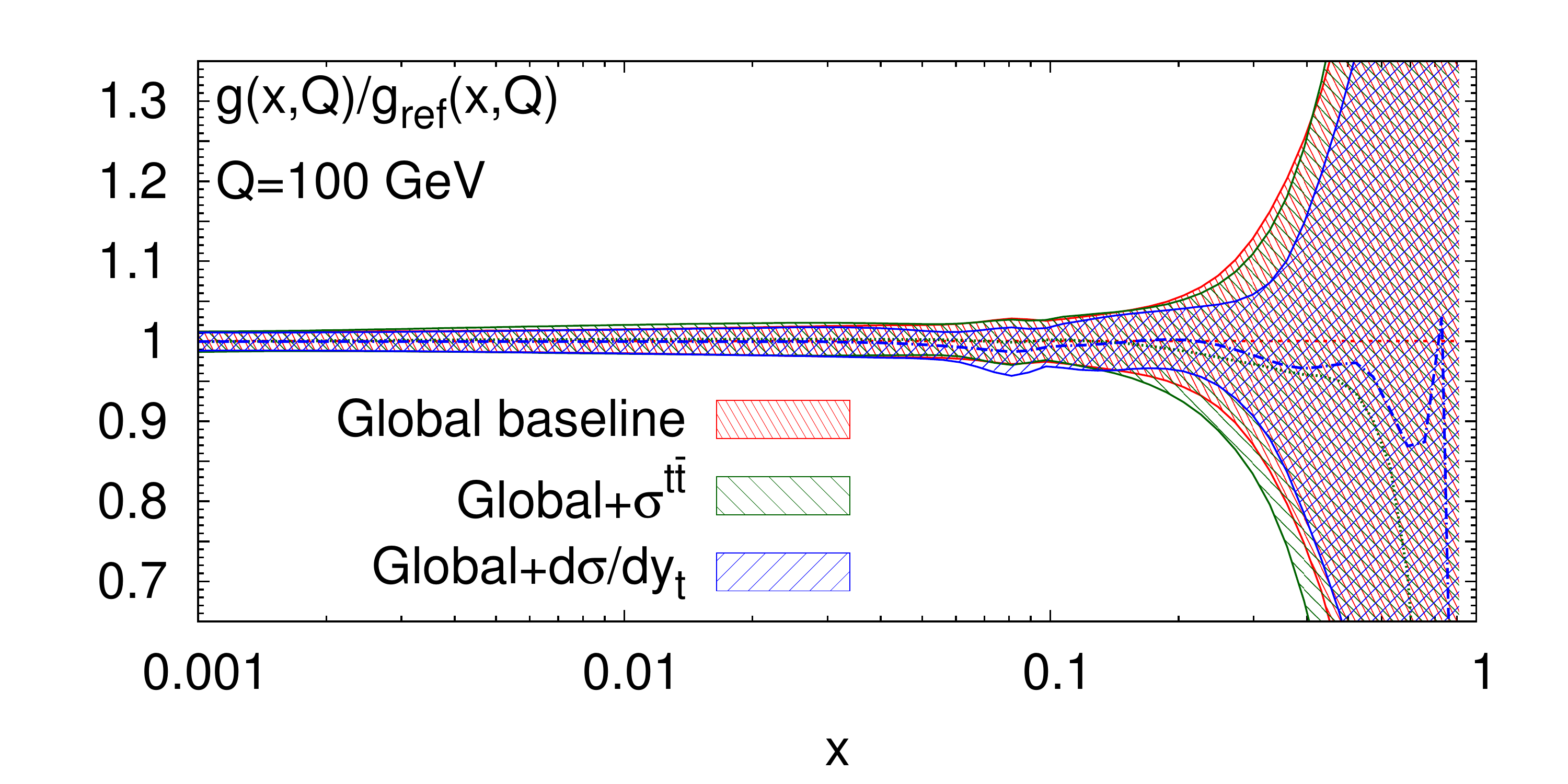}
\includegraphics[scale=0.26,angle=270]{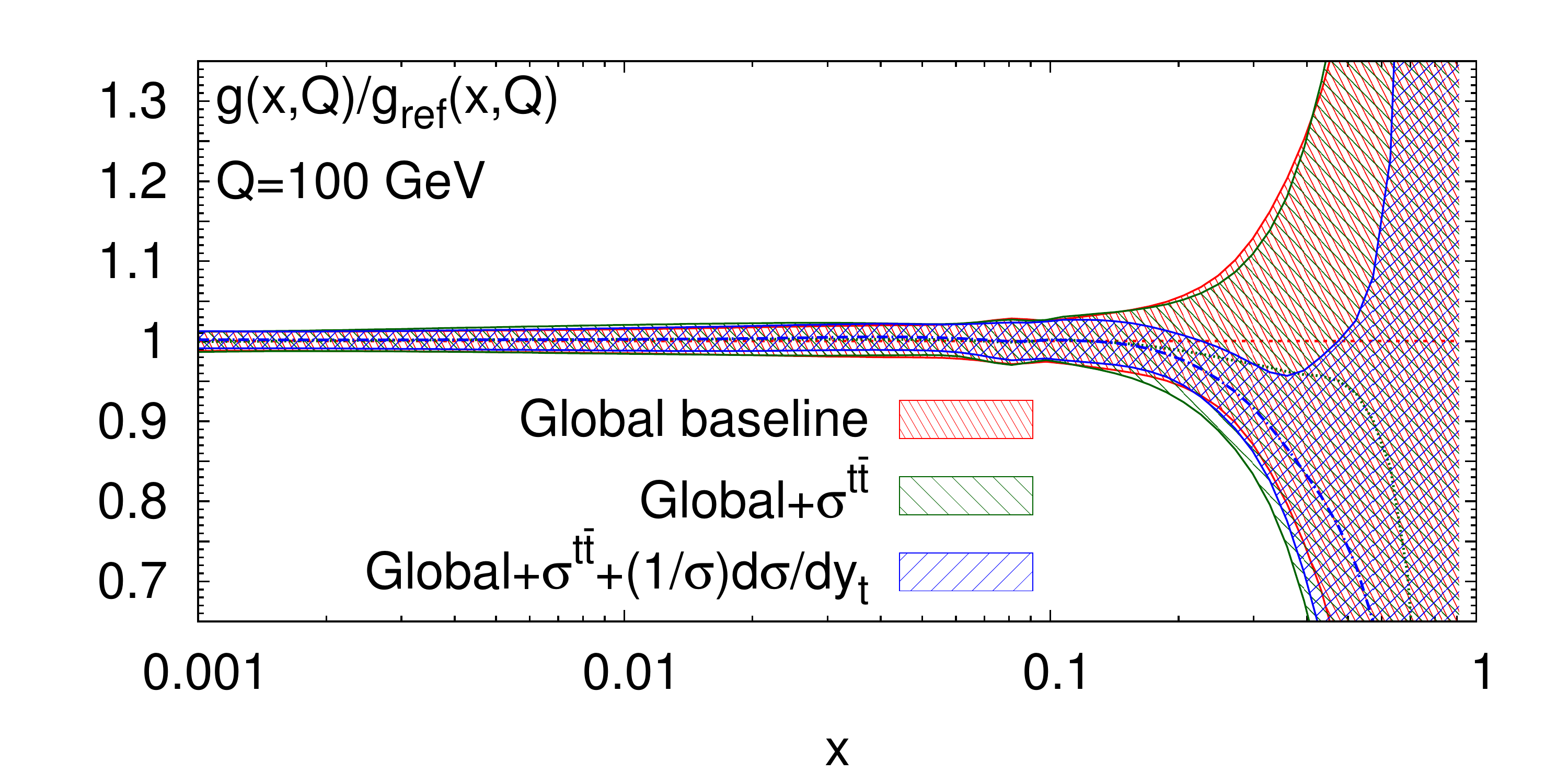}\\
\includegraphics[scale=0.26,angle=270]{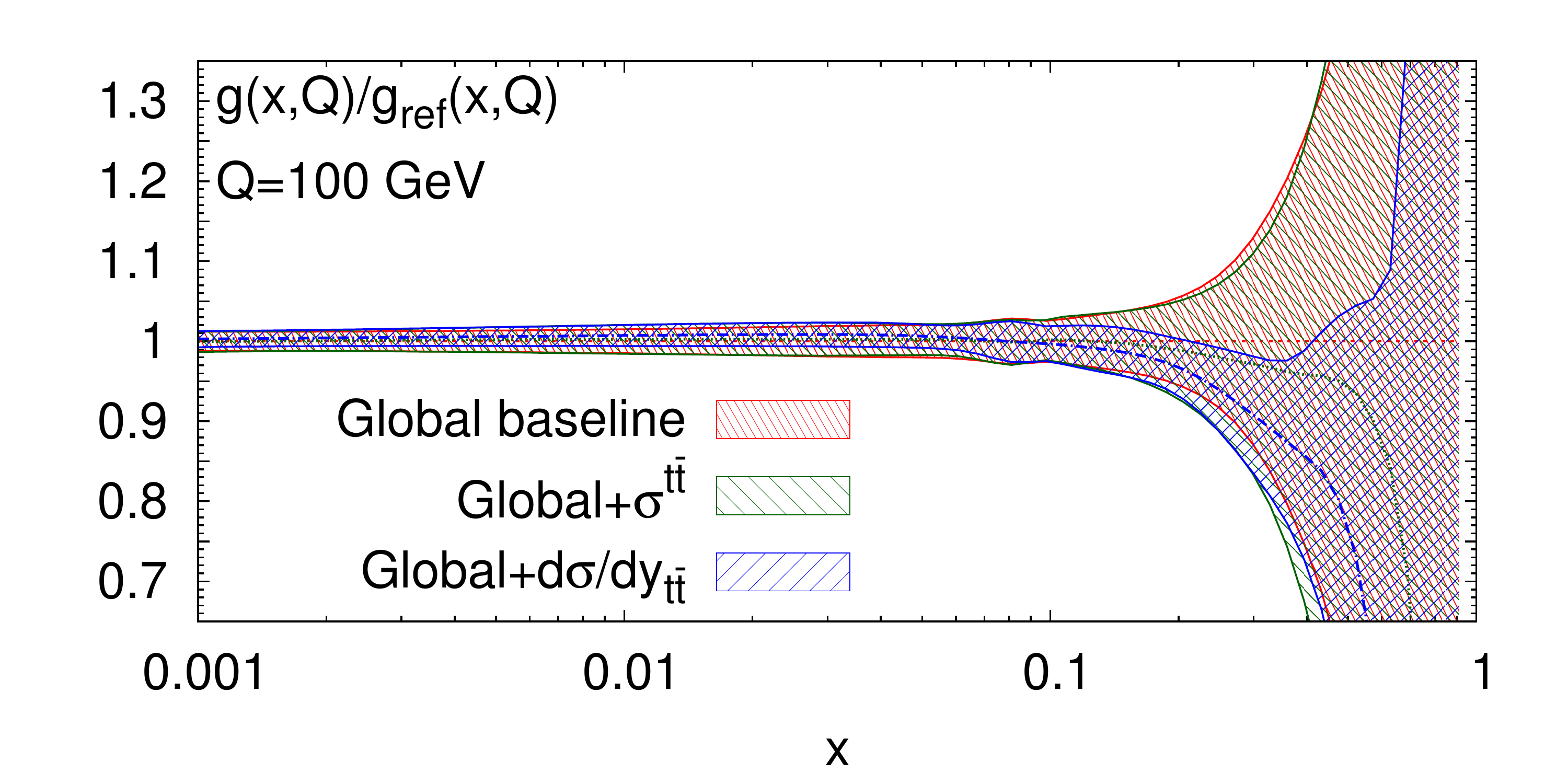}
\includegraphics[scale=0.26,angle=270]{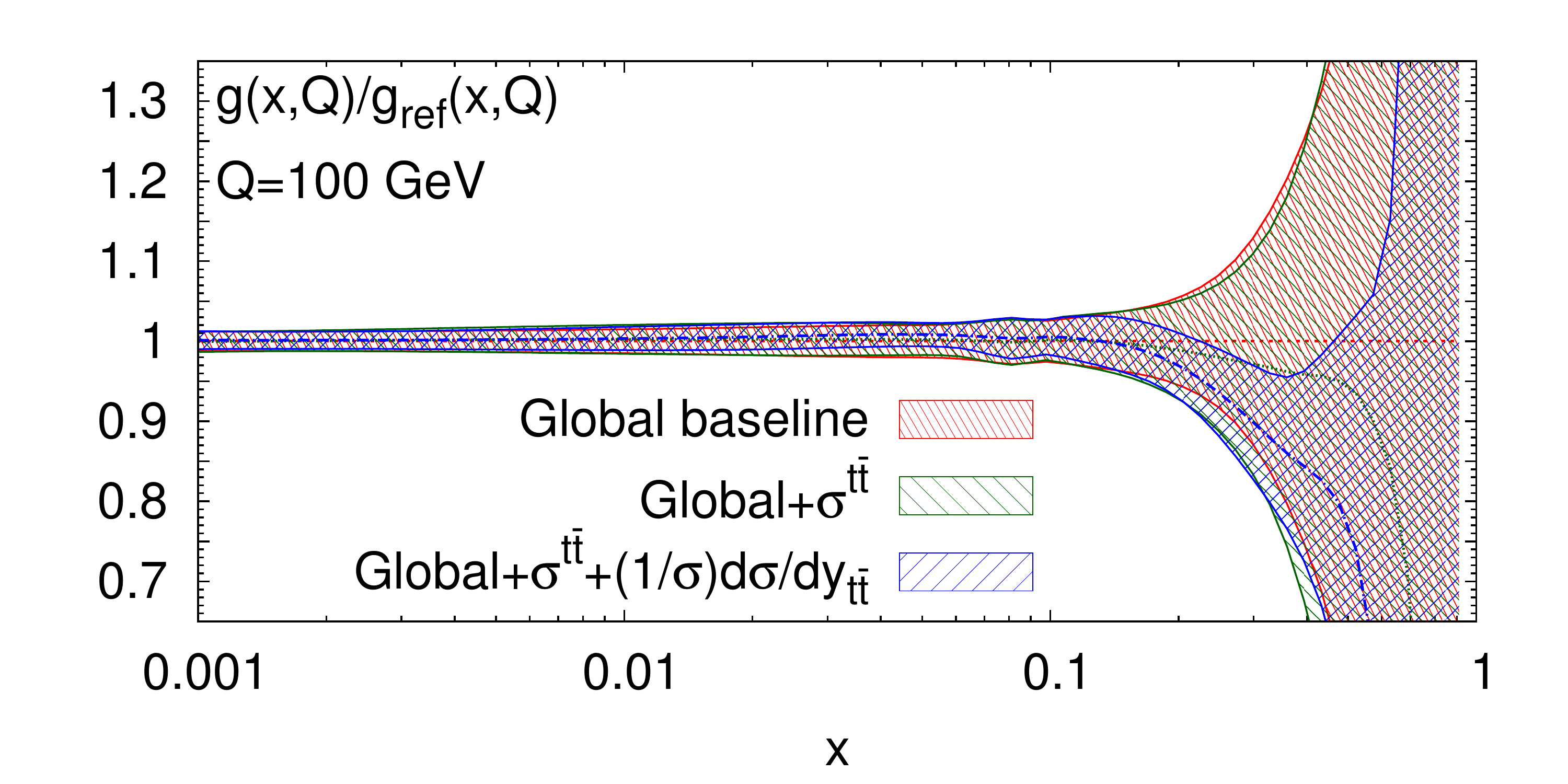}\\
\includegraphics[scale=0.26,angle=270]{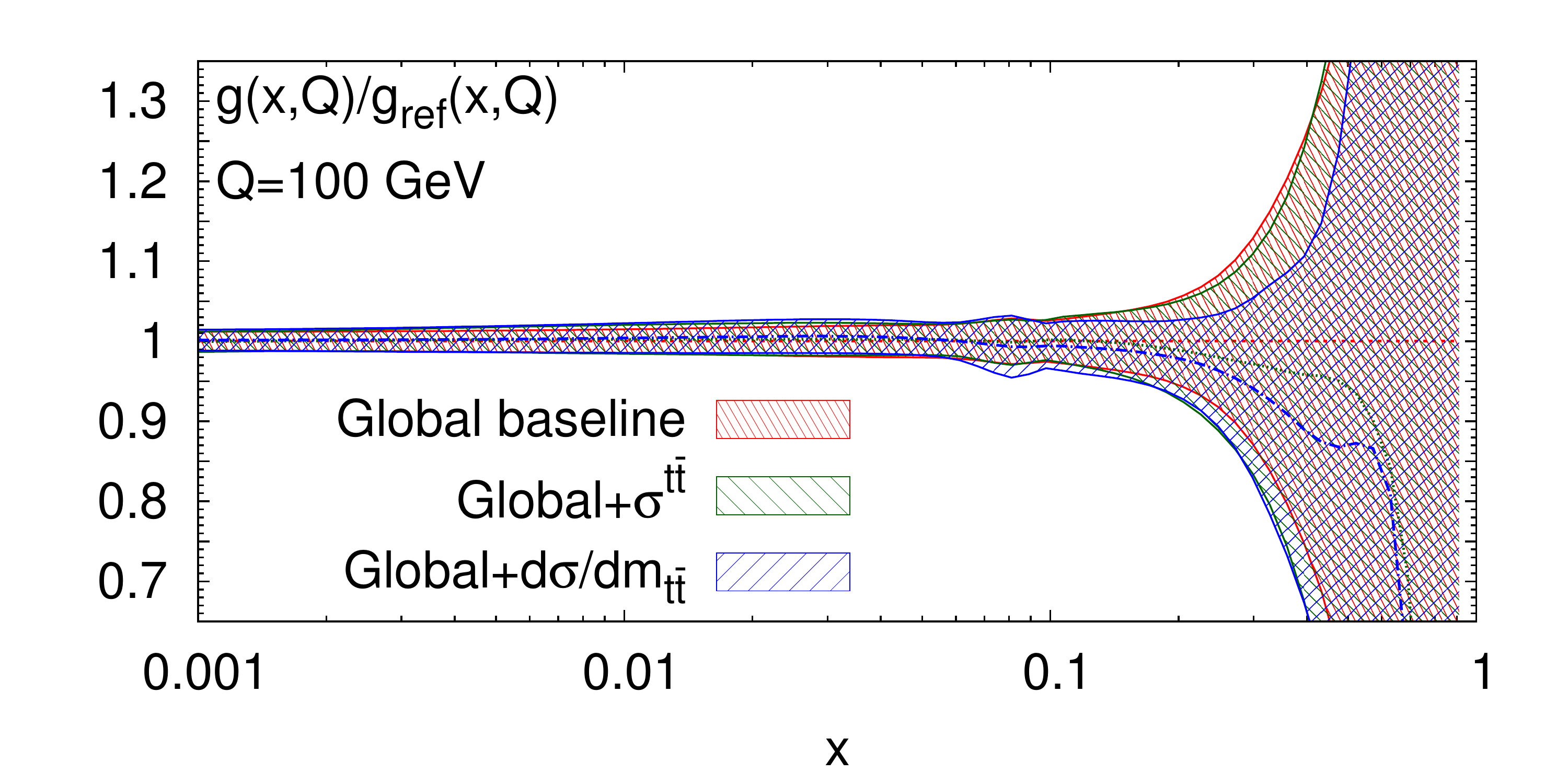}
\includegraphics[scale=0.26,angle=270]{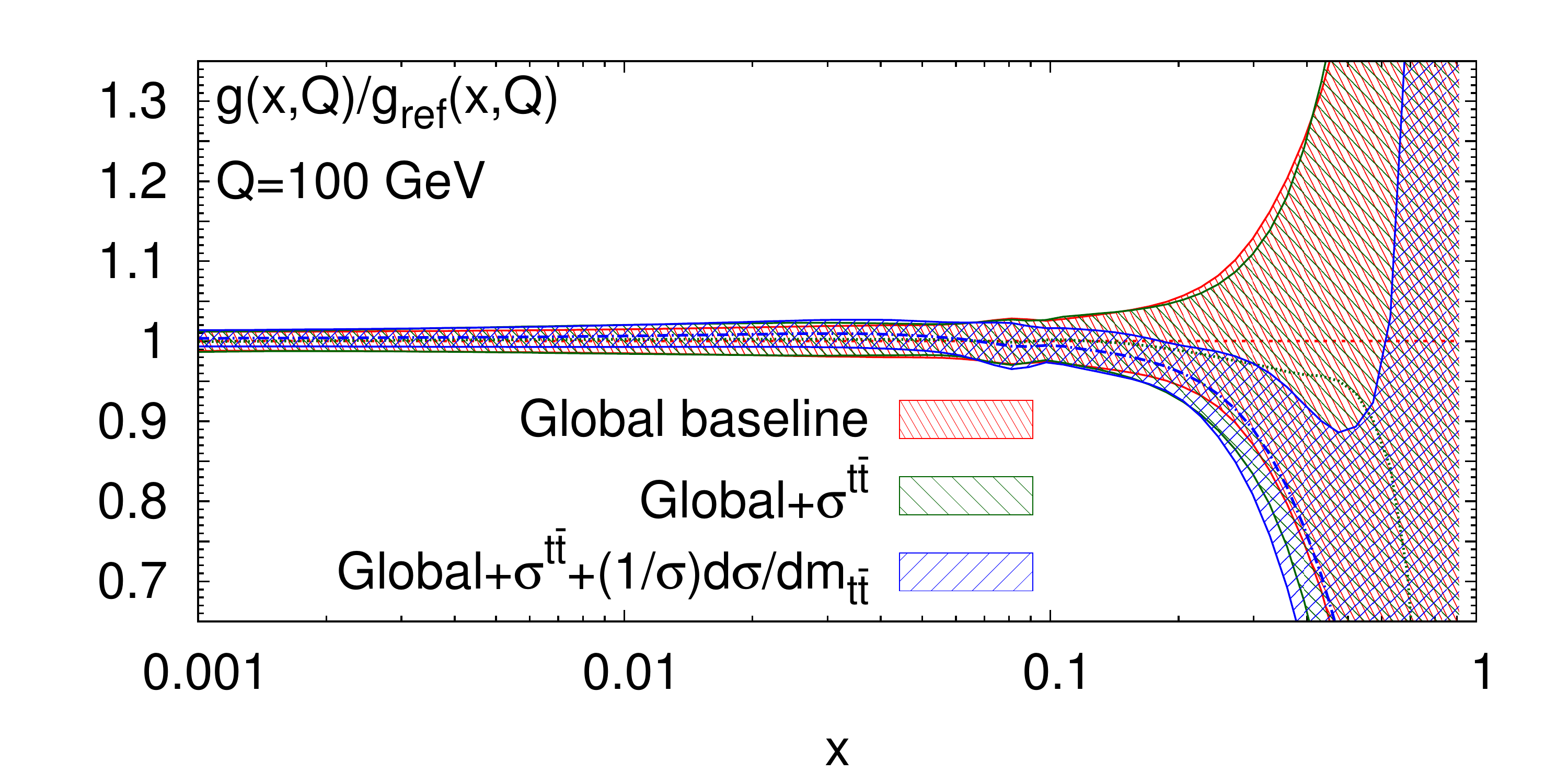}\\
\caption{\small Same as Fig.~\ref{fig:gluonfit}
  for the global fits.
  }
\label{fig:gluonfitGlobal}
\end{figure}

The results for the impact on the large-$x$ gluon of adding
top-quark pair differential data in the global fits are shown in 
Figs.~\ref{fig:gluonexplicit}-\ref{fig:gluonfitGlobal}.
Similarly to the case of the HERA-only fits, we find that
the four differential distributions, as well as the total
cross-section data, have a similar pull on the  central value
of large-$x$ gluon.
Reassuringly, this trend is shared
in both absolute and normalized distributions:
for $x\gsim 0.2$, the LHC top data prefers a softer gluon as
compared to the baseline fit.
In all cases, the fits with top data are contained
within the one-sigma PDF uncertainty band of the baseline fit.
A comparison between Fig.~\ref{fig:gluonfit} and 
Fig.~\ref{fig:gluonfitGlobal}
suggests that the relative effect on the gluon PDF is more pronounced in the
case of the global fits than in the case of the HERA-only fits. 
This is explained by observing, first, that the central value of the gluon PDF 
is smaller for the baseline HERA-only fit than for the baseline global fit, 
and, second, that one-sigma uncertainties are always larger for the pool of
HERA-only fits than for the pool of global fits 
(see Fig.~\ref{fig:gluonexplicit}). 
As a consequence, the error bands displayed in 
Figs.~\ref{fig:gluonfit}-\ref{fig:gluonfitGlobal}, {\it i.e.}
the ratio of the one-sigma uncertainty to the central 
value (of either the HERA-only or the global baseline fit), is larger in the 
HERA-only case than in the global case.

From Fig.~\ref{fig:gluonfitGlobal} we also observe that, when
added to a global dataset, normalized distributions tend to exhibit a higher
constraining power than the corresponding absolute data.
This is especially marked for the $y_t$ and $m_{t\bar{t}}$ distributions,
while in the case of the $p_T^t$ and $y_{t\bar{t}}$ distributions
the differences between the impact
of the absolute and normalized data turns out to be small.
A significant reduction of
the large-$x$ gluon PDF uncertainty is observed for the 
normalized distributions, which can be more than a factor of two
for $x\gsim 0.3$, thus demonstrating
the constraining power of
top-quark differential measurements for global PDF fits.
The exception is the top-quark $p_T^t$ measurement, which leads instead to
a smaller impact on the gluon.

Concerning the impact of the inclusive cross-section data 
(in the fits that do not include differential measurements), we find that
their pull on the central value of the gluon is the same as that of
normalized distributions.
On the other hand, Fig.~\ref{fig:gluonfitGlobal} also shows that,
unlike the case of HERA-only fits, 
the resulting PDF uncertainties are almost unchanged.
We note, however, that a direct comparison with the 
results of~\cite{Czakon:2013tha}
is not straightforward.
Firstly, because here we use a smaller number
of cross-section data points (only two at 8 TeV, and ignore the
7 TeV and 13 TeV data).
Secondly, the dataset that constitutes
the present baseline fit is different from that used in~\cite{Czakon:2013tha}, 
NNPDF2.3~\cite{Ball:2012cx}.
In addition,
the results of~\cite{Czakon:2013tha} were based
on the Bayesian reweighting method~\cite{Ball:2010gb,Ball:2011gg},
while in the present work
top quark measurements are included by means of direct refitting.

\subsection{Impact on the large-$x$ gluon}
\label{sec:impact}

With these studies at hand,
we may now determine a suitable combination of the 
ATLAS and CMS top-quark pair differential measurements that
maximizes 
the constraints on the large-$x$ gluon while, at the same time, leads
to a good agreement between data and theory.
First of all, an inspection of Fig.~\ref{fig:gluonfitGlobal} 
highlights the fact that, in the global fit, normalized
distributions supplemented
with the total cross-section have
superior constraining power 
than the corresponding absolute distributions. This is especially the case 
for the $y_t$ and $m_{t\bar{t}}$ distributions.
Secondly, since each distribution provides different kinematic 
coverage of the gluon, one would like to include in the fit a given 
distribution from ATLAS and a different one from CMS.
Moreover, in order to avoid distortions in the fit due to potential
inconsistencies between ATLAS and CMS, it is advisable to 
include only distributions that can be satisfactory 
described ($\chi^2\simeq 1$) when both ATLAS 
and CMS data are simultaneously included.
Finally, the selected distributions should be among the ones leading
to the largest reduction of the PDF uncertainty of the large-$x$ gluon.

\begin{figure}
\centering
\includegraphics[scale=0.23,angle=270]{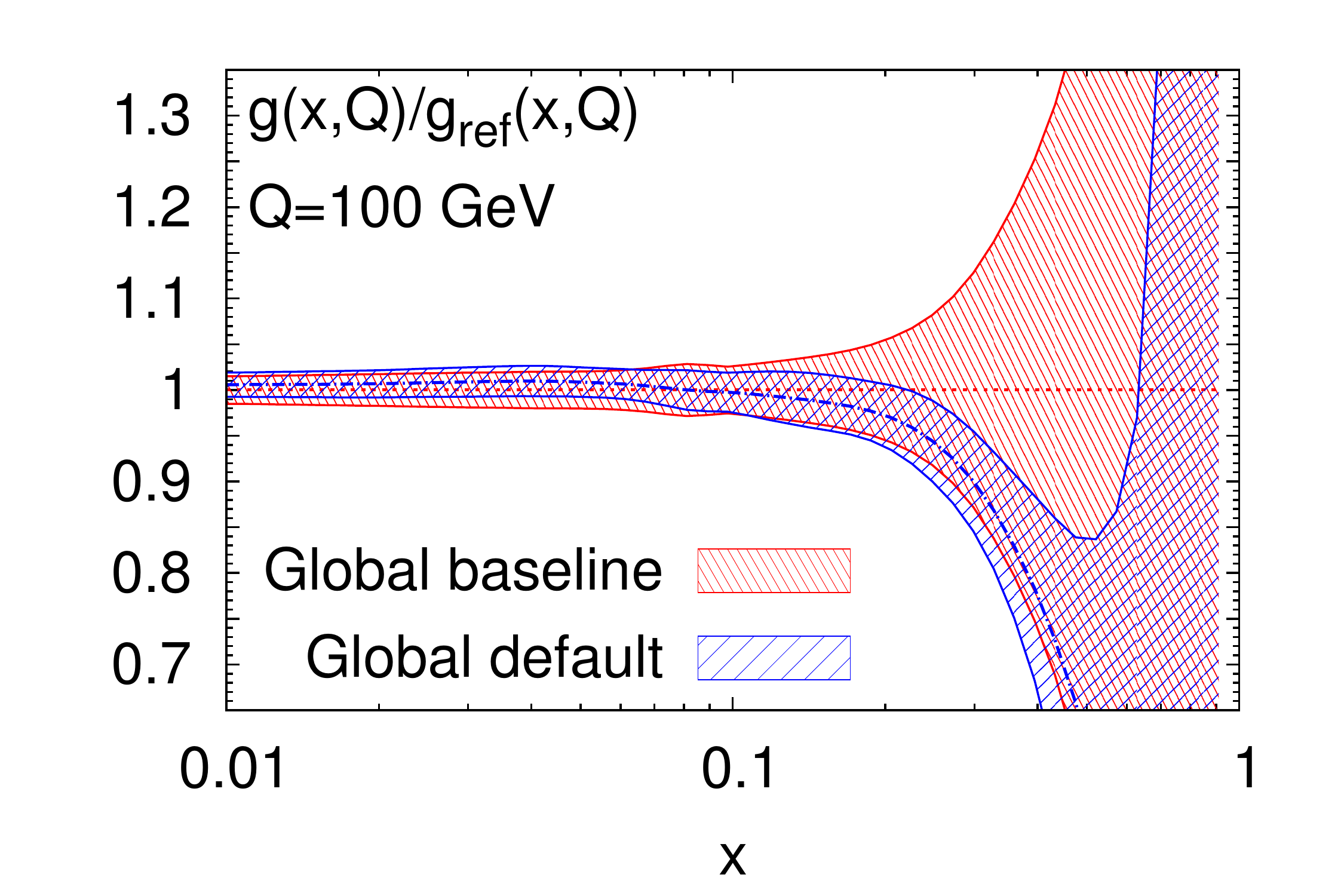}
\includegraphics[scale=0.23,angle=270]{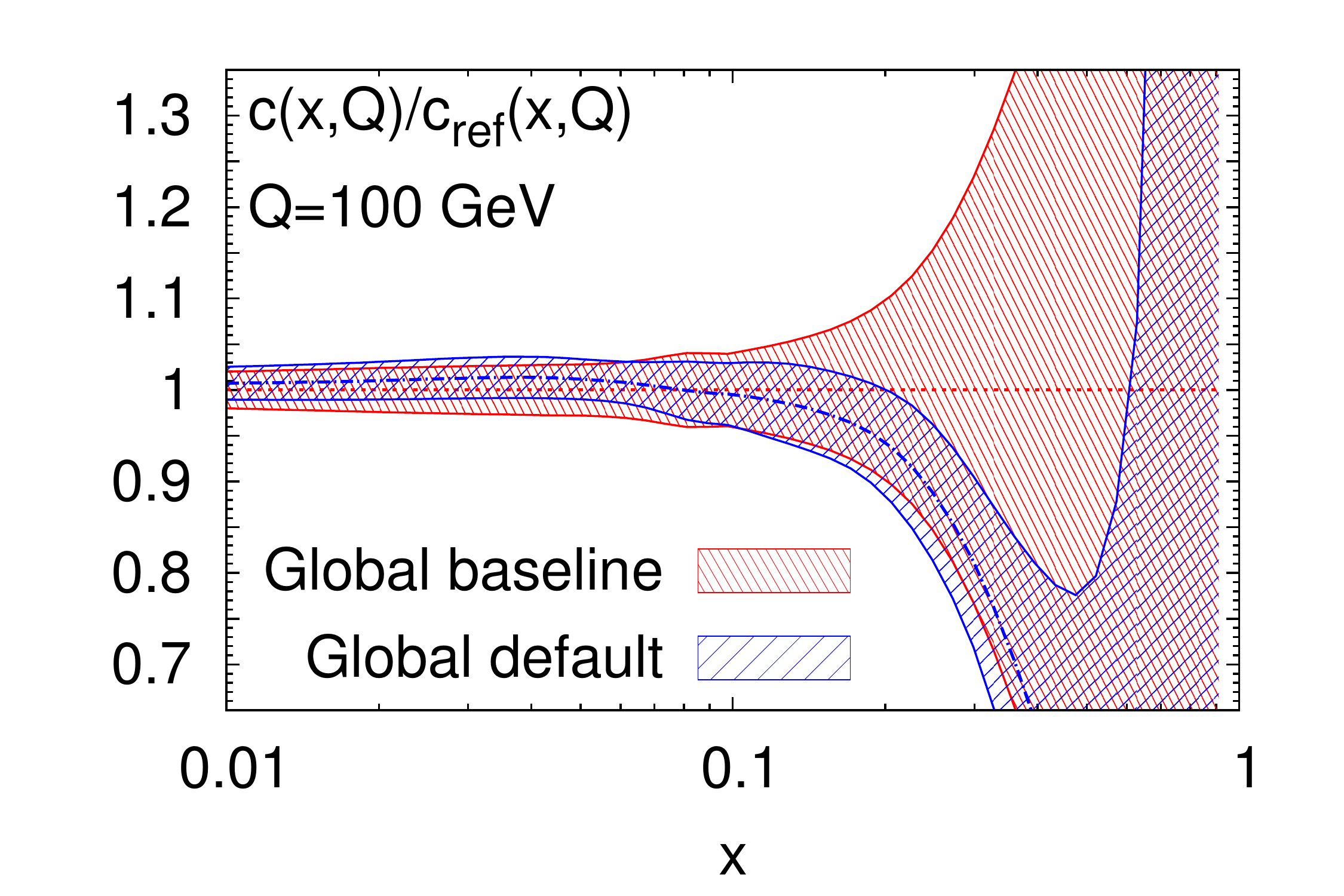}
\includegraphics[scale=0.23,angle=270]{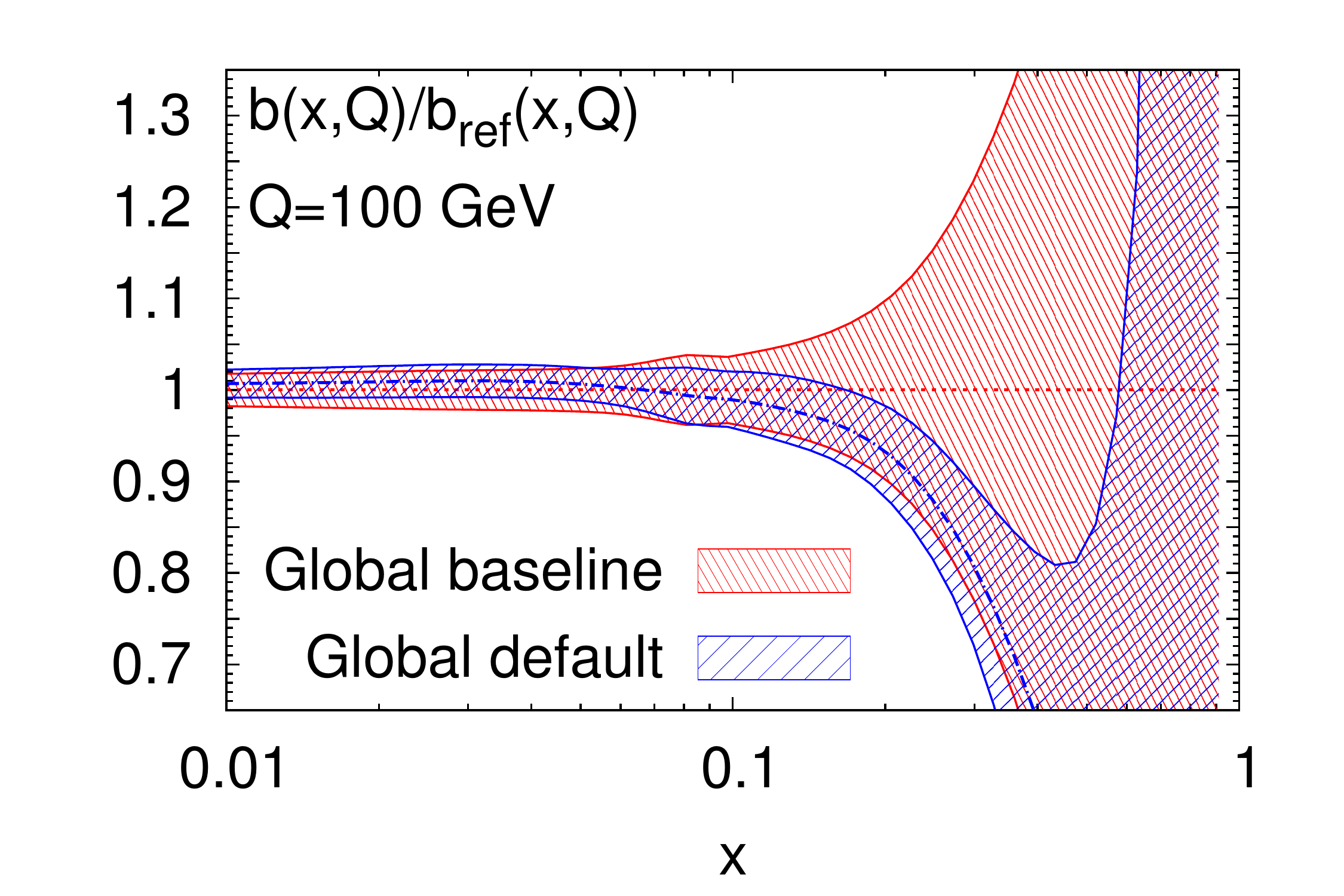}\\
\caption{\small The gluon, charm and bottom PDFs from the global baseline fit 
compared to the optimal fit including our optimal combination of LHC top-quark 
data.}
\label{fig:pdfsdefault}
\end{figure}
\begin{figure}[t]
\centering
\includegraphics[scale=0.38]{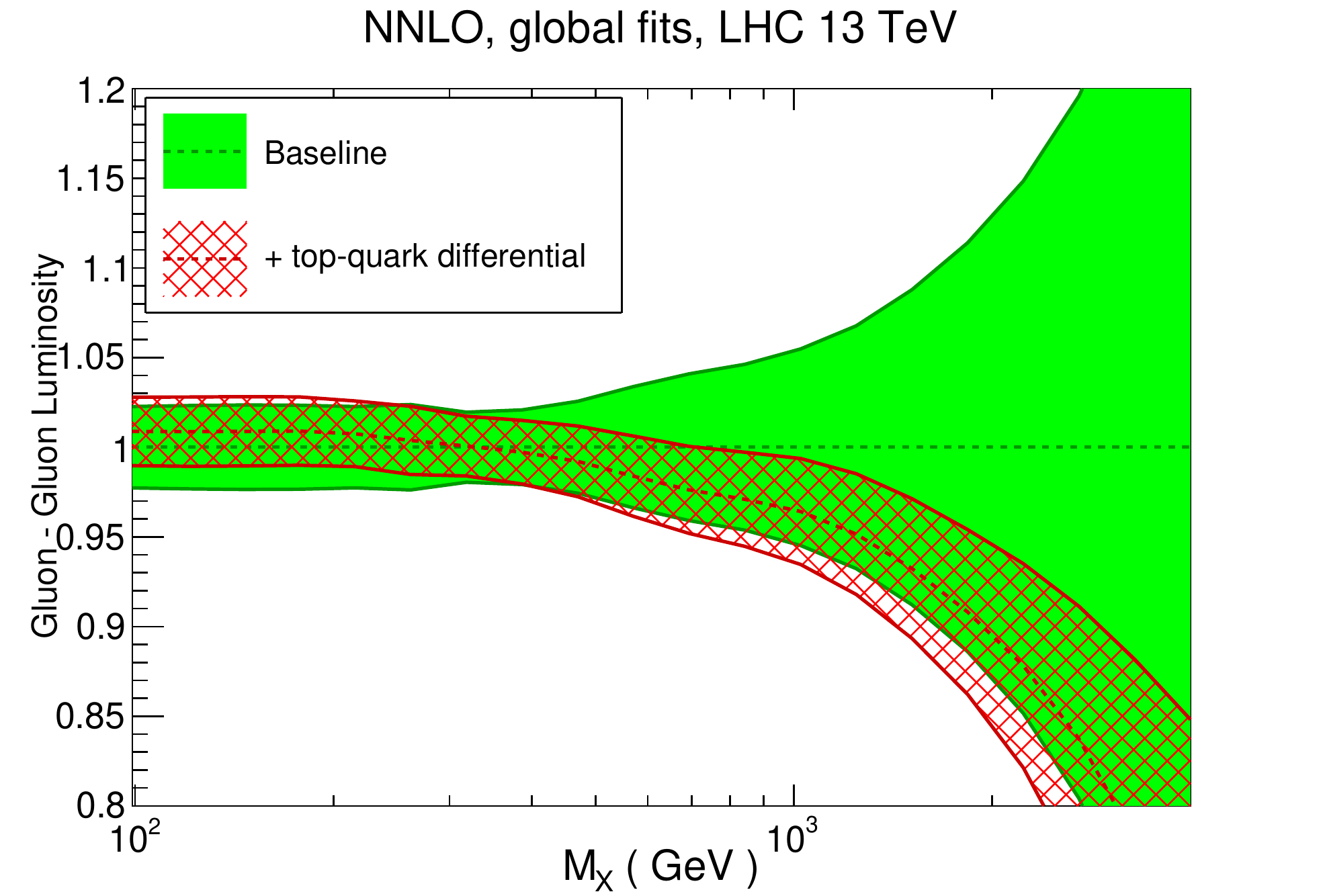}
\includegraphics[scale=0.38]{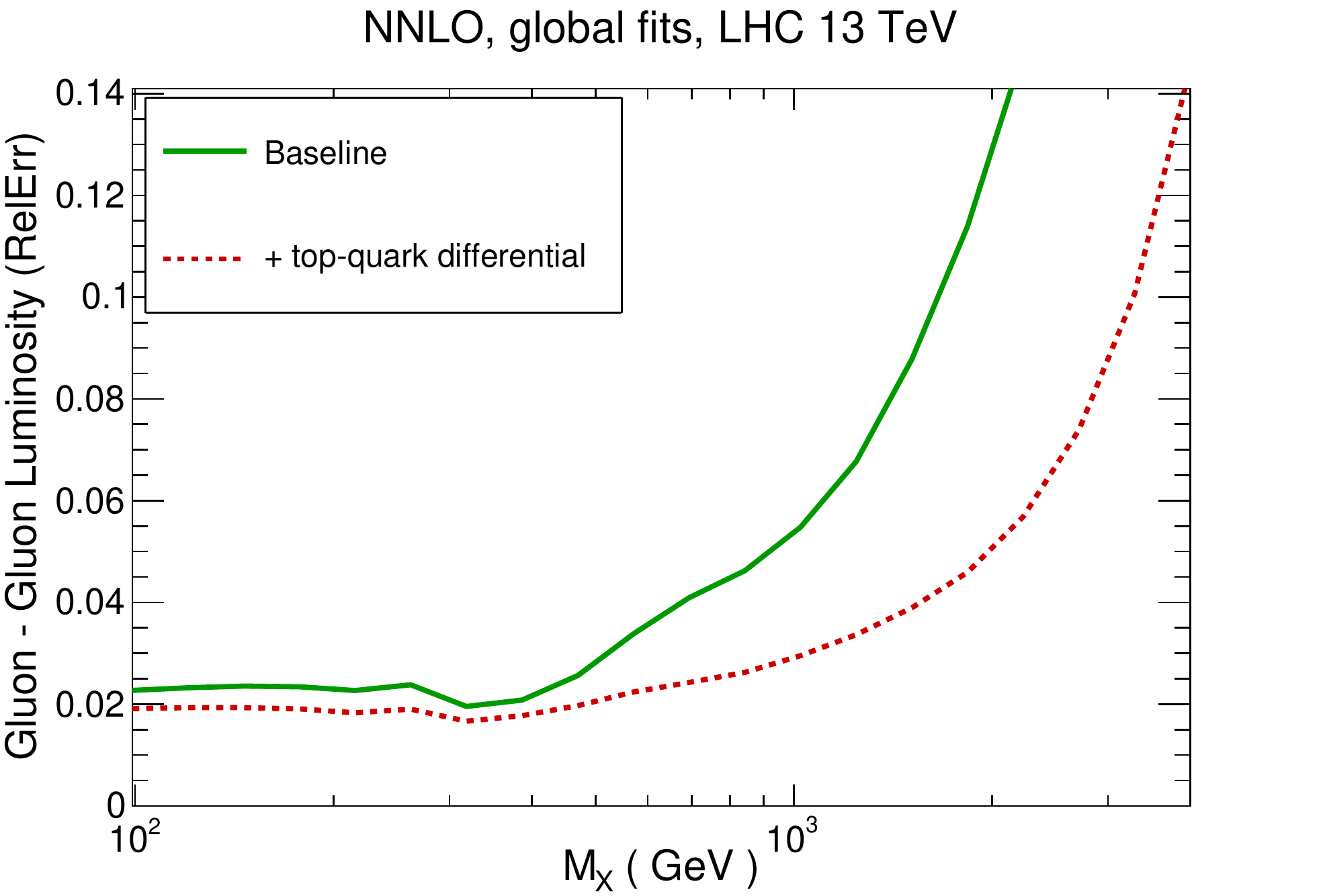}\\
\includegraphics[scale=0.38]{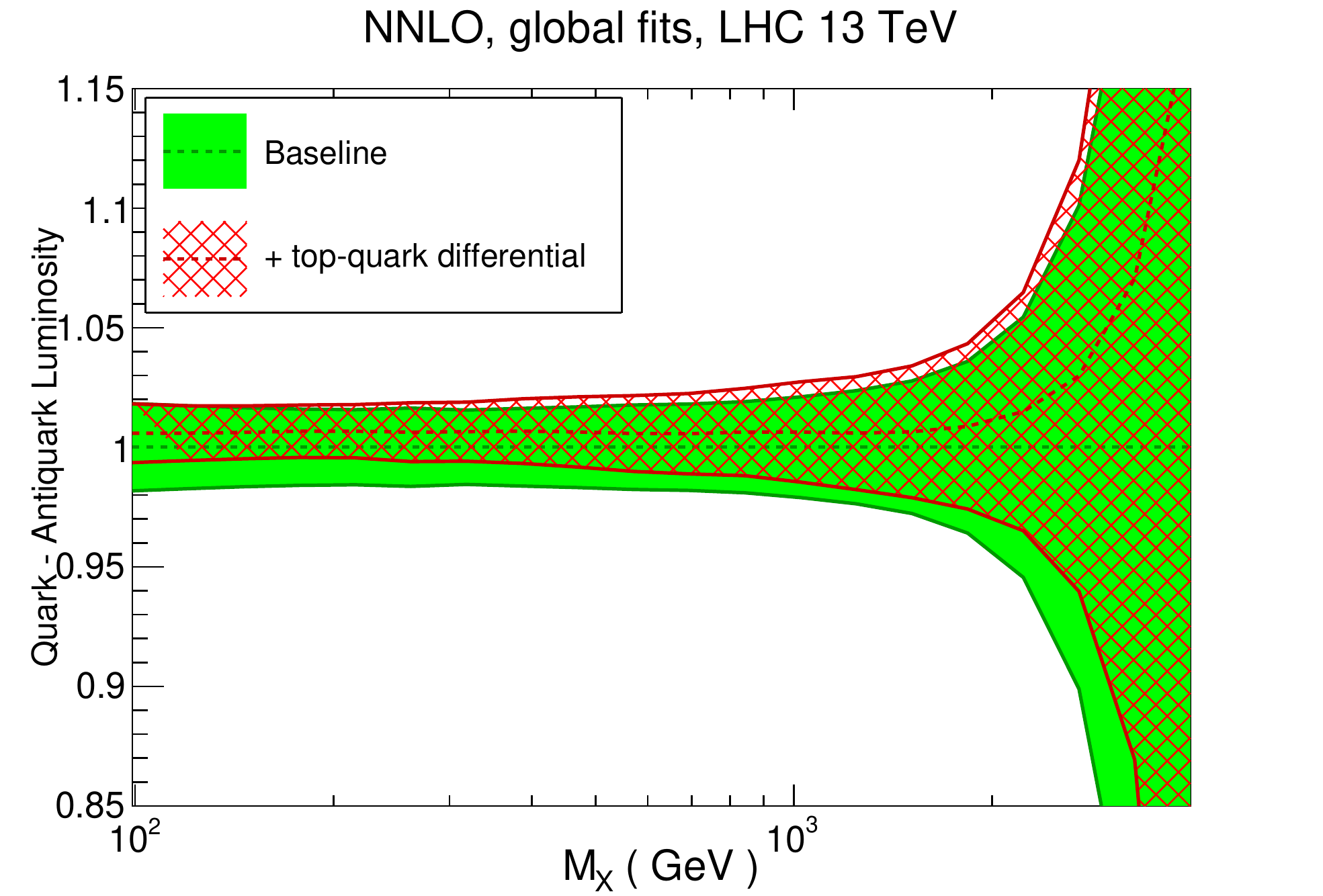}
\includegraphics[scale=0.38]{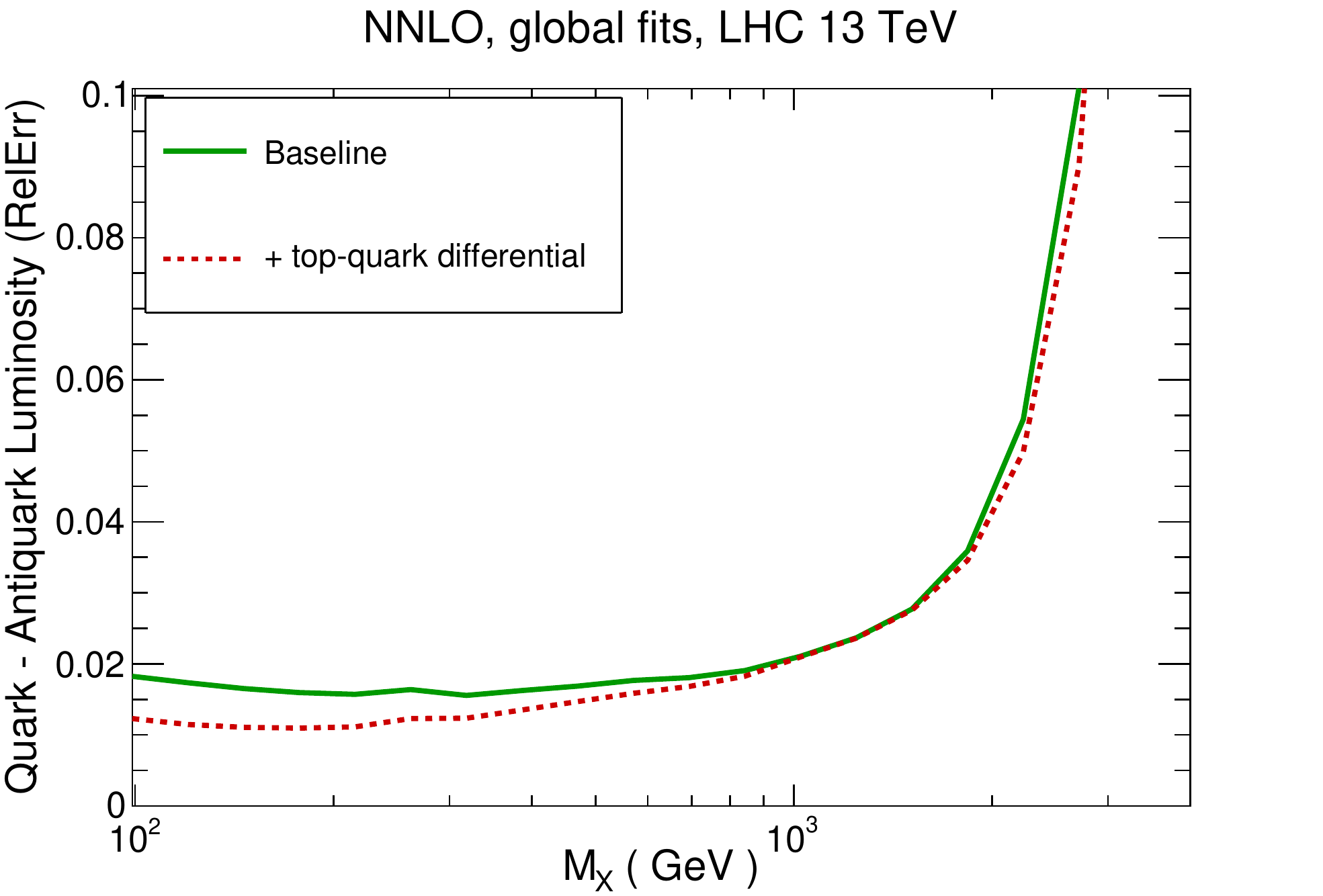}\\
\caption{\small The gluon-gluon (upper)
and quark-antiquark (lower)
NNLO luminosities (left) and their relative
1-$\sigma$  PDF uncertainties (right)
at the LHC with $\sqrt{s}=13$ TeV.
We compare the global baseline fit with the fit including
the optimal combination of LHC top-quark pair differential data.
}
\label{fig:PDFlumis1}
\end{figure}

Taking these guidelines into account, we suggest the following optimal 
combination:
\begin{itemize}
  \setlength\itemsep{-0.2em}
\item the normalized top-quark rapidity distribution $(1/\sigma)d\sigma/dy_t$ from ATLAS;
\item the normalized top-quark pair rapidity distribution
  $(1/\sigma)d\sigma/dy_{t\bar{t}}$ from CMS;
\item and the total inclusive cross-section $\sigma_{t\bar{t}}$ from ATLAS and CMS
  at $\sqrt{s}=8$ TeV.
\end{itemize}
From the results of Fig.~\ref{fig:gluonfitGlobal} it also follows that other possible
choices, consistent with the above guidelines,
would not lead to significantly different results, as 
the pull of the ATLAS and CMS measurements on the large-$x$ gluon 
is consistent among all distributions.

We have therefore performed a final global PDF fit using this optimal
combination of LHC top data, and checked explicitly its features.
The values of the $\chi^2$ per data point 
for each dataset included in the fit are collected in the last column 
of Tab.~\ref{tab:chi2val2}. 
The central value and one-sigma 
uncertainty of the corresponding gluon PDF are displayed in 
Fig.~\ref{fig:gluonexplicit} (thick dashed line).
In Fig.~\ref{fig:pdfsdefault}, we show the gluon, the charm and bottom quark 
PDFs from our global baseline fit and from our optimal fit including our
optimal choice of top-quark data. Results are computed at $Q=100$ GeV
and are normalized to the global baseline fit. Other quark and antiquark PDFs
are marginally affected by top data, as expected, and hence are not shown in 
Fig.~\ref{fig:pdfsdefault}.
We now explore the impact of the new fit both on luminosities
and on kinematic distributions not included in the fit.

First of all, we compute the PDF luminosities at $\sqrt{s}=13$ TeV
for this fit as a function of the invariant mass $M_X$ of the produced
final state.
The factorization scale is set to $\mu_F=M_X$.
In Fig.~\ref{fig:PDFlumis1} we show the $gg$
and the $q\bar{q}$ luminosities
comparing the global baseline fit with the fit
including LHC top data, together with the
corresponding one-sigma PDF uncertainties.
For the $gg$ luminosity, the results of Fig.~\ref{fig:PDFlumis1} confirm
the substantial PDF uncertainty reduction reported in 
Fig.~\ref{fig:pdfsdefault}, which now translates into a reduction of the 
uncertainty for large invariant masses $M_X\gsim 600$ GeV.
For example, in the production of a final state with invariant mass 
$M_X\simeq 2$ TeV (3 TeV),
PDF uncertainties are reduced from 12\% (20\%) down to around 5\% (8\%). 
Such a reduction has clear implications for BSM searches involving top quarks.
The quark PDF uncertainties are also reduced, essentially as a consequence of 
the improved determination of heavy quarks, which follows in turn from a
better determination of the gluon PDF.
For the $q\bar{q}$ luminosity, for example, we observe only a moderate 
uncertainty reduction in the region with $M_X\gtrsim 1$ TeV, while PDF
uncertainties are reduced from $2\%$ to $1\%$ around $M_X\sim 100$ GeV.
  
\begin{figure}[t]
\centering
\includegraphics[scale=0.29,angle=270,clip=true,trim=2cm 0cm 2cm 0cm]{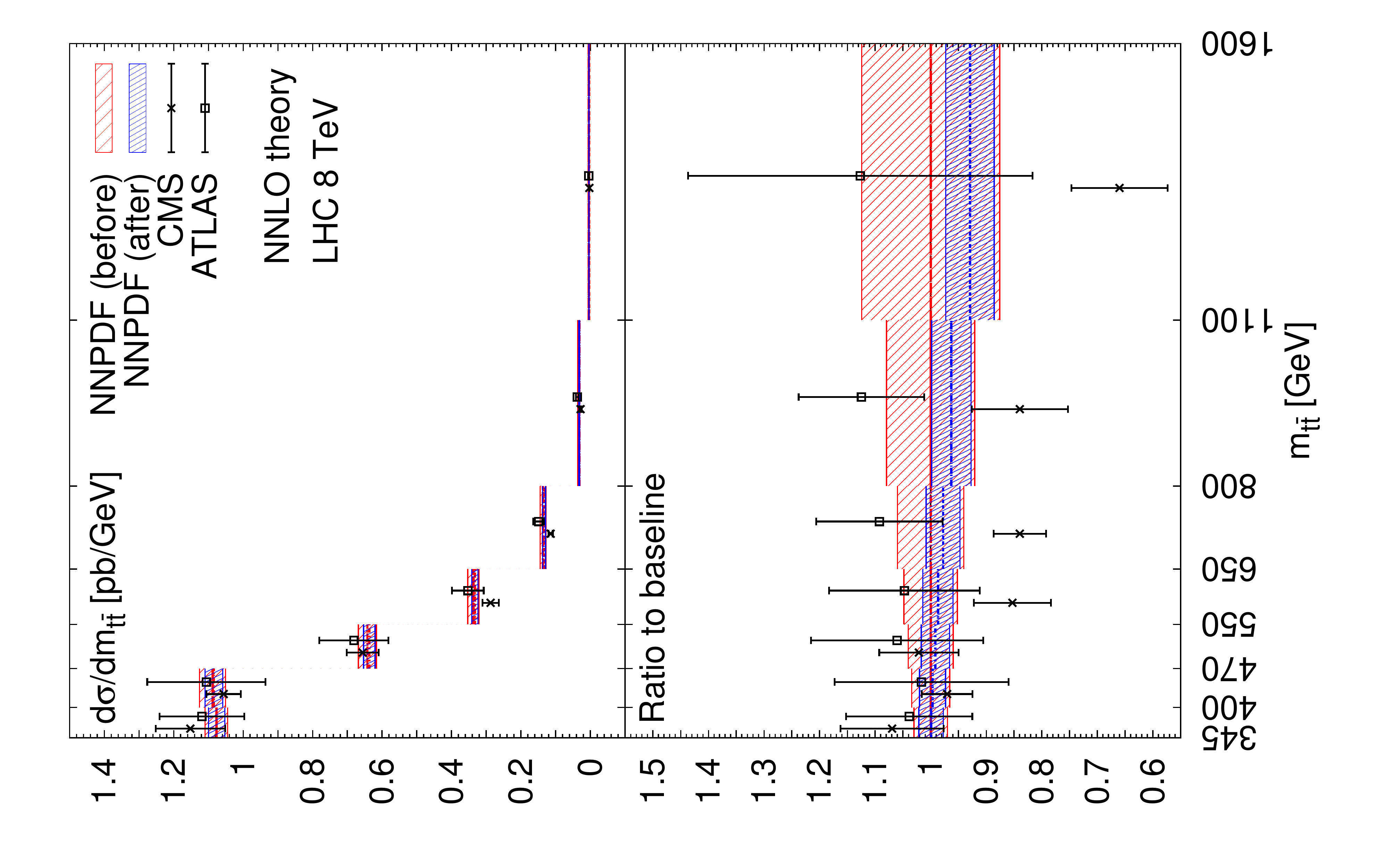}
\includegraphics[scale=0.29,angle=270,clip=true,trim=2cm 0cm 2cm 0cm]{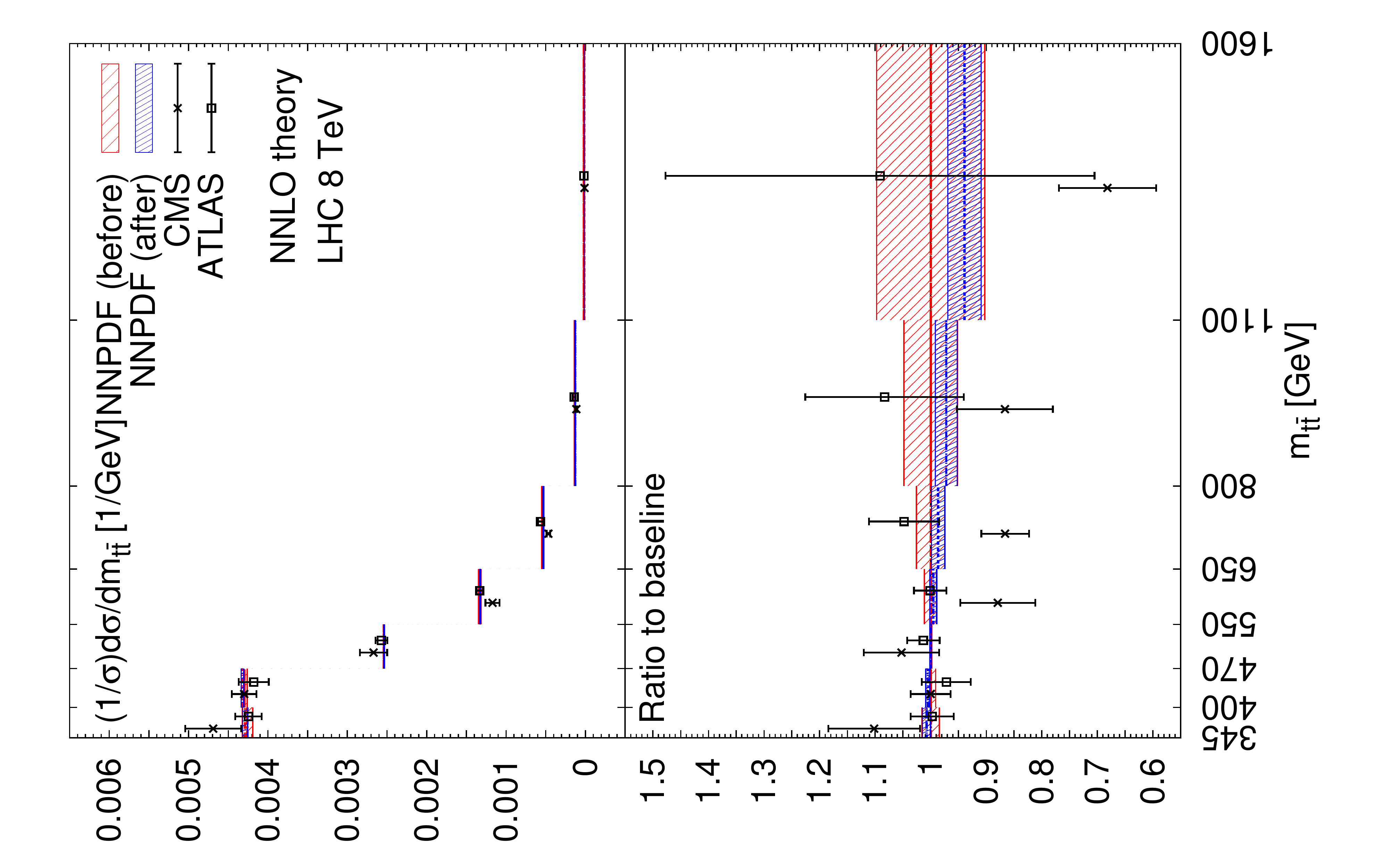}\\
\caption{\small The NNLO theoretical predictions for
  the absolute (left) and normalized (right) 
  $m_{t\bar{t}}$ distributions at the LHC 8 TeV,
  obtained from
  the global PDF fit before and 
  after the optimal combination of top data has been included.
   The theory predictions include only the 1--$\sigma$ PDF uncertainty band, while
  scale uncertainties are not shown.
  The lower panels show the results as a ratio to the
   baseline fit.
}
\label{fig:predglobalfit}
\end{figure}

\begin{figure}[t]
\centering
\includegraphics[scale=0.29,angle=270,clip=true,trim=2cm 0cm 2cm 0cm]{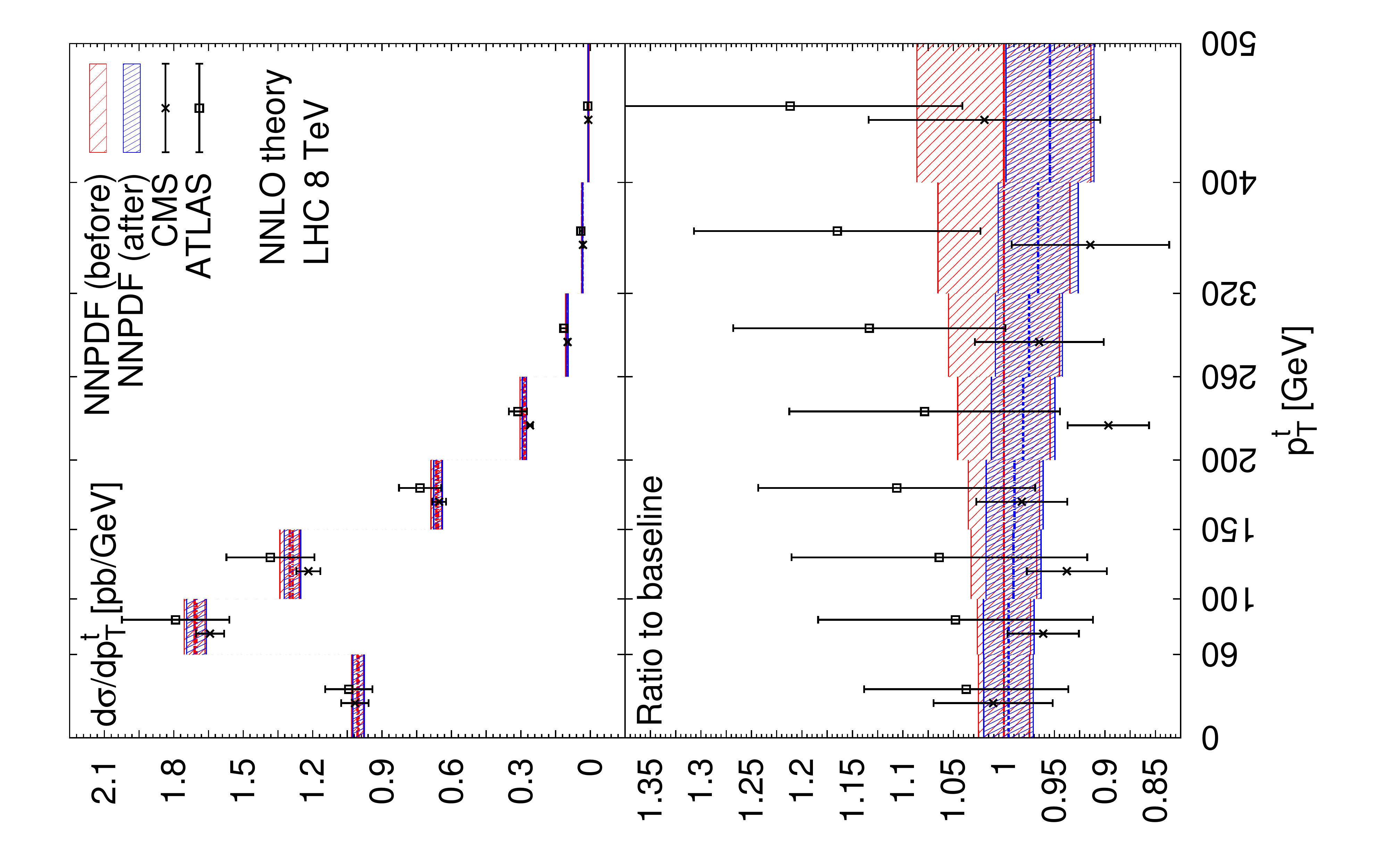}
\includegraphics[scale=0.29,angle=270,clip=true,trim=2cm 0cm 2cm 0cm]{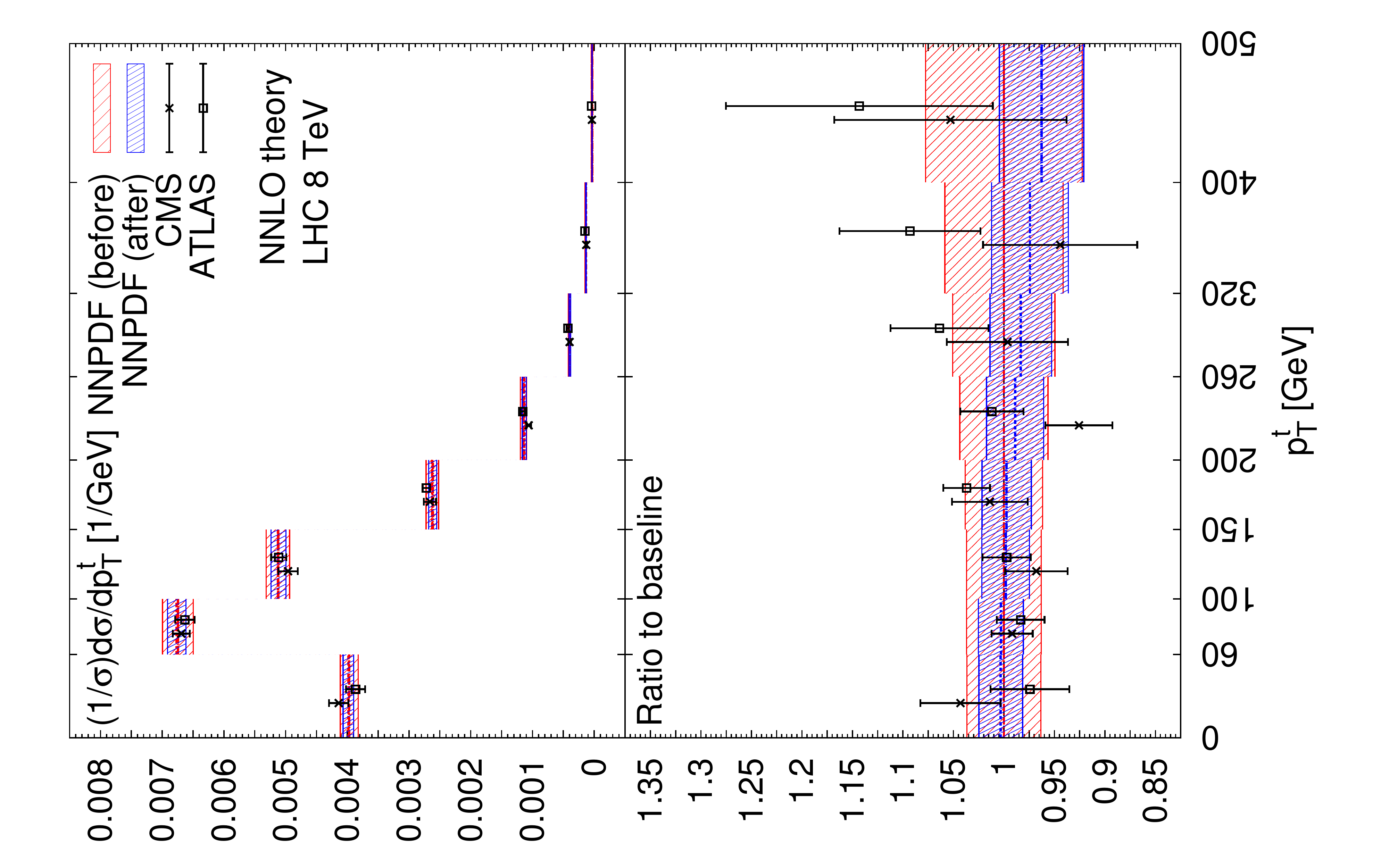}\\
\caption{\small Same as Fig.~\ref{fig:predglobalfit} for
  the top quark pair $p_T^t$ distribution.}
\label{fig:predglobalfit2}
\end{figure}

Next, we study how the theoretical
predictions are modified for those top-quark pair differential
distributions not included in the fit. 
In Figs.~\ref{fig:predglobalfit}
and~\ref{fig:predglobalfit2} we show the
NNLO calculations for the absolute and normalized 
$m_{t\bar{t}}$ and $p_T^t$ distributions, respectively,
obtained from the global PDF fit before and 
after the LHC top-quark data has been included.
In the lower panels, we show the results normalized to the baseline fit.
Note that none of the ATLAS and CMS data shown in Figs.~\ref{fig:predglobalfit}
and~\ref{fig:predglobalfit2} has actually been used
as input in the fit.

The quality of the description of the $p_T^t$ and $m_{t\bar{t}}$ data improves in 
most cases, both for absolute and normalized distributions, as quantified by 
the decrease in the values of the $\chi^2$ per data point collected in 
Tab.~\ref{tab:chi2val2}: for ATLAS absolute (normalized) $p_T^T$ distribution, 
the $\chi^2$ drops down from $2.37$ ($2.93$) to $2.19$ ($2.49$); 
for CMS absolute (normalized) $p_T^T$ distribution from $3.50$ ($4.31$)
to $2.91$ ($3.33$); for CMS absolute (normalized) $m_{t\bar{t}}$ distribution
from $7.07$ ($12.0$) to $4.77$ ($8.05$). 
An exception is represented by ATLAS absolute (normalized) 
$m_{t\bar{t}}$ distribution, where instead the $\chi^2$ increases from 
$4.27$ ($2.30$) to $5.01$ ($4.55$). 
Indeed, the fit tends to move towards the CMS data, which is more precise than
the ATLAS data, but in clear tension with the latter.


In comparison to the global baseline fit, theoretical predictions for the 
$m_{t\bar{t}}$ and $p_T^t$ distributions are more precise in the optimal fit 
with our optimal choice of top-quark data included. 
This is a direct consequence of the large-$x$ gluon constraints
derived from fitting the $y_t$ and $y_{t\bar{t}}$ distributions.
For the top-quark pair invariant mass distributions, the PDF uncertainties
in the rightmost bin, a region which is crucial for BSM searches,
are reduced by more than a factor of two.
This
reduction would be even more pronounced for larger
$m_{t\bar{t}}$, as can be inferred from the $gg$
luminosity in Fig.~\ref{fig:PDFlumis1}.
For the case of the top quark
$p_T^t$ distribution, we also observe a sizable PDF uncertainty reduction
in the entire range probed by the LHC measurements,
which can be again as large as a factor of two for $p_T^t\simeq$ 500 GeV.

Figs.~\ref{fig:predglobalfit} and~\ref{fig:predglobalfit2}
highlight the potential of
a comprehensive program of measurements of top-quark
pair production to achieve a self-consistent reduction of theoretical
uncertainties with the subsequent
improvement of the prospects for BSM searches.
In the specific case studied in this work, we have shown how
the inclusion in the global fit of the normalized $y_t$
and $y_{t\bar{t}}$ distributions
leads to improved theory predictions for ATLAS and CMS $p_T^t$ distributions
and for CMS $m_{t\bar{t}}$ distributions. 
A corresponding improvement in the 
ATLAS $m_{t\bar{t}}$ distributions is not observed, though it might become
evident once the apparent tension between ATLAS and CMS data will be understood.
Similar improvements will apply for other LHC processes, either in the SM or
beyond, that are driven by the $gg$ luminosity at large invariant masses.

It is important to emphasize that,
with our choice of top-quark distributions to be used in the PDF fit,
the possibility for contamination in the resulting gluon 
from BSM effects is reduced.
The reason for this is that heavy new resonances are likely to be 
kinematically suppressed in the rapidity distributions but not in the tails 
of the $m_{t\bar{t}}$ and $p_T^t$ distributions.
Therefore, constraining the gluon from the $y_t$ and $y_{t\bar{t}}$ measurements
and using the resulting PDF to predict the $m_{t\bar{t}}$ and $p_T^t$ distributions
represents a robust strategy in the context of BSM searches.

\subsection{Comparison with the constraints from jet data}
\label{sec:jets}

As discussed in Sect.~\ref{sec:fitsettings}, the global
dataset used for the
baseline fits
excludes the jet production measurements from the Tevatron and the LHC
that were part of NNPDF3.0.
The rationale for this choice is that the NNLO calculation for jet production 
has become available only very recently~\cite{Currie:2016bfm}, and we aim 
at providing a fully consistent determination of the large-$x$ gluon at NNLO.

It is anyway instructive to
assess how the PDF uncertainty reduction
on the large-$x$ gluon driven by
top-quark data in the global fits (Fig.~\ref{fig:gluonfitGlobal}) 
compares with that from inclusive jet measurements.
This way, it is possible
to ascertain whether available differential
top measurements provide competitive constraints as compared
to those from jet production.
To address this question, we have performed a NNLO
fit where now the global baseline dataset is supplemented with collider
inclusive jet production measurements, without any 
top-quark data.
For these fits, theoretical calculations of the inclusive jet cross-section
have been performed with NNLO DGLAP evolution
and $\alpha_s$ running, but NLO matrix elements.
This approximation is justified here since we are not interested in the shift
in the central value of the large-$x$ gluon as a result of the inclusion of the jet 
data, but only in the relative reduction of the PDF uncertainties.

In particular, we have added the inclusive jet production cross-sections from
CDF Run II ($k_t$ algorithm)~\cite{Abulencia:2007ez};
from ATLAS at $\sqrt{s}=2.76$ TeV~\cite{Aad:2013lpa} and 
7 TeV~\cite{Aad:2011fc}, in the latter
case from the 2010 run, together with their cross-correlations;
and from CMS  at 7 TeV~\cite{Chatrchyan:2012bja} from the 2011 data-taking 
period.
These four datasets were already part of the NNPDF3.0 fits.
Moreover, we have added two additional inclusive jet measurements, from CMS at 
$\sqrt{s}=2.76$ TeV
and from ATLAS at 7 TeV from the 2011 run~\cite{Aad:2014vwa}.\footnote{Details
  on the implementation of this two new datasets
  will be discussed in a forthcoming publication~\cite{NNPDF31}.}
The resulting inclusive jet cross-sections
add up to $76$ points for CDF, 180 for ATLAS, and 214 from CMS, 
for a total of $N_{\rm dat}=470$ points.

\begin{figure}[t]
\centering
\includegraphics[scale=0.39]{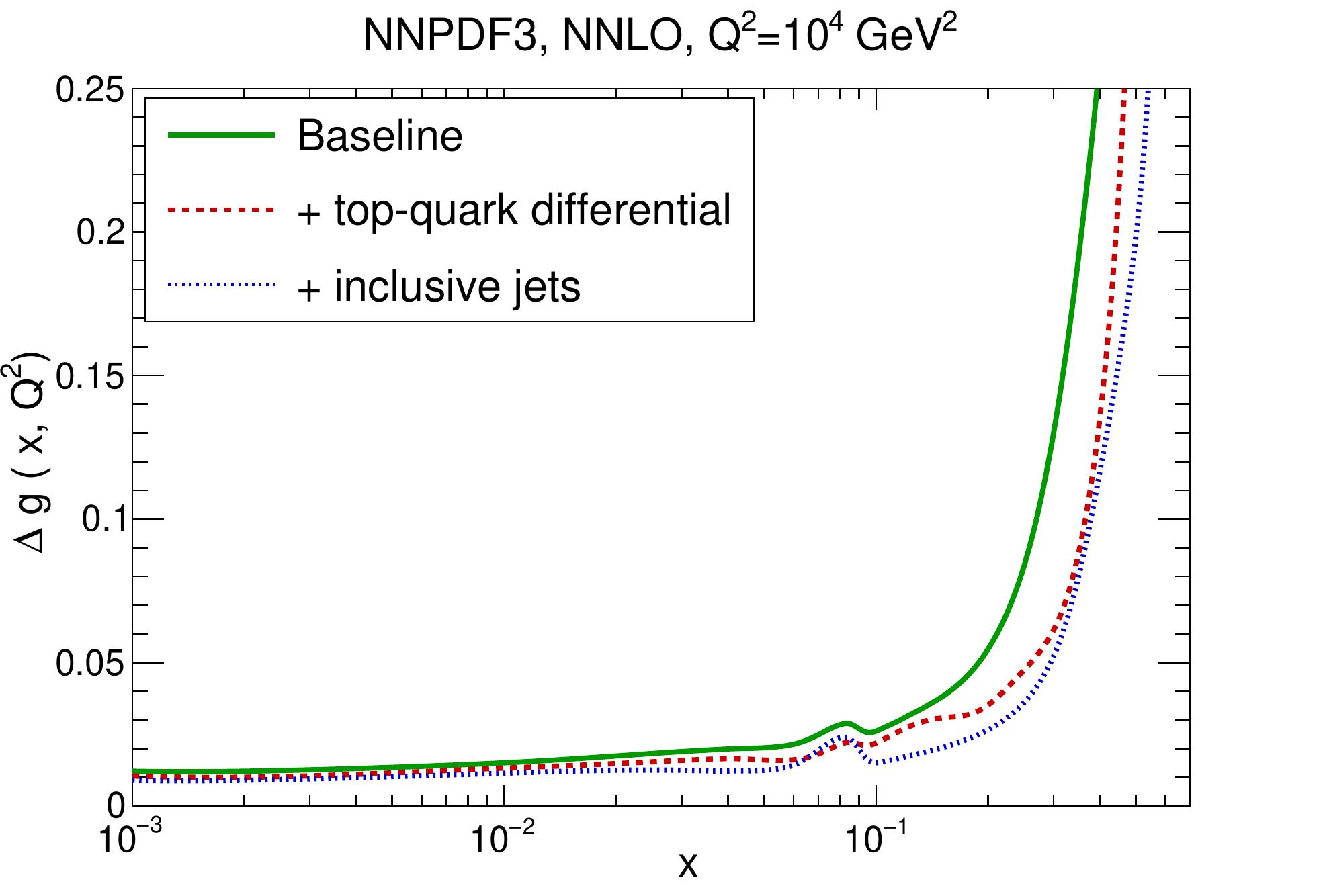}
\includegraphics[scale=0.39]{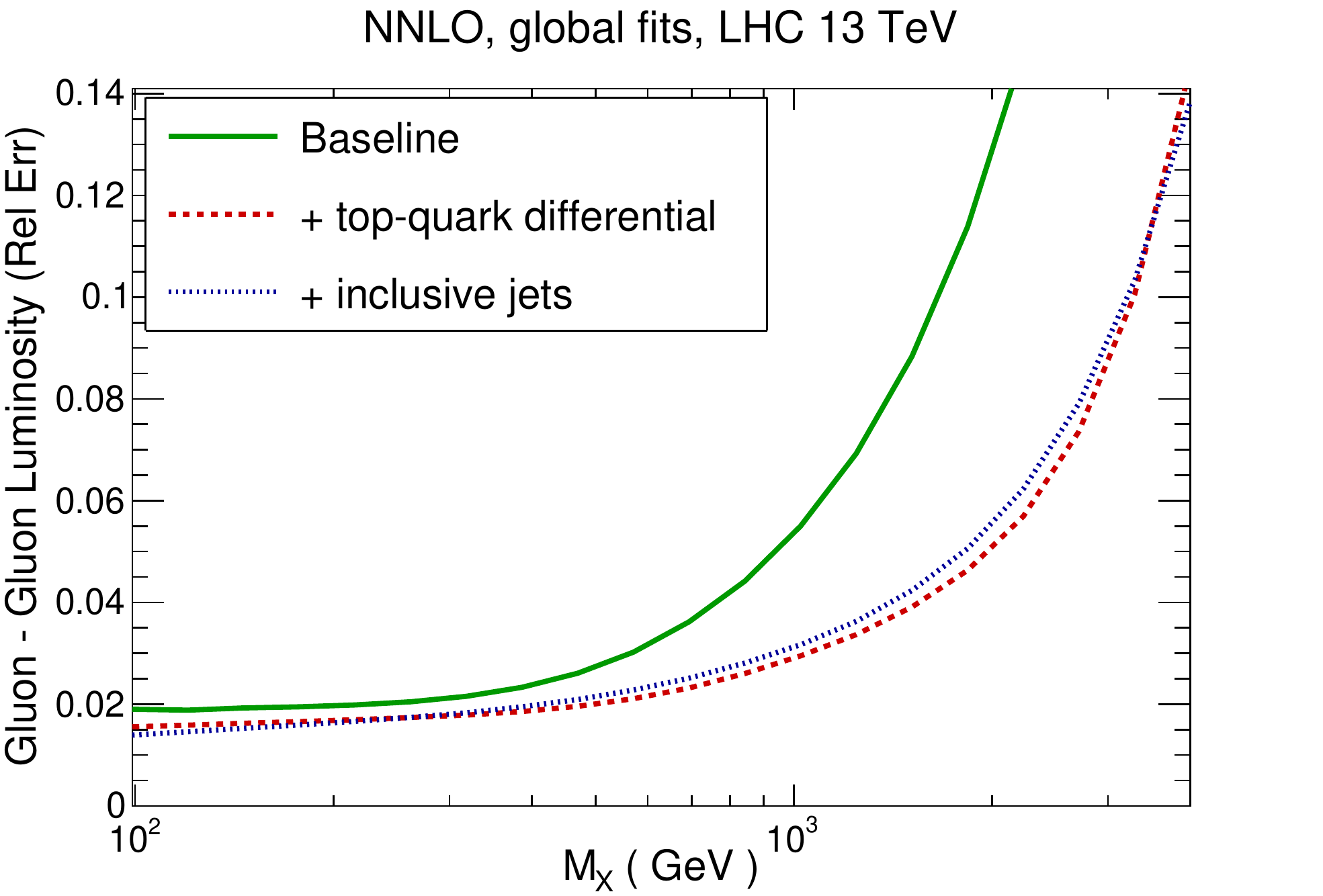}\\
\caption{\small The relative PDF uncertainty on the large-$x$ gluon (left) and on the
gluon-gluon luminosity at large values of $M_X$ (right plot) in the global baseline fit,
compared with the corresponding fits including either top-quark pair differential
measurements or jet production cross-sections.
\label{fig:gluon-ttbar-jets}
}
\end{figure}

In Fig.~\ref{fig:gluon-ttbar-jets}
we show the relative PDF uncertainty on the large-$x$ gluon (left) and on the
$gg$ luminosity at large values of $M_X$ (right) in the global baseline fit,
compared to the corresponding fits including either top-quark pair differential
measurements or jet production cross-sections.
Interestingly, we find that the constraints on the large-$x$ gluon from collider
jet measurements turn out to be similar to those from the LHC top differential data.
This result is particularly remarkable since, as indicated in 
Table~\ref{tab:unc},
the LHC data included in these fits amounts to $N_{\rm dat}=17$ data points
(including the total cross-section
measurements), while the collider jet dataset is substantially larger, $N_{\rm dat}=470$ points.
On the other hand, while jet production is sensitive to the
$qg$ luminosity, and can have a large contribution for $qq$ 
luminosity at high $p_T$,
top quark production is driven instead by the
$gg$ one, which partly explains the comparable impact
on the large-$x$ gluon despite the different number of points.
Note that PDF uncertainties in the $gg$ luminosity at high masses are 
slightly reduced in the fits with top data than in the fits with jet data, 
despite the fact that for the gluon PDF itself the situation is opposite. 
This indicates that the top data induces a somewhat more stringent 
correlation between different $x$ regions of the gluon as compared to jet data, 
thereby leading to smaller fluctuations in the $gg$ luminosity as compared to 
those observed in $g(x,Q^2)$

The results in Fig.~\ref{fig:gluon-ttbar-jets} indicate that the 
constraining power of top-quark pair differential distributions 
at 8 TeV on the large-$x$ gluon is already similar to that of 
collider jet production measurements.
Moreover, accounting for additional measurements at 8 TeV 
in other final states and with boosted kinematics, as well as 
 available and upcoming 13 TeV measurements,
will further strengthen the conclusions and make top-quark
data even more competitive.
On the other hand, Fig.~\ref{fig:gluon-ttbar-jets}
also indicates that ultimate accuracy on the large-$x$ gluon
can only be achieved by means of the simultaneous
inclusion in the global analysis of both top and jet data.

\section{Conclusions and outlook}
\label{sec:summary}

Recent developments in higher-order QCD calculations of LHC processes
require parton distributions with matching accuracy.
PDFs in general, and the limited knowledge of the gluon at large $x$ in particular, 
are often the dominant source of theory uncertainty for top-quark pair 
differential distributions~\cite{Czakon:2016dgf}.
This motivates a self-consistent two-step program where top-quark pair data
from the LHC is first used to constrain the gluon within a NNLO global analysis, 
and then the improved gluon PDF is used to provide updated predictions 
for other top-quark observables or gluon-driven processes.
This way one achieves a significant reduction
of theory uncertainties, improving the prospects of
both precision SM measurements and of BSM searches.

In this work we have quantified
the impact on the large-$x$ gluon of ATLAS and CMS
$\sqrt{s}=8$ TeV measurements of top-quark pair differential distributions 
using the NNPDF framework.
We have shown how differential measurements
can improve PDFs by extending the constraints on the 
gluon obtained from total-cross-section data.
We have also studied the consistency between the
ATLAS and CMS measurements,
identifying certain tension among them.
While the origin of this tension is still not understood,
when the ATLAS and CMS data are included
separately in the fits we find
an improved agreement with NNLO theory for most kinematical distributions.

Our analysis indicates that normalized distributions, supplemented
with the total inclusive cross-sections, have in general
better constraining power than absolute ones.
We have determined a suitable combination of ATLAS and CMS data to
use as input to NNLO fits. This dataset has both high constraining power
and leads to a good agreement between theory and data for the two
experiments.
Based on this analysis, our recommendation concerning the use of 
  LHC top-quark pair production measurements into PDF fits would be to include:
  \begin{itemize}
    \setlength\itemsep{-0.2em}
\item the normalized $y_t$ distribution from ATLAS at $\sqrt{s}=8$ TeV (lepton+jets channel),
\item the normalized $y_{t\bar{t}}$ distribution from CMS at $\sqrt{s}=8$ TeV (lepton+jets channel),
\item total inclusive cross-sections  at $\sqrt{s}=7$, 8 and 13 TeV (all available data).
  \end{itemize}
\noindent
Differential distributions should be included using NNLO theory, while
inclusive cross-sections should be consistently computed at either NNLO+NNLL if fixed scales are used (as is done in {\tt top++}),
or at NNLO if dynamic scales are used.
Future studies should be able to indicate which of the other available top-quark 
differential measurements, in particular in the dilepton channel at 
$\sqrt{s}=8$ TeV and in the lepton+jets and dilepton channels at $\sqrt{s}=13$ TeV, can 
be used to complement the above list.
  
 We have performed a global fit including
 this optimal combination of LHC top-quark data,
 and found that the uncertainty of the large-$x$
gluon is substantially reduced in comparison to the baseline fit.
As an illustration, the PDF error of the $gg$ luminosity at $\sqrt{s}=13$ TeV decreases
from 6\% (11\%) down to 3\% (5\%) at $m_X=1$ TeV (2 TeV),
with an even larger reduction for yet higher values of $m_X$.
We have then shown that the constraints on the large-$x$ gluon provided by top-quark differential data 
are comparable to those derived from inclusive jet
production, despite that the top data is based on a much smaller number of data points.
Our results, therefore, provide a strong motivation for the
inclusion of present and future LHC top-quark pair differential distributions 
into the next generation of PDF analyses.

In this work we have assumed the current world average of the top
mass, $m_t=173.3$ GeV.
However, the spread
among individual $m_t$ measurements leaves open the possibility of a 
future ${\cal O}$(1 GeV) shift in the $m_t$ central value.
Such a shift would impact on the shape of normalized distributions, potentially
affecting the resulting PDF fits.
The optimal combination of LHC top-quark measurements used in our PDF 
fits is based on the $y_t$ and $y_{t\bar{t}}$ distributions, which turn out to be those
with the smallest shape sensitivity to $m_t$ variations.
Therefore, our results should be
robust against future ${\cal O}$(1 GeV) shifts in the central value of $m_t$.

Another important property of the top-quark distributions that we have used as input 
to the PDF fits is that, in general, 
they reduce the risk of a
possible contamination in the gluon from BSM effects in top-quark pair production.
For example, heavy resonances would be kinematically suppressed
in the rapidity distributions, but not in the tails of the
$m_{t\bar{t}}$ and $p_T^t$ ones, where most searches are instead performed.
Therefore, the gluon fitted from data on $y_{t}$ and $y_{t\bar{t}}$  
is safer to be used in BSM searches employing $m_{t\bar{t}}$ and $p_T^t$ distributions.

The studies presented in this work could be extended in several
directions.
First of all, the inclusion of
LHC measurements at 13 TeV with increased
statistics and reduced
systematic uncertainties will improve both the kinematic reach and
the constraining power of top-quark pair data
in PDF fits.
Another avenue worth exploring is to quantify the impact
on the gluon PDF of boosted top quark production, with invariant masses
$m_{t\bar{t}}$ in the multi-TeV
region.
This program requires the inclusion of higher-order
QCD and electroweak effects~\cite{Pagani:2016caq} as well
as the photon PDF.
The latter has been recently calculated in terms of
DIS structure functions~\cite{Manohar:2016nzj},
improving on previous model-independent
estimates~\cite{Ball:2013hta} and reducing the impact of 
photon-initiated contributions in top-quark production.

Another important direction for future work would be
 the exploitation of particle-level distributions in top-quark pair production
 for PDF fits, which however requires
  NNLO calculations with top quark decays.
This would be particularly useful in view of the 
 reported tension between the ATLAS and CMS
 measurements of top-quark level distributions,
 and would remove the need to resort to theory-driven
 extrapolations in top-quark measurements, which
 introduce model dependence with associated 
 uncertainties and biases that are difficult to quantify.

Ultimately, the best constraints on the large-$x$ gluon
will be obtained from the consistent combination of 
inclusive jet and dijet data with top-quark pair production measurements.
The recent NNLO calculation of inclusive jet production~\cite{Currie:2016bfm}
will make it possible in the near future.
This way, it should be possible to achieve an even greater
reduction in the gluon PDF uncertainty, providing a milestone 
contribution to the precision LHC phenomenology program.

\subsection*{Acknowledgements}
We are grateful to Maria Aldaya, Frederic Deliot,
Andrea Gianmanco, Alison Lister, Andreas Meyer, Mark Owen,
Pedro Silva and Francesco Spano for continous assistance with the
ATLAS and CMS inclusive and differential
top-quark measurements as well as for many illuminating
discussions and comments on this project.
We thank David Heymes for related collaboration.
We acknowledge extensive discussions on top quark production and PDF fits
with the members of the NNPDF Collaboration.

M.~C. is
supported in part by grants of the DFG and BMBF.
N.~H. and  J.~R. are
supported by an European Research Council Starting Grant ``PDF4BSM".
The work of A.~M. is supported by the
UK STFC grants ST/L002760/1 and ST/K004883/1 and by the European Research Council Consolidator Grant ``NNLOforLHC2".
E.~R.~N. is supported by the
UK STFC grant ST/M003787/1.

\appendix

\section{On the compatibility between the ATLAS and CMS  data}
\label{sec:compatibility}

One of the most puzzling aspects of the comparison between the NNLO 
theoretical calculations and the
ATLAS and CMS top-quark pair differential cross-sections reported
in this work is the apparent tension between
some of the distributions from the two experiments.
This tension was first observed in the
comparisons between data and theory of
Sect.~\ref{sec:comparison}, and then further quantified by the $\chi^2$ 
analysis from
the HERA-only and global fits in Sect.~\ref{sec:results}.
There we found that achieving a good simultaneous description of several
of the ATLAS and CMS distributions was not possible.

In this appendix we study further the  issue of
the compatibility between the ATLAS and CMS data
by performing additional PDF fits where the two experiments are
included separately.
Our aim is to  disentangle a genuine tension between the ATLAS and CMS 
measurements from alternative explanations of the poor $\chi^2$ reported in 
Sect.~\ref{sec:results}, for instance,
the inadequacy of  NNLO theory to describe the LHC data,
or tension between the top-quark
data and other experiments included in the global fit.
To find out which is the correct explanation, we have repeated the
HERA-only fits, as well as a selection of the global fits,
but now adding the ATLAS and CMS distributions separately.
These fits should
lead to improved $\chi^2$ values as compared to Tables~\ref{tab:chi2val1}
and~\ref{tab:chi2val2}, provided that NNLO QCD
is accurate enough to describe the experimental data, and that, in the case of the global fits,
there are no tensions with other experiments.

\begin{table}[t]
\footnotesize
\centering
\begin{tabular}{lcccccccc}
\toprule
Dataset & \multicolumn{8}{c}{Fit ID}\\
HERA only +&  3* & 4*  & 5*  & 6*  & 7*  & 8*  & 9*  & 10* \\
\midrule
ATLAS\, $d\sigma/dp_T^t$ 
          & {\bf 0.44} & 3.01 & 2.98 & 2.29 
          & 1.06 & 2.66 & 3.76 & 2.60 \\
ATLAS\, $d\sigma/dy_t$ 
          & 1.27 & {\bf 0.47} & 0.75 & 3.01  
          & 2.17 & 0.50 & 0.75 & 1.56 \\
ATLAS\, $d\sigma/dy_{t\bar{t}}$ 
          & 1.75 & 2.12 & {\bf 0.43} & 7.06  
          & 3.95 & 1.77 & 0.44 & 2.32 \\
ATLAS\, $d\sigma/dm_{t\bar{t}}$  
          & 2.37 & 4.03 & 3.85 & {\bf 0.39}  
          & 1.96 & 4.26 & 4.07 & 3.01 \\
ATLAS\, $(1/\sigma)d\sigma/dp_T^t$  
          & 3.06 & 3.86 & 4.10 & 3.16 
          & {\bf 0.60} & 3.40 & 3.55 & 2.15 \\
ATLAS\, $(1/\sigma)d\sigma/dy_t$  
          & 3.55 & 1.09 & 1.60 & 2.65  
          & 16.6 & {\bf 0.75} & 1.25 & 11.3 \\
ATLAS\, $(1/\sigma)d\sigma/dy_{t\bar{t}}$  
          & 3.32 & 5.00 & 2.49 & 4.82 
          & 2.48 & 3.94 & {\bf 0.45} & 14.5 \\
ATLAS\, $(1/\sigma)d\sigma/dm_{t\bar{t}}$  
          & 5.00 & 7.46 & 10.1 & 2.65  
          & 2.61 & 8.29 & 7.13 & {\bf 0.55} \\
ATLAS\, $\sigma_{\rm t\bar{t}}$  
          & 2.76 & 2.60 & 3.96 & 0.10  
          & {\bf 0.99} & {\bf 0.88} & {\bf 1.02} & {\bf 0.71} \\
\bottomrule
\toprule
Dataset & \multicolumn{8}{c}{Fit ID}\\
HERA only +& 3** & 4**  & 5**  & 6**  & 7**  & 8**  & 9**  & 10** \\
\midrule
CMS\, $d\sigma/dp_T^t$ 
          & {\bf 0.82} & 2.96 & 2.36 & 1.83  
          & 0.60 & 2.82 & 3.09 & 1.93 \\
CMS\, $d\sigma/dy_t$ 
          & 3.80 & {\bf 1.30} & 3.05 & 5.17  
          & 6.20 & 1.25 & 3.19 & 6.36 \\
CMS\, $d\sigma/dy_{t\bar{t}}$ 
          & 1.29 & 3.88 & {\bf 0.74} & 2.51  
          & 3.16 & 3.46 & 0.66 & 3.69 \\
CMS\, $d\sigma/dm_{t\bar{t}}$  
          & 3.69 & 5.47 & 3.81 & {\bf 1.28}  
          & 2.67 & 5.50 & 5.13 & 0.78 \\
CMS\, $(1/\sigma)d\sigma/dp_T^t$  
          & 1.46 & 5.62 & 3.28 & 2.13  
          & {\bf 0.85} & 5.67 & 4.83 & 2.29 \\
CMS\, $(1/\sigma)d\sigma/dy_t$  
          & 5.83 & 1.82 & 4.46 & 8.33 
          & 8.98 & {\bf 1.70} & 4.05 & 9.55 \\
CMS\, $(1/\sigma)d\sigma/dy_{t\bar{t}}$  
          & 1.61 & 5.40 & 0.94 & 3.05  
          & 3.71 & 4.95 & {\bf 0.75} & 4.32 \\
CMS\, $(1/\sigma)d\sigma/dm_{t\bar{t}}$  
          & 5.69 & 9.42 & 6.15 & 1.41  
          & 4.10 & 9.41 & 8.90 & {\bf 0.92} \\
CMS\, $\sigma_{\rm t\bar{t}}$  
          & 5.53 & 1.91 & 4.41 & 5.73 
          & {\bf 0.57} & {\bf 0.79} & {\bf 0.70} & {\bf 0.80} \\
\bottomrule
\end{tabular}
\caption{\small Same as Table~\ref{tab:chi2val1} for the fits
where the ATLAS and CMS data are included separately.
}
\label{tab:chi2val1APP}
\end{table}

The results of the fits to HERA data supplemented with ATLAS (CMS)
top-quark pair differential distributions, with CMS (ATLAS) data excluded,
are summarized in the upper (lower) part of Table~\ref{tab:chi2val1APP}.
As in Table~\ref{tab:chi2val1}, we indicate the
values of the $\chi^2$ 
obtained from each fit, with numbers in boldface indicating
the datasets that have been included in each fit.
A comparison with Table~\ref{tab:chi2val1}
shows that when the ATLAS or CMS measurements are included in the
HERA-only fit separately, a better agreement between data and theory
is obtained for all the kinematic
distributions, both absolute and normalized.
Note that this good agreement is not guaranteed: in
several cases, the $\chi^2$ for individual
kinematical distributions is poor unless they are
used in the fit, even when other top-quark
distributions are being fitted.
This behaviour reflects the fact that each 
distribution contains independent information on the gluon PDF.

The inclusion of perturbative corrections beyond NNLO, if they were
known, would be unlikely to improve this picture. First, the size of the 
$\mathcal{C}$-factors in Fig.~\ref{fig:cfacts1}, which can be taken as a 
measure of the perturbative convergence, is approximately the same for all
distributions. This suggests that they all converge with similar rapidity.
Second, we explicitly checked that the size of the ratio of NNLO to NLO
corrections is smaller than the size of the relative uncertainties of the 
data. This suggests that the data will be hardly sensitive to beyond-NNLO 
perturbative corrections within its present precision.

Therefore, when the ATLAS and CMS measurements are included separately 
in the HERA-only fit, we find
no evidence of a tension between data and NNLO theory, 
indicating that the poor
values of $\chi^2$ 
in Table~\ref{tab:chi2val1}
arise from a genuine incompatibility between the two experiments.
One particularly illustrative example of this improvement
is provided by the invariant mass $m_{t\bar{t}}$ normalized
distribution.
In this case, from Table~\ref{tab:chi2val1} we find that, for the
fits including both experiments, the $\chi^2$ is
3.03 and 6.26 for ATLAS and CMS respectively, while
from Table~\ref{tab:chi2val1APP} we see
that the corresponding values are  0.55 and 0.92 when each experiment
is included separately.

In the case of the global fits, in Table~\ref{tab:chi2val2APP}
we show the $\chi^2$ values for a selection of global fits
with the ATLAS and CMS data included separately.
In particular, the fits shown include either the
$m_{t\bar{t}}$ or the $y_{t}$ normalized distributions from one of the two
experiments.
As before, the numbers in boldface indicate the specific 
distributions included in each case.
By comparing with Table~\ref{tab:chi2val2}, we find a picture 
that is qualitatively similar to the case of HERA-only fits.
In general, also in this case improved  $\chi^2$ values are found
when the ATLAS and CMS distributions
are fitted separately.

\begin{table}[t]
\footnotesize
\centering
\begin{tabular}{lcccclcccc}
\toprule
Dataset & \multicolumn{4}{c}{Fit ID} &
Dataset & \multicolumn{4}{c}{Fit ID}\\
Global + & 7 * & 8* & 9 * & 10* &  
Global + & 7** & 8** & 9** & 10**\\
\midrule
ATLAS\, $d\sigma/dp_T^t$                & 2.25 & 2.25 & 2.20 & 2.33 & 
CMS\, $d\sigma/dp_T^t$                  & 2.58 & 3.34 & 2.86 & 2.36 \\
ATLAS\, $d\sigma/dy_t$                  & 1.17 & 0.64 & 0.77 & 1.35 &
CMS\, $d\sigma/dy_t$                    & 3.91 & 2.37 & 3.99 & 7.83 \\
ATLAS\, $d\sigma/dy_{t\bar{t}}$           & 2.82 & 1.80 & 1.22 & 3.19 & 
CMS\, $d\sigma/dy_{t\bar{t}}$             & 0.84 & 2.06 & 0.89 & 3.21 \\
ATLAS\, $d\sigma/dm_{t\bar{t}}$           & 4.12 & 5.12 & 5.06 & 4.03 &
CMS\, $d\sigma/dm_{t\bar{t}}$             & 4.67 & 7.06 & 5.24 & 2.69 \\
ATLAS\, $(1/\sigma)d\sigma/dp_T^t$      & {\bf 2.38} & 5.10 & 4.80 & 2.71 &
CMS\, $(1/\sigma)d\sigma/dp_T^t$        & {\bf 3.03} & 4.10 & 3.40 & 3.73 \\
ATLAS\, $(1/\sigma)d\sigma/dy_t$        & 8.07 & {\bf 1.11} & 1.32 & 9.74 &
CMS\, $(1/\sigma)d\sigma/dy_t$          & 4.58 & {\bf 2.66} & 4.54 & 11.0 \\
ATLAS\, $(1/\sigma)d\sigma/dy_{t\bar{t}}$ & 12.2 & 3.94 & {\bf 2.12} & 14.7 & 
CMS\, $(1/\sigma)d\sigma/dy_{t\bar{t}}$   & 0.93 & 2.85 & {\bf 0.93} & 4.44 \\
ATLAS\, $(1/\sigma)d\sigma/dm_{t\bar{t}}$ & 2.11 & 4.85 & 4.23 & {\bf 1.88} &  
CMS\, $(1/\sigma)d\sigma/dm_{t\bar{t}}$   & 7.92 & 12.1 & 8.93 & {\bf 4.12} \\
ATLAS\, $\sigma_{\rm t\bar{t}}$            & {\bf 0.78} & {\bf 0.11} & {\bf 0.45} & {\bf 0.15} & 
CMS\, $\sigma_{\rm t\bar{t}}$              & {\bf 0.23} & {\bf 0.17} & {\bf 0.51} & {\bf 1.52}\\
\bottomrule
\end{tabular}
\caption{\small Same as Table~\ref{tab:chi2val2}, but for global fits including
normalized distributions only.}
\label{tab:chi2val2APP}
\end{table}

On the other hand, even for the global fits
which include separately the ATLAS and CMS data,
the description of some of the top-quark
distributions is still not optimal.
For instance,
when the two experiments are included simultaneously, we find that the
$\chi^2$ values of the
normalized $y_t$ ($m_{t\bar{t}}$) distributions
 for ATLAS and CMS are 1.68~(2.98) and 4.76~(7.27), respectively.
When each experiment is included separately, the corresponding
$\chi^2$ values are instead 1.11~(1.88) and 2.66~(4.12).
Therefore, while there is a significant improvement,
the $\chi^2$ values tend to be worse than those from the
corresponding HERA-only fits in Table~\ref{tab:chi2val1APP},
especially for CMS data.
This behaviour might be related to a tension between
some top-quark distributions and other
experiments (see Sect.~\ref{sec:fitsettings})
included in the global fit. 
Some additional insight on this issue is provided in 
appendix~\ref{sec:reduceddatasets}.

\section{Fitting top data with non-global datasets}
\label{sec:reduceddatasets}

A second puzzling issue, which we have encountered in this study, is the 
apparent tension between top-pair differential distributions, both absolute
and normalized, and the rest of the dataset included in the global fits. 
Indeed, the quality of the description of top-pair data is significantly 
worse in the global fits (presented in Sec.~\ref{sec:global}) than in the 
HERA-only fits (presented in Sec.~\ref{sec:heraonlyPDFfits}). 
This is apparent by comparing the values of the $\chi^2$ in 
Table~\ref{tab:chi2val2} with their counterparts in Table ~\ref{tab:chi2val1}.
Such a discrepancy persists even when ATLAS and CMS distributions, which
were demonstrated to show some signs of tension in 
appendix~\ref{sec:compatibility}, are included separately in the fits.

In order to identify the data which originates the tension with top-pair
differential distributions in the global fit, we compute the $\chi^2$ for 
the experiments included in the global fits, but not in the HERA-only fits, 
based on the outcome of the HERA-only fits performed in 
appendix~\ref{sec:compatibility}. The results are collected in 
Tables~\ref{tab:HeraotherATLAS}-\ref{tab:HeraotherCMS}, which 
integrate the information contained in Table~\ref{tab:chi2val1APP}.
We have explicitly checked that the $\chi^2$ obtained for HERA 
inclusive and charm reduced cross-sections does not 
significantly change from the $\chi^2$ obtained in a simultaneous fit
to ATLAS and CMS top-pair data (reported in Table~\ref{tab:chi2val1}).

\begin{table}[!t]
\footnotesize
\centering
\begin{tabular}{lcccccccc}
\toprule
Dataset & \multicolumn{8}{c}{Fit ID}\\
        & 3* & 4* & 5* & 6* & 7* & 8* & 9* & 10* \\
\midrule
NMC
        & 9.38 & 9.39 & 8.56 & 8.76 & 11.1 & 10.1 & 9.80 & 9.95 \\
\midrule
SLAC
        & 2.04 & 2.13 & 2.35 & 1.91 & 2.37 & 2.18 & 2.14 & 2.07 \\
\midrule
BCDMS
        & 6.90 & 5.58 & 5.75 & 7.09 & 6.61 & 6.08 & 6.42 & 6.47 \\
\midrule
CHORUS
        & 7.37 & 23.5 & 22.3 & 7.76 & 6.58 & 19.3 & 29.6 & 13.4 \\
\midrule
NuTeV   & 109  & 22.9 & 28.0 & 47.4 & 52.2 & 26.4 & 29.9 & 20.4 \\
\midrule
E866
        & 371  & 440  & 776  & 35.6 & 68.1 & 612  & 163  & 33.4 \\
\midrule
E605    & 1.35 & 3.23 & 4.77 & 2.99 & 1.29 & 4.02 & 8.15 & 3.18 \\
\midrule
CFD $Z$ rapidity
        & 2.97 & 3.37 & 3.76 & 3.71 & 3.95 & 2.90 & 3.60 & 4.16 \\
\midrule
D0 $Z$ rapidity
        & 1.84 & 1.74 & 1.79 & 1.99 & 2.15 & 1.62 & 1.93 & 2.16 \\
\midrule
LHCb $W$, $Z$ rapidity
        & 3.07 & 1.91 & 1.82 & 1.94 & 2.13 & 2.19 & 1.44 & 2.09 \\
\midrule
ATLAS $W$, $Z$ 2010
        & 6.55 & 4.03 & 3.84 & 3.95 & 5.78 & 4.83 & 3.81 & 3.90 \\
ATLAS high-mass DY
        & 1.41 & 1.46 & 1.53 & 1.13 & 1.10 & 1.48 & 1.30 & 1.37 \\
\midrule
CMS $W$ electron asy
        & 17.0 & 24.7 & 10.2 & 29.3 & 12.7 & 36.0 & 4.91 & 15.2 \\
CMS $W$ muon asy
        & 141  & 79.1 & 52.2 & 74.9 & 108  & 90.3 & 68.3 & 77.3 \\
CMS 2D DY 2011
        & 2.19 & 2.22 & 2.41 & 2.08 & 2.18 & 2.26 & 2.16 & 2.12 \\
\bottomrule
\end{tabular}
\caption{\small The $\chi^2$ per data point of the experiments not included in 
the HERA-only fits performed in appendix~\ref{sec:compatibility} (see also 
Table~\ref{tab:chi2val1APP}), computed with the outcome of the corresponding 
HERA-only fits.}
\label{tab:HeraotherATLAS}
\end{table}

\begin{table}[!t]
\footnotesize
\centering
\begin{tabular}{lcccccccc}
\toprule
Dataset & \multicolumn{8}{c}{Fit ID}\\
        & 3** & 4** & 5** & 6** & 7** & 8** & 9** & 10** \\
\midrule
NMC
        & 10.3 & 11.0 & 10.7 & 7.66 & 9.33 & 10.5 & 10.3 & 8.25 \\
\midrule
SLAC
        & 2.45 & 2.46 & 2.25 & 1.85 & 2.42 & 2.47 & 2.37 & 1.94 \\
\midrule
BCDMS
        & 6.90 & 7.07 & 7.91 & 6.26 & 6.71 & 6.10 & 5.97 & 6.19  \\
\midrule
CHORUS
        & 6.29 & 13.5 & 23.4 & 17.1 & 9.28 & 7.48 & 18.0 & 10.1 \\
\midrule
NuTeV
        & 75.9 & 39.6 & 24.5 & 15.8 & 38.2 & 48.3 & 21.9 & 33.2 \\
\midrule
E866
        & 33.2 & 986  & 83.7 & 272  & 268  & 296  & 718  & 695  \\
\midrule
E605
        & 1.35 & 8.53 & 11.9 & 10.2 & 2.85 & 5.42 & 4.00 & 19.0 \\
\midrule
CFD $Z$ rapidity
        & 2.72 & 4.12 & 3.97 & 3.22 & 2.12 & 3.51 & 3.75 & 2.68 \\
\midrule
D0 $Z$ rapidity
        & 1.65 & 2.17 & 2.07 & 1.61 & 2.05 & 1.84 & 1.92 & 1.48 \\
\midrule
LHCb $W$, $Z$ rapidity
        & 2.30 & 1.41 & 1.73 & 1.92 & 2.59 & 1.26 & 1.39 & 1.43 \\
\midrule
ATLAS $W$, $Z$ 2010
        & 4.90 & 5.99 & 4.10 & 3.64 & 4.64 & 5.49 & 4.26 & 2.94 \\
ATLAS high-mass DY
        & 1.83 & 0.93 & 1.28 & 1.45 & 3.04 & 0.95 & 1.04 & 1.26 \\
\midrule
CMS $W$ electron asy
        & 24.4 & 24.4 & 24.0 & 24.1 & 33.2 & 28.8 & 21.7 & 24.1 \\
CMS $W$ muon asy
        & 82.2 & 108  & 71.5 & 59.1 & 82.5 & 104  & 86.2 & 45.2 \\
CMS 2D DY 2011
        & 2.17 & 2.14 & 2.16 & 2.18 & 2.21 & 2.10 & 2.05 & 2.10 \\
\bottomrule
\end{tabular}
\caption{Same as Table~\ref{tab:HeraotherATLAS}, but for the HERA-only fits 
including CMS top-pair differential distributions.}
\label{tab:HeraotherCMS}
\end{table}

From Tables~\ref{tab:HeraotherATLAS}-\ref{tab:HeraotherCMS}, it is apparent 
that the HERA-only fits provide a very poor description of most of the 
data not included in them, especially of those sets which are expected to 
constrain individual quark flavours at large $x$. 
This can be understood since the HERA data provides
information only on the total quark singlet, and only very little on quark 
flavour separation. A rearrangement of the quark flavour
separation is then needed in the global fits to obtain a good 
description of the whole data set. Such an improved description can be 
achieved, as proven by the values of the $\chi^2$ collected in 
Tab.~\ref{tab:chi2val2} (very similar values are obtained in the global 
fits to ATLAS and CMS top-pair data separately). However, this makes the gluon
PDF less flexible in accommodating the top-pair data, which is then 
described worse in the global fits than in the HERA-only fits.
We note that in principle some datasets will require PDFs which are genuinely 
incompatible with HERA plus top data, while some other dataset will not.
For instance, the strange quark can be presumably modified to fit the NuTeV 
data, which is mostly sensitive to it, without causing much change in the fit 
to HERA plus top data.

In order to further investigate this issue, we have performed a series of fits 
to a reduced dataset, which we defined as the global dataset except
all fixed-target DIS data; one top-pair normalized distribution,
separately from ATLAS or from CMS, and the corresponding total cross-section
have been retained on top of the reduced
dataset. In all cases, we have found that the quality of the description of 
the ATLAS and CMS top-pair data significantly improves, with respect to the 
corresponding global fits, and becomes comparable to that obtained in the 
HERA-only fits. The relevant values of the 
$\chi^2$ are similar to those reported in boldface
in Table~\ref{tab:chi2val1APP}. For example, in the case of the
normalized $p_T^t$ ($m_{t\bar{t}}$) distribution, they are $0.79$ ($0.61$) 
for ATLAS (in the fits including the corresponding top-pair distributions 
only from ATLAS, in addition to the reduced data set)
and $0.90$ ($1.01$) for CMS (in the fits including the corresponding top-pair 
distributions only from CMS, in addition to the reduced data set). 
The corresponding values in the case
of the HERA-only fits are, from Table~\ref{tab:chi2val1APP},
$0.60$ ($0.55$) for ATLAS and $0.85$ ($0.92$) for CMS.

These studies indicate that most of the tension between some of the top-pair
differential distributions and the rest of the dataset in the global fits
can be alleviated by removing the fixed-target DIS data. 
Of course, our studies do not indicate whether the tension comes from a 
specific fixed-target DIS experiment, or from the general constraint applied 
by fixed-target DIS data at a particular $x$. In principle, the information 
collected in Table~\ref{tab:chi2val2} would have provided some insight into 
this issue, if the top data had carried enough weight to result in a 
significant deterioration in fit quality to the data in tension.

It is however beyond the scope of this paper to draw a definite conclusion from 
this fact, since in principle all distributions have a similar correlation to 
the underlying gluons and quarks, as shown in 
Figs.\ref{fig:correlations}-\ref{fig:correlations_quarks}.
A comprehensive disclosure of the origin of the tension between top data and 
fixed-target DIS data can be addressed by 
performing a series of additional fits in which one fixed-target DIS
experiment is removed at a time from the global data set.
However, such an exercise will require a non negligible amount of extra 
computational effort and is therefore left to future study.
Future comparisons between theory and LHC data for particle-level observables, 
as well as with the $\sqrt{s}=13$ TeV measurements, might also shed more light 
on this issue.

\bibliography{ttbardiff}

\end{document}